%% file: small_spin.tex
\documentclass[11pt]{article}
\usepackage{jheppub}

\usepackage[usenames,dvipsnames,table]{xcolor}
\usepackage{graphicx,amsmath,amssymb,multirow,array,bm,mathrsfs}
\usepackage{epsf,amsfonts}
\usepackage[numbers,sort&compress]{natbib}

\usepackage{slashed}
\usepackage[yyyymmdd,hhmmss]{datetime}

\usepackage{color}
\usepackage{hyperref}
\hypersetup{hidelinks}
\usepackage{leftidx}
\usepackage{subcaption}
\usepackage[utf8]{inputenc}
\usepackage[section]{placeins}
\usepackage{physics}

\newcommand{\be}{\begin{equation}}
\newcommand{\ee}{\end{equation}}
\newcommand{\ba}{\begin{aligned}}
\newcommand{\ea}{\end{aligned}}
\newcommand{\DeltaP}{\Delta_{\phi}}
\newcommand{\cO}{\mathcal{O}}

\newcommand{\eps}{\varepsilon}
\newcommand{\zb}{\bar{z}}
\def\hb{\bar{h}}

\def\Res{\operatorname*{Res}}

\title{The leading trajectory in the 2+1D Ising CFT}

\author[a]{Simon Caron-Huot,}
\emailAdd{schuot@physics.mcgill.ca}
\author[a]{Yan Gobeil,}
\emailAdd{yan.gobeil@mail.mcgill.ca}
\author[a]{Zahra Zahraee,}
\emailAdd{zr.zahraee@physics.mcgill.ca}

\affiliation[a]{Department of Physics, McGill University, Montr\'{e}al, QC, Canada}

\abstract{
We study the scattering of lumps in the 2+1-dimensional Ising CFT,
indirectly, by analytically continuing its spectrum
using the Lorentzian inversion formula.
We find evidence that the intercept of the model is below unity: $j_*\approx 0.8$,
indicating that scattering is asymptotically transparent corresponding to a negative Lyapunov exponent.
We use as input the precise spectrum obtained from the numerical conformal bootstrap.
We show that the truncated spectrum allows the inversion formula
to reproduce the properties of the spin-two stress tensor to $10^{-4}$ accuracy and
we address the question of whether the spin-0 operators of the model lie on Regge trajectories.
This hypothesis is further supported by analytics in the large-N O(N) model.
Finally, we show that anomalous dimensions of heavy operators
decrease with energy at a rate controlled by $(j_*-1)$, implying regularity of the heavy spectrum.
}

\begin{document}
\maketitle
\flushbottom

\def\be{\begin{equation}}
\def\ee{\end{equation}}
\renewcommand{\abstractname}{\vspace{-\baselineskip}}
\def\nl{\nonumber\\}
\def\ket#1{\big| #1\big\rangle}
\def\l{\langle}
\def\r{\rangle}
\def\MM{\mathcal{M}}
\def\AA{\mathcal{A}}
\def\phib{\bar{\phi}}
\def\eps{\epsilon}

\newmuskip\pFqmuskip
\newcommand*\pFq[6][8]{%
  \begingroup 
  \pFqmuskip=#1mu\relax
  \mathcode`\,=\string"8000
  \begingroup\lccode`\~=`\,
  \lowercase{\endgroup\let~}\pFqcomma
  {}_{#2}F_{#3}{\left[\genfrac..{0pt}{}{#4}{#5};#6\right]}%
  \endgroup
}
\newcommand{\pFqcomma}{\mskip\pFqmuskip}

\section{Introduction}

Recent results in conformal field theories open the possibility to answer questions about
their real time dynamics \cite{Rattazzi:2008pe}.
Thanks to the venerable Wick rotation,
a $d$-dimensional Euclidean CFT is equivalent to  a $(d{-}1)+1$ dimensional one with a time direction,
a map which can be used in either direction.
Intuition about real-time processes, in particular lightcone
and high-energy limits, underlies many recent
analytic results about CFTs \cite{Komargodski:2012ek,Fitzpatrick:2012yx, Alday:2015ota,Simmons-Duffin:2016wlq,Caron-Huot:2017vep}.
On the other hand, the currently most precise numerical results rely on Euclidean methods.
In this paper we attempt to use these numerical results to learn
about the real-time dynamics of the 2+1-dimensional Ising CFT.

\begin{figure}
\centering{\def\svgwidth{8cm}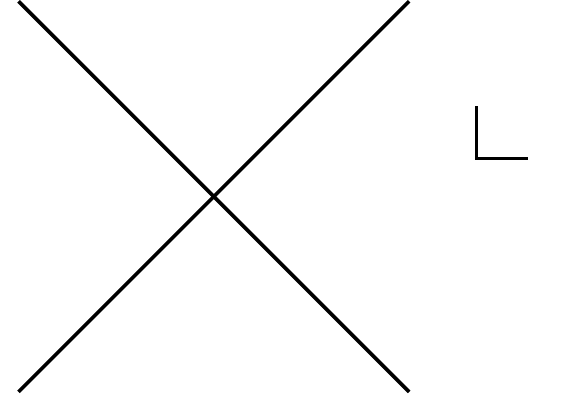}
\caption{Scattering of lumps.
We probe this process by correlating 
four local measurements.
\label{fig:lumps}
}
\end{figure}

The most basic question we would like to answer is whether high-energy scattering
in this theory is transparent or opaque.
A typical physical experiment we have in mind consists of preparing a pair of lumps,
regions of positive spins and some given transverse size, to which we apply a large boost,
see fig.~\ref{fig:lumps}. Do the lumps pass through each other, or disperse into oblivion?

This information is contained in the Regge limit of the
four-point correlator of the spin field $\sigma$:
\be
\langle \sigma_4\sigma_2\sigma_3\sigma_1 \rangle - \langle\sigma_2\sigma_1\rangle\langle\sigma_4\sigma_3\rangle
\propto G-1 \,.
\ee
CFT four-point functions depend on two real variables.
The Regge limit is attained by applying a large relative boost between $(1,2)$ and $(3,4)$, 
and the two variables represent respectively the boost factor and impact parameter (see \cite{Costa:2012cb}).
Note that correlations near the lightcone are sensitive to the physics of scattering,
even if operators 1 and 2 are spacelike-separated, as depicted in the figure.~\ref{fig:lumps}.
In the Regge limit we expect exponential dependence in the boost:
\be
\lim_{{\rm boost}\ \eta\to\infty} (G-1) \propto e^{(j_*-1)\eta} \label{big boost}
\ee
The Regge intercept $j_*$ is interpreted as the spin of an effective Reggeized particle exchanged between the lumps.
It is known that $j_*\leq 1$ in any unitary CFT \cite{Caron-Huot:2017vep}: the correlator is asymptotically bounded.
Let us review some general expectations about this limit, whose study has a long history;
see \cite{Amaldi:2015jhq} for a historical overview combining experiment and theory.

There is a sharp qualitative distinction between the cases of $j_*<1$ and $j_*=1$:
scattering can be asymptotically transparent or opaque, respectively.
Transparency, for the 2+1-dimensional Ising CFT,
would mean that highly boosted lumps pass through each other without interacting.
This is to be contrasted with the strong interactions (which is
not a CFT, but high-energy forward scattering can be discussed very generally) where protons appear increasingly opaque
at high energies, as witnessed experimentally by the increasing elastic and inelastic cross sections. 
For CFTs, since high-energy scattering can be viewed as late-time evolution in Rindler space,
the question of transparency versus opacity
is equivalent to the question of whether the theory thermalizes on Rindler space,
transparency meaning lack of thermalization, see section 9 of  \cite{Murugan:2017eto}.

In the opaque case, one may expect to see transient exponential growth: $j_{*}^{\rm transient}>1$.
The bound of chaos states that, in any unitary theory, $j_{*}^{\rm transient}\leq 2$ \cite{Maldacena:2015waa}
(the precise interpretation of $(j_*^{\rm transient}-1)$ as a Lyapunov exponent is discussed in Appendix A there).
A plausible, standard scenario
is that in theories with $j_*^{\rm transient}>1$, opacity is first reached at small impact parameters, leading
to a black disc whose radius grows with energy.

Examples of theories with either type of behavior exist.  To give a few examples,
two-dimensional minimal models have $j_*<1$ \cite{Caputa:2016tgt}.
Holographic CFTs have $j_*\approx 2$, thus nearly saturating the chaos bound.
This reflects graviton exchange in the dual gravitational picture \cite{Brower:2006ea, Shenker:2013pqa}.
This feature is also observed in the Sachdev-Ye-Kitaev (SYK) model \cite{Maldacena:2016hyu}.
The two-dimensional version of SYK studied in \cite{Murugan:2017eto}
also exhibits transient growth, albeit with a non-maximal exponent: $1<j_*^{\rm transient}<2$.
For QCD, fits of hadron scattering data suggest a  Pomeron intercept $j_*^{\rm transient} \approx 1.09$ \cite{Menon:2013vka}.
In the perturbative regime of a four-dimensional perturbative gauge theory,
the famous BFKL analysis shows, very generally, that $j_*^{\rm transient}\approx 1+\cO(\alpha_s)>1$.
This conclusion is very much tied to the gluon having spin 1, and
in weakly coupled quantum field theories without vector bosons, we thus generally expect $j_*<1$. 

In general, it can be difficult to determine which category a given theory fits in.
On the one hand, for many purposes the 3D Ising CFT appears to be ``close" to a perturbative
scalar theory, as witnessed by the successful approximation of the spectrum by
the $\varepsilon$-expansion around $d=4$.
This would suggest transparency.
On the other hand, the theory lacks a tuneable coupling constant, and it is unclear whether $d=3$ is ``close enough" to $d=4$
for this argument to be convincing.
The main goal of this paper is to study this question using numerical data on the excited states of the 3D Ising CFT.
We will find numerical evidence that the model is indeed in the category of vector-free perturbative theories:
$j_*\approx 0.8<1$. 
More generally, we study the leading Regge trajectory of the model, $j_*(\Delta)$, which reduces to the intercept
at a special point. 

Let us briefly review the 3D Ising CFT.  It is characterized by having $\mathbb{Z}_2$ symmetry
and only two relevant operators, called $\sigma$ and $\epsilon$, which are respectively odd and even under $\mathbb{Z}_2$. 
(From the bootstrap perspective, this \emph{defines} the theory.)
They are scalars and their scaling dimensions and OPE coefficients have been determined using Monte Carlo simulations and the numerical bootstrap. The best numbers available, including the errors, are
\be\ba & \Delta_{\sigma}=0.5181489(10)\ ,\qquad \Delta_{\epsilon}=1.412625(10)\ , \\ & f_{\sigma\sigma\epsilon}=1.0518537(41)\ , \qquad f_{\epsilon\epsilon\epsilon}=1.532435(19)\ . \ea\ee
The spectrum also contains multi-twist families made out of these operators.
The leading trajectory can be viewed as composites $[\sigma\sigma]_{0,J}$ (defined below),
which can be identified unambigously for $J\geq 2$.
We will also study the leading odd trajectory $[\sigma\epsilon]_{0,J}$.
We will benefit from the high-accuracy data and analysis for these families and other operators
reported in \citep{Simmons-Duffin:2016wlq}.
Note that the stress tensor is a member of the leading trajectory: $T=[\sigma\sigma]_{0,2}$.

The intercept is but one point on a continuous curve, $j_*\equiv j_*(\Delta=\frac32)$.
Our main tool to study the full curve will be the Lorentzian inversion formula,
which reconstructs the dimensions and OPE coefficients in one channel, as a continuous function of spin,
in terms of operators exchanged in cross-channels.
A well-understood large-spin expansion has been known to work well even down to $J=2$
\cite{Komargodski:2012ek,Fitzpatrick:2012yx,Alday:2015ota,Simmons-Duffin:2016wlq}.
We will approach the intercept in two steps: First we will establish numerical convergence of the operator sum
by reproducing the known stress tensor dimension, $\Delta_*(2)=3$, to high accuracy.
From there we will gradually reduce $\Delta$.

In addition to the leading trajectory and intercept, we will discuss the following
simple question: do the spin-0 operators $\sigma$ and $\epsilon$ lie on Regge trajectories?
We will find numerical evidence that $\sigma$ lies on the shadow of the leading odd trajectory.
Within the $\varepsilon$-expansion, it is know that $\epsilon$ resides on an analytically continued branch of the leading trajectory \cite{Alday:2017zzv}; we will find that the 3D numerical data, while compatible with this hypothesis,
does not add to the evidence.

This paper is organized as follows. In section \ref{sect:inversion} we first review the Lorentzian inversion formula and how it can be used to extract low twist CFT data in a general theory. Then we focus on three dimensions and discuss the accurate numerical evaluation of the formula using the method of dimensional reduction \citep{Albayrak:2019gnz, Hogervorst:2016hal}, comparing the result with large-spin approximations.
In section \ref{sect:ising-equal}  and \ref{sect:ising-unequal}  we apply this method to the 3d Ising model.
We specifically work on the $\expval{\sigma\sigma\sigma\sigma}$ and the $\expval{\sigma\epsilon\epsilon\sigma}$ correlators to extract data for spin-two operators in the $[\sigma\sigma]_0$ and $[\sigma\epsilon]_0$ families,
the intercept, and we describe our attempts to reach spin 0.
In section \ref{sect:discussion} we discuss various aspects relevant to the interpretation of the results.
In subsection \ref{sect:CF} we comment on general distinctions between theories with intercept above and below 1.
In subsection \ref{sect:critical} we analyze the leading trajectories of the critical $O(N)$ model at large $N$
in both bilinears $[\phi_i\phi_j]_0$ and $[\phi_i S]_0$, which we will find to be analogous to the $[\sigma\sigma]_0$ and
 $[\sigma\epsilon]_0$ trajectories in 3D Ising.
Finally, in subsection \ref{sect:predictions} we propose a formula which
relates the intercept being less than unity to regularity of the heavy spectrum.
Section \ref{sect:conclusions} contains our concluding remarks.
Appendix~\ref{app:integrals} contains explicit inversion integrals utilized in the paper,
appendix~\ref{app:cheap} records compact approximations to large-spin operators,
and appendix~\ref{app:convexity} provides a short proof that the leading trajectory is convex.
{\bf Note:} While this work was being completed, closed related methods have been applied to the critical
O(2) model \cite{Liu:2020tpf}.

\section{Review of Lorentzian inversion formula and other ingredients}
\label{sect:inversion}

We consider a correlation function of 4 scalar primary operators
\be \expval{\phi_1(x_1)\phi_2(x_2)\phi_3(x_3)\phi_4(x_4)}=\frac{1}{(x_{12}^2)^{\frac{\Delta_1+\Delta_2}{2}}
(x_{34}^2)^{\frac{\Delta_3+\Delta_4}{2}}}\left(\frac{x_{14}^2}{x_{24}^2}\right)^a\left(\frac{x_{14}^2}{x_{13}^2}\right)^b\mathcal{G}(z,\zb), \ee
where $x_{ij}=x_i-x_j$, $a=\frac{\Delta_2-\Delta_1}{2}$, $b=\frac{\Delta_3-\Delta_4}{2}$ and the conformal cross-ratios $z$, $\zb$ are defined as
\be\label{eq:cross-ratios} z\zb=\frac{x_{12}^2 x_{34}^2}{x_{13}^2 x_{24}^2}\ ,
\qquad (1-z)(1-\zb)=\frac{x_{23}^2 x_{14}^2}{x_{13}^2 x_{24}^2} 
\,. \ee
We can use the OPE for operators 1 and 2 together and for operators 3 and 4 to decompose the correlator in s-channel conformal blocks as follows
\be \mathcal{G}(z,\zb)=\sum_{\Delta,J}f_{12\cO}f_{43\cO}G^{(a,b)}_{\Delta,J}(z,\zb) \, ,\ee
where $f_{ij\cO}$ is the OPE coefficient and $G^{(a,b)}_{\Delta,J}(z,\zb)$ is the s-channel conformal block,
which resums the contribution of the primary with dimension $\Delta$ and spin $J$ and all of its descendants. More explicitly, conformal blocks are special functions that are the eigenfunctions of the quadratic and quartic Casimir equation. They admit closed form in even spacetime dimension, for instance the conformal blocks in $d=2$ dimension can be written as follows:
\begin{equation}
\label{eq:2d-block}
G_{\Delta,J}^{(a,b)}(z,\zb)=\frac{k^{(a,b)}_{\Delta-J}(z)k^{(a,b)}_{\Delta+J}(\zb)+k^{(a,b)}_{\Delta+J}(z)k^{(a,b)}_{\Delta-J}(\zb)}{1+\delta_{J,0}}\ ,\qquad (d=2)\ ,
\end{equation}
where $k$ is the hypergeometric function
\be\label{2F1}
k^{(a,b)}_{\beta}(z)=z^{\beta/2}\ {}_2F_1(\beta/2+a,\beta/2+b,\beta,z).
\ee
Here $\beta=\Delta+J$ is the conformal spin. We also introduce $\tau=\Delta-J$ which we refer to as twist.
We will use $\Delta,J,\beta,\tau$ in different contexts to specify the operators in the spectrum.

Conformal blocks do not accept a simple
closed-form expression in odd spacetime dimensions and one must resort to various approximations.
The main approximation we will use is to write 3d blocks as sums over 2d blocks \cite{Hogervorst:2016hal},
as reviewed in appendices~\ref{appen:dimensional-reduction}. 

In general, we normalize the blocks so that: $\lim_{z\ll \zb\ll 1} G_{\Delta,J}(z,\zb)= z^{\frac{\tau}{2}}\zb^{\frac{\beta}{2}}$. The leading term as $z,\zb\to0$ is then
\be\begin{aligned}\label{limit block}
\lim_{z,\zb\to0} G^{(a,b)}_{\Delta,J}(z,\zb) &= (z\zb)^{\frac{\Delta}{2}} C_J\left(\frac{z+\zb}{2\sqrt{z\zb}}\right),
\\
\mbox{where } C_J(\eta) &\equiv \frac{\Gamma\big(\tfrac{d-2}{2}\big)\Gamma(J+d-2)}{\Gamma(d-2)\Gamma\big(J+\tfrac{d-2}{2}\big)}
 \ {}_2F_1\big(-J,j+d-2,\tfrac{d-1}{2},\tfrac{1-\eta}{2}\big)\,.
\end{aligned}\ee
The function $C_J(\eta)$ is a multiple of a Gegenbauer function, $C_J(\eta)\propto C^{(d/2-1)}_J(\eta)$,
satisfying $\lim_{\eta\to\infty} C_J(\eta)= (2\eta)^J$.

 \subsection{Lorentzian inversion formula}
 
A good starting point for analytics is an alternate form of the OPE
in which one integrates over operators dimensions along the principal series, but
where spin is still discrete and needs to be summed over:
\begin{equation}
\label{eq:ope-principal-series}
\mathcal{G}(z,\zb)=\sum_{J=0}^{\infty}\int^{d/2+i\infty}_{{d/2-i\infty}}\frac{d\Delta}{2\pi i}c(\Delta,J)F^{(a,b)}_{\Delta,J}(z,\zb)+
\mbox{(non-normalizable)}\ .
\end{equation}
where non-normalizable modes describe operators with $\Delta< \frac{d}{2}$ (which includes, notably, the identity).
The CFT data is then encoded in the poles of the analytic function $c(\Delta,J)$. These are located at the position of the physical operators in the conformal block expansion, and the residues give the OPE coefficients in the following way
\be f_{12\cO}f_{43\cO}=-\Res_{\Delta'=\Delta}c(\Delta',J) \label{f from poles}\,.\ee
The harmonic function $F_{\Delta,J}$ is a single-valued, shadow-symmetric combination of the block and its shadow \cite{SimmonsDuffin:2012uy}:
\begin{equation}
F_{\Delta,J}^{(a,b)}(z,\zb)=\frac{1}{2}\left(G^{(a,b)}_{\Delta,J}(z,\zb)+\frac{K^{(a,b)}_{d-\Delta,J}}{K^{(a,b)}_{\Delta,J}}G^{(a,b)}_{d-\Delta,J}(z,\zb)\right),
\end{equation}
where 
\be K^{(a,b)}_{\Delta,J}=\frac{\Gamma(\Delta-1)}{\Gamma\left(\Delta-\frac{d}{2}\right)}\kappa^{(a,b)}_{\Delta+J}, \qquad \kappa^{(a,b)}_{\beta}=\frac{\Gamma\left(\frac{\beta}{2}-a\right)\Gamma\left(\frac{\beta}{2}+a\right)\Gamma\left(\frac{\beta}{2}-b\right)\Gamma\left(\frac{\beta}{2}+b\right)}{2\pi^2\Gamma(\beta-1)\Gamma(\beta)} \,.
\ee

The functions $F_{\Delta,J}$ satisfy an orthogonality relation which allows to read off the OPE data from the correlator
(Euclidean inversion formula).
The Lorentzian inversion formula reconstructs the same data using less information, the double discontinuity \cite{Caron-Huot:2017vep,Simmons-Duffin:2017nub,Kravchuk:2018htv}:
\be
\label{eq:inversion-integral} c^t(\Delta,J)=\frac{\kappa^{(a,b)}_{\Delta+J}}{4}\int_0^1\int_0^1dzd\zb\,\mu(z,\zb)\,G^{(-a,-b)}_{J+d-1,\Delta+1-d}(z,\zb)\,\text{dDisc}\left[\mathcal{G}(z,\zb)\right], \ee
which needs to be summed with the contribution of the u-channel to give the full coefficients:
\be
\label{t-pm-u} c(\Delta,J)=c^t(\Delta,J)+(-1)^J c^u(\Delta,J). \ee
$c^u(\Delta,J)$ is obtained from $c^t(\Delta,J)$ by exchanging the operators 1 and 2. The measure is\footnote{
This form agrees with ref.~\cite{Caron-Huot:2017vep} using the identity: $G^{(-a,-b)}(z,\zb)_{J,\Delta} = ((1-z)(1-\zb))^{a+b}
G^{(a,b)}(z,\zb)_{J,\Delta}$.}
\be \mu(z,\zb)= \frac{1}{(z\zb)^2} \left|\frac{z-\zb}{z\zb}\right|^{d-2} \,.\ee
The double discontinuity of the correlator is a certain linear combination of analytic continuation around $\zb=1$
which computes the expectation value of a double commutator \cite{Caron-Huot:2017vep}:
\be\label{eq:dDisc} \mbox{dDisc}\left[\mathcal{G}(z,\zb)\right]=\cos[\pi(a+b)]\mathcal{G}(z,\zb)-\frac{1}{2}e^{i\pi(a+b)}\mathcal{G}^{\circlearrowright}(z\zb)-\frac{1}{2}e^{-i\pi(a+b)}\mathcal{G}^{\circlearrowleft}(z\zb) \,.\ee
This combination  is positive definite and is analogous to the absorptive (imaginary) part of a scattering amplitude.
The coefficient function which comes out of Lorentzian inversion is automatically shadow-symmetric:
\begin{equation}
\label{eq:ope-shadow}
\frac{c(\Delta,J)}{K^{(a,b)}_{\Delta,J}}=\frac{c(d-\Delta,J)}{K^{(a,b)}_{d-\Delta,J}}\ .
\end{equation}

Along the principal series, ${\rm Re}(\Delta)=\frac{d}{2}$,
convergence of the Lorentzian inversion formula is controlled by the Regge limit
and requires $J>j_*$, where $j_*$ is the intercept defined in eq.~(\ref{big boost}).
The Lorentzian inversion formula then manifests the analyticity of the spectrum in spin,
giving an organizing principle for operators of spin $J> j_*$. In a unitary CFT this always
include all operators with $J\geq 2$.

In this paper we will focus on the leading trajectory $\Delta_*(J)$. For
integer $J\geq 2$, eq.~(\ref{f from poles}) shows that this is the pole
nearest to the principal series, and positivity of the integrand (for real $\Delta$)
implies convergence in a strip: $d-\Delta_*(J)< {\rm Re}(\Delta)< \Delta_*(J)$.
For non-integer spin, the leading trajectory answers a simple question: when does the integral (\ref{eq:inversion-integral}) converge?

The resulting smooth curve can also be parametrized as $j_*(\Delta)$, where convergence is
satisfied for $J>j_*(\Delta)$. With this definition, it is easy to show using
positivity of the dDisc that $j_*(\Delta)$ is a real and convex function, see \cite{Costa:2017twz},
extending the integer-spin convexity proved in ref.~\cite{Komargodski:2012ek,Kundu:2020gkz}
using Nachtmann's theorem (we give an alternative proof in  appendix \ref{app:convexity}).
Since the leading trajectory is also manifestly shadow-symmetrical, $j_*(\Delta)=j_*(d-\Delta)$, it follows that
its minimum, the \emph{intercept} must be at the symmetrical point: $j_*\equiv j_*(\frac{d}{2})$.
This agrees with the physical definition of the intercept given earlier in eq.~(\ref{big boost}) since
convergence of the Lorentzian inversion formula at that point is controlled by the Regge limit of correlator. 

Two practical points worth mentioning are as follows: first, at the cost of a factor of two,
we can restrict the integration range in the Lorentzian inversion formula to $z<\zb$.
Second, when we are interested in extracting s-channel data from poles at $\Delta> d/2$,
we can decompose the s-channel blocks follows and restrict to the first term, $g^{\text{pure}}$
(defined to have a single tower of terms in the limit $0\ll z\ll \zb\ll1$):
\begin{equation}
\label{eq:block-decomp}
G^{(a,b)}_{J+d-1,\Delta+1-d}(z,\zb)=g^{(a,b)\text{pure}}_{J+d-1,\Delta+1-d}(z,\zb)+\frac{\Gamma(\Delta-1)\Gamma(-\Delta+\frac{d}{2})}{\Gamma(\Delta-\frac{d}{2})\Gamma(-(\Delta+1-d))}g^{(a,b)\text{pure}}_{J+d-1,-\Delta+1}(z,\zb).
\end{equation}
This is because the second term does not contribute to the poles $\Delta> d/2$ and just ensure shadow symmetry. However, when one is interested in extracting data in the vicinity of the intercept
$\Delta\sim d/2$ (as we will do in section \ref{sec:intercept}) one cannot use this decomposition.

\subsection{Extracting low-twist OPE data}

For generic $\beta$, we will only be interested in the poles and residues of $c(z,\beta)$,
which will come from the small-$z$ limit of the integrand.  In particular, for the pole corresponding
to the operator of smallest twist it suffices to take $z\to 0$ in the inversion formula (\ref{eq:inversion-integral}):
\be c^t(\Delta,J)=\int_0^1 \frac{dz}{2z}z^{-\frac{\tau}{2}}C^t(z,\beta) + \mbox{(collinear descendents)}\, \label{ct coll} ,\ee
where we have defined a generating function $C^t(z,\beta)$:
\be \label{gen func}
 C^t(z,\beta)=\int_z^1 \frac{d\zb}{\zb^2}\kappa_{\beta}k^{(-a,-b)}_{\beta}(\zb)\ \mbox{dDisc}\left[\mathcal{G}(z,\zb)\right].
\ee
The generating function encodes the spectrum through power laws. More precisely,
if we expand it as
\be\label{eq:generating_sum} C(z,\beta)=\sum_m C_m(\beta) z^{\frac{\tau_m}{2}}\ , \ee
it is easy to see that each power will produce in eq.~(\ref{ct coll}) a pole $\frac{C_m}{\tau_m-\tau}$,
interpreted as an operator of twist $\tau_m$ following eq.~(\ref{f from poles}).
Note that eq.~(\ref{ct coll}) does not subtract collinear descendants, since neglected
corrections by integer powers of $z$ affect the residues at shifted valued
$\tau_m+2, \tau_m+4, \ldots$.  In this paper we will restrict ourselves to the lowest twist family for which $m=0$ and collinear descendants play no role (for more information on how to treat higher twist families see \cite{Alday:2017vkk}).

The exponents $\tau_m$ give the twist of operators in the spectrum.
The coefficient $C_m(\beta)$ are related to OPE coefficients
but the relation is slightly subtle because
eq.~(\ref{f from poles}) requires residues computed at constant spin $J$,
whereas $C_m(\beta)$ gives residues at constant $\beta$. 
The exact relation includes a Jacobian \cite{Alday:2015eya,Simmons-Duffin:2016wlq,Caron-Huot:2017vep}:
\be\label{eq:jacobian} f_{12\cO}f_{34\cO}=\left(1-\dv{\tau_m(\beta)}{\beta}\right)^{-1}\eval{C_m(\beta)}_{\beta-\tau=2J}. \ee

Our strategy to gain knowledge from the inversion formula is to insert the t-channel decomposition of the correlator (obtained from the s-channel by swapping operators 1 and 3, equivalent to fusing 1 with 4 and 2 with 3) into the generating function
in eq.~(\ref{gen func}):
\be
\label{eq:generating_int} C^t(z,\beta)=\sum_{\Delta',J'}f_{14\cO'}f_{23\cO'}c^{\Delta_1\cdots \Delta_4}_{\Delta',J'}(z,\beta)
\ee
where
\be \label{generating_int c}
c^{\Delta_1\cdots \Delta_4}_{\Delta',J'}(z,\beta)
\equiv \int_z^1 \frac{d\zb}{\zb^2}
\kappa^{(a,b)}_{\beta}k_{\beta}^{(-a,-b)}(\zb)\mbox{dDisc}\left[\frac{(z\zb)^{\frac{\Delta_1+\Delta_2}{2}}}{[(1{-}z)(1{-}\zb)]^{\frac{\Delta_2+\Delta_3}{2}}}G^{(a',b')}_{\Delta',J'}(1{-}z,1{-}\zb)\right]. 
\ee
where $a'=\frac{\Delta_2-\Delta_3}{2}$, $b'=\frac{\Delta_1-\Delta_4}{2}$.
To obtain the $u$-channel generating function $C^u$
we interchange $\Delta_1$ with $\Delta_2$ wherever they appear in eq.~(\ref{eq:generating_int}).

We only know closed-form expressions for the integral (\ref{eq:generating_int}) in special cases.
An important one is the $t$-channel identity (which can only be
physically realized when $\Delta_2=\Delta_3$ and $\Delta_1=\Delta_4$):
\be\label{eq:generating_id} C(z,\beta)\Big|_{t-\rm channel\ identity}
= \frac{z^{\frac{\Delta_1+\Delta_2}{2}}}{(1-z)^{\frac{\Delta_2+\Delta_3}{2}}}
\mathcal{I}^{(\Delta_1,\Delta_2)}(\beta),
\ee
where
\be\ba\label{eq:identity-contribution} \mathcal{I}^{(\Delta_1,\Delta_2)}(\beta) & =\int_0^1 \frac{d\zb}{\zb^2}
\kappa_{\beta}k_{\beta}^{(-a,-a)}(\zb)\mbox{dDisc}\left[\frac{\zb^{\frac{\Delta_1+\Delta_2}{2}}}{(1-\zb)^{\Delta_2}}\right] \\ & =\frac{\Gamma\left(\frac{\beta+\Delta_1-\Delta_2}{2}\right)\Gamma\left(\frac{\beta+\Delta_2-\Delta_1}{2}\right)}{\Gamma\left(\Delta_1\right)\Gamma\left(\Delta_2\right)\Gamma(\beta-1)}\frac{\Gamma\left(\frac{\beta+\Delta_1+\Delta_2}{2}-1\right)}{\Gamma\left(\frac{\beta-\Delta_1-\Delta_2}{2}+1\right)}. \ea\ee


Another important analytic result pertains to the case where we insert a two-dimensional block
in the $t$-channel, where the integral reduces to a 1d 6j symbol
\cite{Karateev:2018oml,Cardona:2018dov,Sleight:2018epi}.  Although we are interested in $d=3$, we will
use this result by writing the 3d blocks as sums over 2d blocks. A brief review on this method of dimensional reduction is given in appendix~\ref{appen:dimensional-reduction}, here we quote the final result.
A conformal block in $d$ dimension can be expanded as a sum over $(d-1)$-dimensional blocks as shown in
eq.~(\ref{eq:2d-blocks-expansion}):
 \begin{equation} \label{3dto2d}
G^{(a,b)}_{\Delta,J}(z,\zb;d)=\sum \mathcal{A}^{(a,b)}_{m,n}(\Delta,J)G^{(a,b)}_{\Delta+m,J-n}(z,\zb;d{-}1)
\qquad 0\leq n\leq J, m=0,1,2...
\end{equation}
where the coefficients $\mathcal{A}$ are determined recursively using the Casimir differential equation.
Inserting this expansion into the inversion integral  \eqref{eq:generating_int} for each individual block,
we obtain three-dimensional inversion integrals as a sum over two-dimensional inversion integrals given
analytically in eq.~(\ref{eq:2d-inversion})
\begin{equation}
\label{eq:2dto3d-inversion}
c^{\Delta_1\cdots \Delta_4}_{\Delta',J'}(z,\beta;d{=}3)=
\sum_{m=0} \sum_{n=0}^{J'} \mathcal{A}^{(a',b')}_{m,n}(\Delta',J')
c^{\Delta_1\cdots \Delta_4}_{\Delta'+m,J'-n}(z,\beta;d{=}2).
\end{equation}
In practice, the error in this method can be reduced to zero by including as many terms as needed,
since exponential convergence rapidly sets in.
In our analysis, for most values of $J'$ and $\Delta'$, going to $m=15$ is more than enough.
The analytic formulas for 2d integrals requires the $\zb$ integral in eq.~(\ref{eq:generating_int}) to have lower bound
$0$ instead of $z$; the difference is negligible compared to other sources of error as long as we are away from the intercept (however for completeness in fig.~\ref{sec:intercept} the twist of stress-tensor resulting  from the inversion formula with $\bar{z}> z$ is given as well).
In section \ref{sec:intercept} we will use a different approximation when we approach the intercept.


\subsection{OPE data from truncated spectrum}\label{sect:procedure}

In theory, the exponents in eq.~(\ref{eq:generating_sum}) are obtained by analyzing
the $z\to 0$ limit of the $t$-channel sum (\ref{eq:generating_int}).
In particular, the leading twist and OPE coefficient is equal to the following limit:
\begin{equation}
\label{eq:procedure}
\tau(\beta)=\text{lim}_{z\rightarrow 0}\frac{2z\partial_zC(z,\beta)}{C(z,\beta)}, \qquad C(\beta)=\text{lim}_{z\rightarrow 0}\frac{C(z,\beta)}{z^{\frac{z\partial_zC(z,\beta)}{C(z,\beta)} }}.
\end{equation}
In practice, however, we only have access to a finite number of terms in the $t$-channel sum,
which prevents us from taking $z$ arbitrarily small: the limit
lies at the boundary of convergence of the $t$-channel OPE.
In some previous analyses, a convenient value of $z$ was simply fixed \cite{Simmons-Duffin:2016wlq,Albayrak:2019gnz};
one could also consider fitting the $z$-dependence to a power law.

Our approach in this paper will be to plot the quantity $\tau=\frac{2z\partial_zC(z,\beta)}{C(z,\beta)}$ as a function of $z$
and look for a plateau.
If $z$ is chosen too large, we expect errors due to neglected higher-twist $s$-channel operators,
while if $z$ is too small, we expect truncation errors from the $t$-channel sum.
By restricting our attention to a plateau region we hope to simultaneously minimize both sources of error
(in addition to getting rough error estimates).

\subsection{Relation to large spin expansion and its accuracy}
\label{subsec:L-S Expansion}


The effectiveness of the analytical bootstrap in extracting large spin data is well established \cite{Komargodski:2012ek, Alday:2015ota, Fitzpatrick:2012yx}.
These results are typically obtained by considering the double lightcone limit $(z,\zb)\to(0,1)$, where one argues
that singularities in the $t$-channel must be reproduced by large-spin-tails in the $s$-channel.
Let us briefly review how these results relate to the formulas just reviewed, highlighting the ways in which our analysis will differ.


The basic physical picture is that large spin (or large $\beta$) pushes the integral (\ref{eq:generating_int})
to the $\zb\to 1$ corner, due to the shape of the $k$-function (defined in eq.~(\ref{2F1})).
At sufficiently large spin, the $t$-channel identity given in eq.~(\ref{eq:generating_id}) thus dominates.
Its particular $z$ dependence then implies the existence of so-called
double-twist families $[\phi_1\phi_2]_{n,J}$, where $J$ denotes the spin of operator. Their twist
approximates the naive dimensional analysis \cite{Fitzpatrick:2012yx, Komargodski:2012ek,Kaviraj:2015cxa}:
\begin{equation}
\tau_{[\phi_1\phi_2]_{n,J}}= \Delta_{\phi_1}+\Delta_{\phi_2}+2n+\gamma,
\end{equation}
where $\gamma$, the anomalous dimension of the operator, vanishes in the large-spin limit.
From the scaling relation $\sqrt{1-\zb}\propto \frac{1}{\beta}$, one can easily see
that the correction, due to exchange of $t$-channel operator of lowest nontrivial twist $\tau_{\rm min}$,
decays like
\begin{equation}
\gamma(n,\ell)\simeq \frac{\gamma_n}{J^{\tau_{\rm min}}}\ .
\end{equation}
These corrections are found by analyzing the collinear $\zb\to 1$ limit of $t$-channel blocks.
\be \label{collinear limit block}
\lim_{\zb\to 1} G^{(a',b')}_{\Delta',J'}(1-\zb,1-z)\rightarrow(1-\zb)^{\frac{\Delta'-J'}{2}}k^{(a',b')}_{\Delta'+J'}(1-z)+O(1-\zb)^2.
\ee
One can readily see that using this approximation
the inversion integral over $\zb$ is greatly simplified and can be performed analytically;
the result is particularly simple if one expands instead in powers of $\frac{1-\zb}{\zb}$, see eq.~(\ref{eq:identity-contribution}).
For the leading trajectory in the Ising CFT, taking the coefficient of $\log z$ in eq.~(\ref{collinear limit block}) then gives
a simple pocket-book formula for large-spin corrections:
\be
\tau_{[\sigma\sigma]_{0,J}} \approx 2\Delta_\sigma - \sum_{\cO=\epsilon,T}
\frac{2 \lambda_{\sigma\sigma\cO}^2\Gamma(\Delta_\sigma)^2}{\Gamma\big(\Delta_\sigma-\tfrac{\tau_{\cO}}{2}\big)^2} \frac{\Gamma(\Delta_{\cO}+J_{\cO})}{\Gamma\big(\tfrac{\Delta_{\cO}+J_{\cO}}{2}\big)^2}
\left(\frac{2}{\beta-1}\right)^{\tau_{\cO}}\,. \label{large spin gamma}
\ee
We have chosen $\beta\mapsto \beta-1$ as our expansion parameter since it is manifest from the exact formulas
that the series proceeds in even powers of $\beta-1$ (the so-called reciprocity relation \cite{Basso:2006nk,Alday:2015eya}).
Numerically, this formula works surprisingly well down to spin $J=2$, although the errors are somewhat difficult
to estimate a priori.  Analogous formulas for OPE coefficients and for $[\sigma\sigma]_1$ and $[\epsilon\epsilon]_0$ trajectories are recorded in appendix \ref{app:cheap}.

\begin{figure}
\centering
  \includegraphics[width=0.75\linewidth]{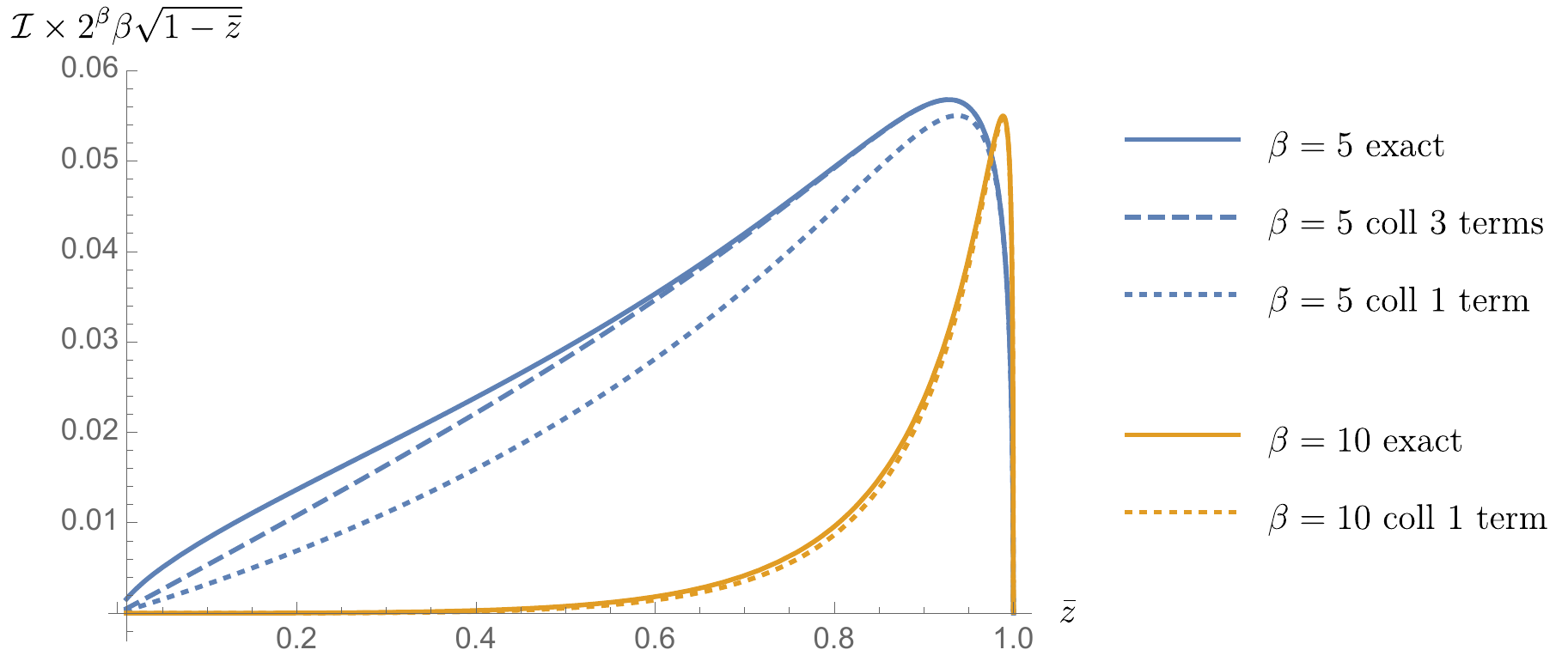}
  \caption{Integrand of the inversion formula (\ref{eq:generating_int}) with $z=10^{-2}$,
  comparing the exact cross-channel block for $\epsilon$-exchange (using the 3d to 2d series)
  with its collinear series in $(1-\zb)$. 
 For $\beta=5$, the collinear limit in eq.~(\ref{collinear limit block}) approximates the dominant region well but
 underestimates the integrand at small $\zb$.  At larger values of $\beta$ this region becomes negligible.
 Note that we rescaled the integrand by $2^\beta \beta\sqrt{1-\zb}$ to make features more visible.
  \label{fig:integrand}}
\end{figure}

One could try to estimate errors by studying further $1/J$ corrections, but let us report here on
a more straightforward exercise which is
to simply compare the approximation in eq.~(\ref{collinear limit block}) with the actual
integrand entering the Lorentzian inversion formula.
We do this here for a single $t$-channel block ($\epsilon$),
reserving discussion of the sum over blocks to the next section.
The $\zb$-dependence of the integrand of eq.~(\ref{eq:generating_int}) comes from two factors:
the $s$-channel block $k_\beta(\zb)$ and the $t$-channel block.
Their product is shown for $\epsilon$-exchange in fig.~\ref{fig:integrand} for $\beta=5$ and $\beta=10$.
We show three approximations for the $t$-channel blocks: the
3d to 2d expansion (called ``exact" since terms beyond the third one are invisible on the plot),
and the collinear series in powers of $(1-\zb)$ whose first two terms are given for reference in eq.~(\ref{eq:coll-block}).

One can see that at the larger value $\beta=10$ (corresponding roughly to $J=4$)
even the leading collinear term matches the integrand very well.
At the integrated level, it underestimates the $\epsilon$ contribution by only $3\%$.
For $\beta=5$ (corresponding to the stress tensor) the error is up to $10\%$,
coming mostly from the region of $\zb$ not close to 1.
Because this multiplies a small coefficient, this corresponds to a $4\times 10^{-3}$ error on the twist of the stress tensor.  
Replacing the power of $(1-\zb)$ by $\frac{1-\zb}{\zb}$ produces similar numbers.  
Including up to the third term in the $(1-\zb)$ series reduces the errors
to $0.3\%$ and $2\%$, respectively.
Since our goal will be to do much better than this, we need to employ formulas which are
valid at all $\zb$. We achieve this in the next section by using the 3d to 2d expansion of blocks,
which converges much faster.

\section{Leading twist $Z_2$-even family}\label{sect:ising-equal}

In this section we study the leading Regge trajectory in the 3D Ising model (i.e., $[\sigma\sigma]_0$ family) by applying the formalism developed in section~\ref{sect:inversion}, focusing on low spins.

We begin with the stress tensor, which is the spin 2 operator of $[\sigma\sigma]_0$ family.  This will serve as a benchmark case: while its twist is known from conservation laws, reproducing it as an infinite sum over cross-channel operators is nontrivial.  We show that in order to get best control over systematic errors, we need to work at significantly lower values of $z$ than previously considered, which is feasible by including subleading families ($[\sigma\sigma]_1$ and $[\epsilon\epsilon]_0$) in the cross-channel and resumming their large-spin tails. We obtain both the twist and OPE coefficient of the stress-tensor with error at the level $10^{-4}$ which  is compatible with the error in the numerical data used in inversion formula (see the numerical error for spin 6 operator of $[\epsilon\epsilon]_0$ family in table.~4 of \cite{Simmons-Duffin:2016wlq}). 


In subsection \ref{sec:intercept} we apply a similar analysis to the intercept, where convergence in the s-channel twist will be found to be slower. 
In subsection \ref{sec:Epsilon} we briefly discuss attempts to reach the operator $\epsilon$ itself through an analytic continuation
of the trajectory close to the intercept.

\subsection{Recovering the stress-tensor}
\label{sec:Benchmark: Stress-Tensor}

Here we calculate the twist and the OPE coefficient of the stress-energy tensor in the $[\sigma\sigma]_0$ family using the inversion formula. Since it is a conserved operator it saturates the generic spin unitarity bound:
\begin{equation}
\label{eq:unitarity}
\Delta=J+d-2.
\end{equation}
So when we are in 3 dimension stress tensor has scaling dimension 3, twist $\tau=1$ and conformal spin $\beta=5$. This operator belong to $[\sigma\sigma]_0$ family with asymptotic twist at large spin equal to
$2\Delta\sigma\approx 1.036298$.  We will thus be looking for a small negative anomalous dimension:
$\gamma_T\approx -0.036298$.

We compute the generating function in eq.~\eqref{eq:generating_int}, analyzing the effect of various truncations of the operators included in the t-channel. The truncations cause errors in the twist and OPE coefficient of the stress tensor. The conformal blocks are computed using the 3d to 2d series mentioned in eq.~(\ref{3dto2d}).
This method was also used in \cite{Albayrak:2019gnz}.
However the main difference between our approach and theirs is that we do not expand the argument of the double discontinuity in eq.~\eqref{eq:generating_int} at small $z$.  This is important and allows us to sum over infinite families.
This is because the $z\rightarrow 0$ limit and OPE sums do not generally commute (beyond the leading trajectory) and
retaining the full $z$ dependence is necessary to accurately cut off the sums at $J'\sim 1/\sqrt{z}$ (see \cite{Simmons-Duffin:2016wlq,Caron-Huot:2017vep}). The 3d to 2d series converges rapidly and we
always include sufficiently many terms that we can neglect this source of error,
effectively treating the blocks as ``exact".
We will comment on the small-$z$ expansion for the exchange of a single operator and its region of validity later in this section.

 
 
\begin{figure}
    \begin{subfigure}[t]{0.5\textwidth}
        \centering
  \includegraphics[width=\linewidth]{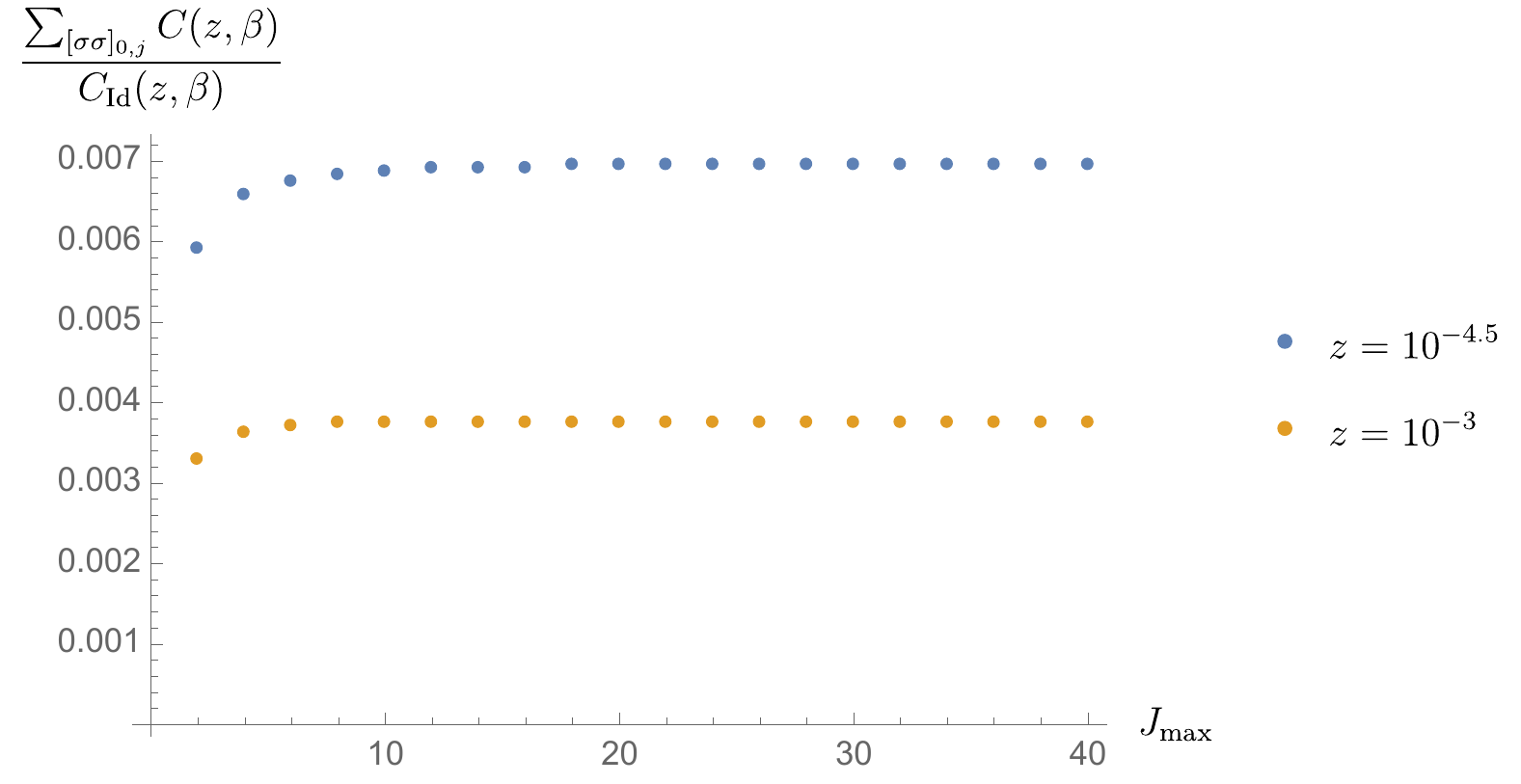}
   \caption{$[\sigma\sigma]_0$ family}
    \end{subfigure}
    \begin{subfigure}[t]{0.5\textwidth}
        \centering
        \includegraphics[width=1\linewidth]{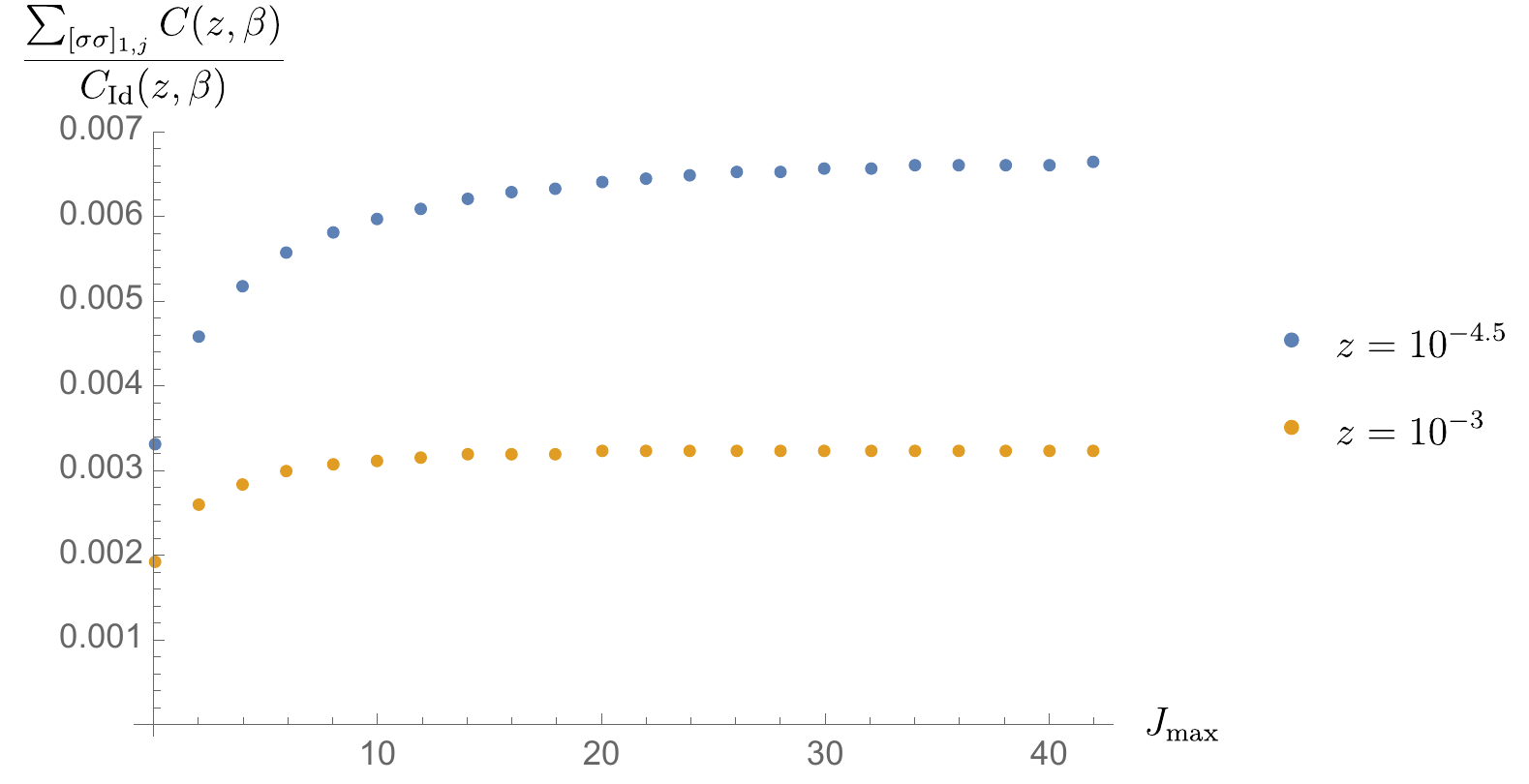}
      \caption{$[\sigma\sigma]_1$ family}
    \end{subfigure}\\[1ex]
	\begin{subfigure}[t]{0.5\textwidth}
   \centering
  \includegraphics[width=1\linewidth]{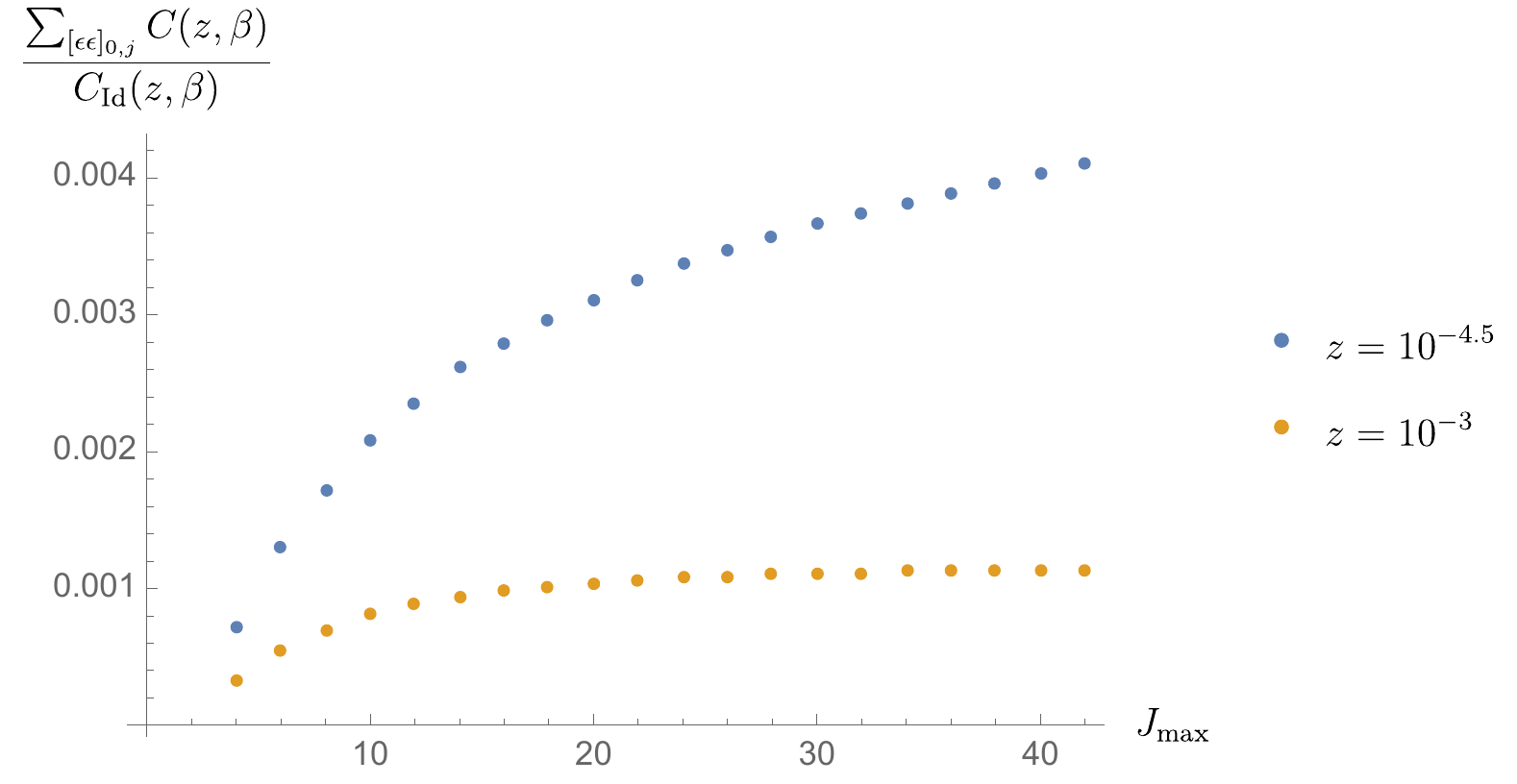}
  \caption{$[\epsilon\epsilon]_0$ family}
    \end{subfigure}
    	\begin{subfigure}[t]{0.5\textwidth}
   \centering
     \includegraphics[width=1\linewidth]{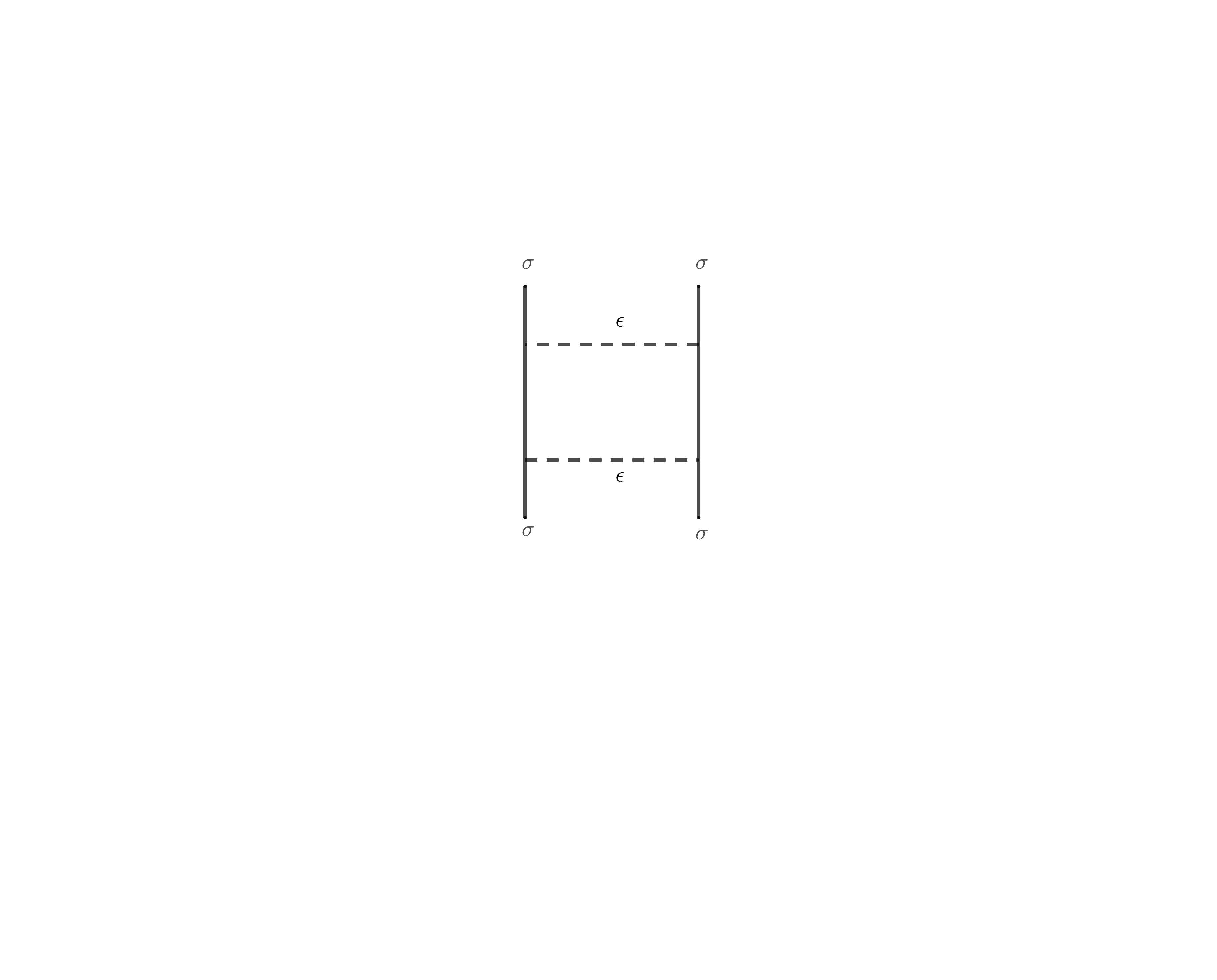}
  \caption{large-spin diagram for $[\epsilon\epsilon]_0$ exchange}
  \end{subfigure}
 
        \caption{Partial sum contributions to $C(z,\beta=5$) for the first three leading families as a function of the maximal spin.
        The $[\sigma\sigma]_0$ family converges for any $z$, but other families, especially $[\epsilon\epsilon]_0$,
        are very sensitive to large spins.  Figure (d) shows how the exchange of $[\epsilon\epsilon]_0$ can contribute as $\log^2 z$
    \label{fig:partial_st}
    }
\end{figure}

First, let us show the effect of various $t$-channel truncations
to twist $\tau= 2\frac{z\partial_z C(z,\beta)}{C(z,\beta)}$, evaluated at $\beta=5$ and for various values of $z$.
This will illustrate the relative importance of subleading families depending on the value of $z$.
The data used for exchanged operators is from tables in the appendix of\cite{Simmons-Duffin:2016wlq}
 \begin{figure}[!htb]
  \centering
  \includegraphics[width=1\linewidth]{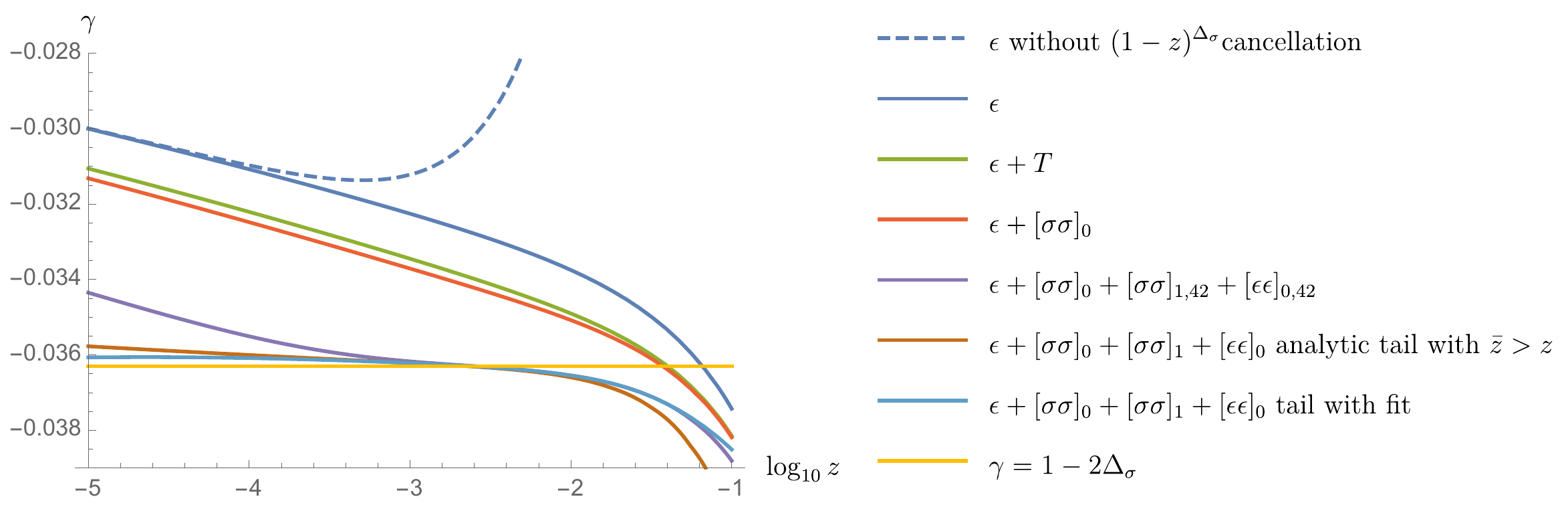}
  \caption{The effect of different $t$-channel truncations in eq.~(\ref{eq:generating_int}) on the extracted
stress-tensor anomalous dimension $\gamma_T=\tau_T-2\Delta_\phi$.  Including more operators enables us to reach lower values of $z$ where we find a stable $z$-independent plateau.
  \label{fig:st-twist1}}
\end{figure}
We see in fig.~\ref{fig:st-twist1} that including the
subleading twist families $[\sigma\sigma]_1$ and $[\epsilon\epsilon]_0$ significantly improves the result. In addition, we also observe once we multiply $C(z,\beta)$ with $(1-z)^{\Delta_{\sigma}}$, the plateau extends to larger value of $z$ for each of truncation in the cross channel. This is illustrated  for $\epsilon$ exchange in fig.~\ref{fig:st-twist1}.
The reason for this is that the region of large $z$ is contaminated both by collinear descendants of the leading trajectory,
and by higher-twist trajectories. Since $\Delta_\sigma$ is close to the unitarity bound, the latter are much smaller,
and the former are largely canceled by multiplying by the mean-field factor $(1-z)^{\Delta_{\sigma}}$.
One salient point is that for sufficiently small $z$ all
the curves eventually depart from the correct stress-tensor twist.
This is because at smaller $z$ the OPE converges more slowly and operators with both higher spin and twist need to be included.
It is also apparent that summing up to a finite spin cutoff is not sufficient to create a plateau, since
operators with quite large spin
are also important (see fig.~\ref{fig:partial_st}). By resumming the higher spin tails in all the families in two different independent way, we bypass this problem and produce two  curves which as can be seen are the  most  successful curves in reproducing the anomalous dimension of stress-tensor, both in terms of accuracy and stability, given the publicly available numerical data.
We will discuss how we performed these resummations and obtained the stable curves in the paragraphs below. 

Fig.~\ref{fig:partial_st} shows that high spin tails are strongly needed for $[\epsilon\epsilon]_0$ family,
which has not yet converged at spin 40 at the shown values of $z$.
In addition, it is also required for $[\sigma\sigma]_1$ family even though this family is converging more quickly.
This is in contrast with $[\sigma\sigma]_0$ family for which the sum is fully convergent for the whole region we are considering, so resumming its high spin operator has a negligible effect as can be seen in fig.~\ref{fig:partial_st}.

A strong tail for the $[\epsilon\epsilon]_0$ family was to be expected physically since, in large-spin perturbation theory,
single-$\epsilon$ exchange produces a $\log(z)$ term which accounts for a large fraction of the stress tensor anomalous dimension (see eq.~(\ref{large spin gamma})).
One may thus expect the box-like diagram in fig.~\ref{fig:partial_st}(d) to contain a $\tfrac12\log^2z$ term which
exponentiates single-$\epsilon$ exchange. Since such a $\log^2z$ cannot be generated by individual $t$-channel operators
and must necessarily come from a large-spin tail \cite{Fitzpatrick:2015qma,Simmons-Duffin:2016wlq}.

\begin{figure}
   \centering
  \includegraphics[width=1\linewidth]{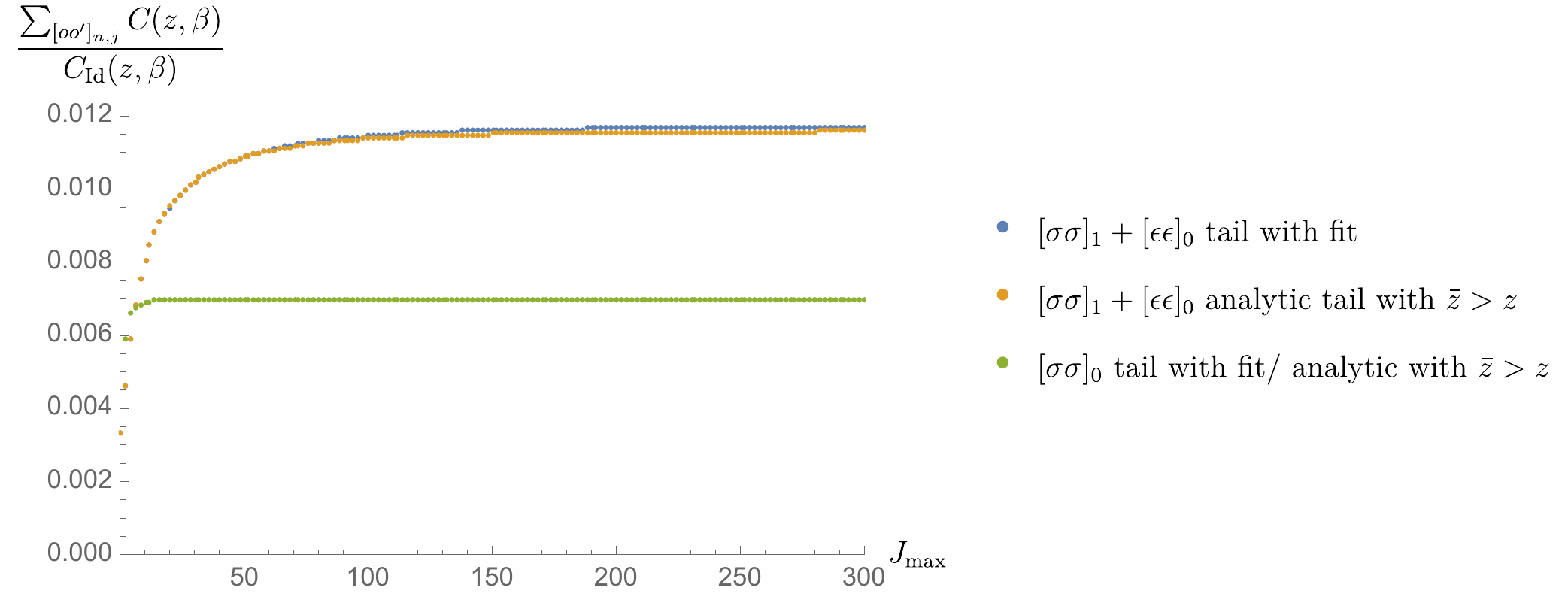}
    \caption{Partial sums contributing to $C(z{=}10^{-4.5},\beta{=}5)$, extrapolated to very large spins.}
    \label{fig:partial_st-hs}
\end{figure}

\begin{figure}[!htb]
  \centering
  \includegraphics[width=1\linewidth]{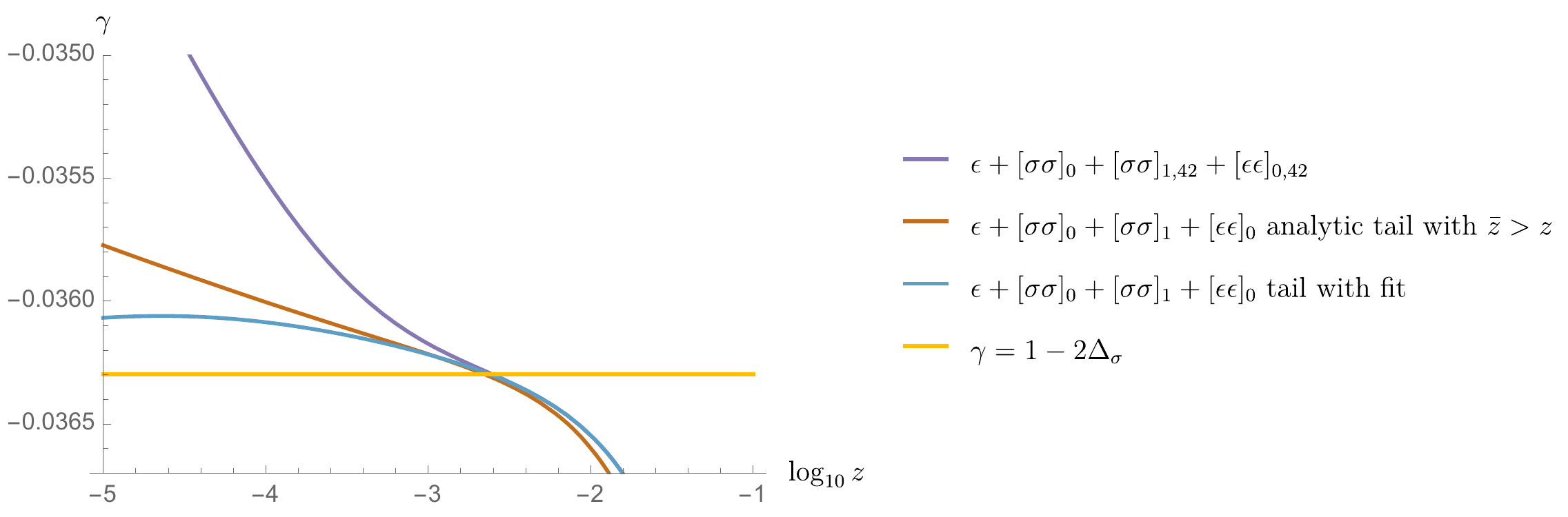}
  \caption{Magnified version of fig.~\ref{fig:st-twist1}, showing the importance of resumming large-spin tails to
  extract a $z$-independent stress-tensor twist. The difference between the two tails which is of order $10^{-4}$ and is a result of the difference between blue and orange curves in fig.~\ref{fig:partial_st-hs}, gives an estimate of the error for the tail contribution)}
  \label{fig:hs_st_zoom}
\end{figure}

Accurate numerical data for large spin tails at spins $J>40$ is unavailable. 
In principle one could obtain good analytic approximations for this region using large spin perturbation theory,
where the couplings between $\sigma\sigma$ and $[\epsilon\epsilon]_0$ follow from mixed correlators,
as was also studied in  \cite{Simmons-Duffin:2016wlq}. We derive this analytic approximation with the inversion integrals having lower bound $z$ (this is done by subtracting 0 to $z$ integral from the $2d$ integrals in eq.~\ref{eq:2d-inversion}. See appendix.~\ref{app:Lorentzian inversion in 2d} for more details). However, in the spirit of the data-driven approach followed
in this paper, we adopt a simple modelling and fitting strategy as well. We will compare the two methods for estimation of the error in the tail.
We do not directly fit the OPE data of large-spin operators (twist and OPE coefficients) since all we will need is
their contribution to the $\zb$-integrated double-discontinuity. The important advantage of this method is that the difficulties related to performing the inversion formula for large spin blocks, such as the expensive $3d$ to $2d$ expansion is avoided. The cross channel block has a simple $z$-dependence, as can be seen from
the large spin and small $z$ expansion (see appendix~A in \cite{Fitzpatrick:2012yx}):
\begin{equation}
 G_{\tau',\ell'}(1-\zb,1-z)\rightarrow k_{\beta}(1-z)v^{\tau'/2}F(\tau',\zb)+O(1/\sqrt{\ell},1/\sqrt{z}),
\end{equation}
where $k_{\beta}(1-z)=(1-z)^{\beta/2}{}_2F_1(\beta/2+a,\beta/2/+b,\beta,1-z)$ was defined in eq.~(\ref{2F1})
and the function $F$ won't be important to us.
The prime notation is associated with the cross-channel operators.
The next thing we want to estimate is the large spin expansion of the OPE coefficients. At large spin the OPE coefficient of operators converges to their values in mean field theory \cite{Fitzpatrick:2012yx},
$f_{\sigma\sigma[\sigma\sigma]_{n,\ell}}^2\sim [1+(-1)^\ell]P^{\Delta_\sigma}_{2\Delta_{\sigma}+2n,\ell}$ where
\begin{equation} \label{MFT}
P^{\Delta_\sigma}_{2\Delta_{\sigma}+2n,\ell}\equiv
\frac{(\Delta_{\sigma}-d/2+1)^2_n(\Delta_{\sigma})^2_{n+\ell}}{n !\ell !(\ell{+}d/2)_n(2\Delta_{\sigma}{+}n{-}d{+}1)_n(2\Delta_{\sigma}{+}2n{+}\ell{-}1)_{\ell}(2\Delta_{\sigma}{+}n{+}\ell{-}d/2)_n}
  \end{equation}
and where $(a)_b$ denotes the Pochhammer symbol which is defined as $(a)_b\equiv\frac{\Gamma(a+b)}{\Gamma(a)}$. This allows us to estimate the inversion formula in \eqref{eq:generating_int}  at large spin as follows:
 \begin{equation}
 \label{fit}
 C_{\beta}(z,\beta')\sim C'P^{\Delta_\sigma}_{2\Delta_{\sigma}+2n,\ell}\beta'^{-\tau'}k_{\beta'}(1-z)\frac{z^{\frac{\Delta_1+\Delta_2}{2}}}{(1-z)^{\frac{\Delta_2+\Delta_3}{2}}}
\end{equation}
This fitting can be done for all of the three families included in the cross-channel. However as just explained it only has an impact for subleading families.
To account for the mixing of the $[\sigma\sigma]_1$ and $[\epsilon\epsilon]_0$ family, we will fit their sum for each spin to the function given in the RHS of eq.~\ref{fit}, for which we will have $n=1$ and $\beta=2J+\tau_{[\epsilon \epsilon]_0}$. The fit is done with data having spin 18 and higher for both families . The parameter of the fit and their covariance matrix for each family is given as follows:
\begin{align}
C'_{[\sigma\sigma]_1+[\epsilon\epsilon]_0}&=0.009575\ , & \tau'_{[\sigma\sigma]_1+[\epsilon\epsilon]_0}&=0.9226\ , & COV&=
\begin{pmatrix}
0.00002378 &0.0006099\\
0.0006099& 0.01568
\end{pmatrix}.
\end{align}
Note that $\tau'\sim 2\Delta_{\sigma}$ which is the expected value. Using this fit we can estimate the contribution of the tail of the aforementioned families to $C(z,\beta)$ and make the sum over families a convergent sum as can be seen in the fig.~\ref{fig:partial_st-hs}.
By adding Gaussian noise to the fitted values $C'$ and $\tau'$ with the quoted covariance,
we find branching curves results in an error of order $10^{-4}$ in the final answer for the stress-tensor twist (the size of the branching is compatible with the difference between the analytic tail and the fit manifested in a magnified version fig.~\ref{fig:partial_st-hs} ). We note that the fit uncertainties are highly correlated, and varying
$C'$ and $\tau'$ independently would generate very different curves!
We see that once the contribution of higher spin is taken into account in the sum over families, the flatness of the curve and thus in dependency of $\gamma_T$ from $z$ is restored (see fig.~\ref{fig:hs_st_zoom} and fig.~\ref{fig:st-twist1}). 
 
The order $\sim 10^{-4}$ error in the tail at small $z$  in  fig.~\ref{fig:hs_st_zoom} is comparable with the error on the numerical data of the $[\epsilon\epsilon]_0$ family (see for instance the data for spin 6 operator table.~4 in \cite{Simmons-Duffin:2016wlq}). So one cannot hope to reduce the error just by improving the tail. In addition, the error on the numerical data as opposed to the error caused by truncation do not have a definite sign, this makes it impossible to give an upper bound on the result.

The final issue to be addressed is the range of $z$ accessible to us.
As predicted we cannot get arbitrarily close to zero as the error of the tails would eventually become significant (see fig.~\ref{fig:hs_st_zoom}). 
However, we are allowed to take any $z$ such that in the range depicted in fig.~\ref{fig:hs_st_zoom}. We choose the decade in which the curve has the smallest error (in terms of the standard deviation with respect to the average of the function in the decade) which is $\log_{10}z\in(-3.8,-2.8)$. This would in turn determine the stability of the result.

\begin{equation}
\label{eq:twist-stress-tensor}
\begin{split}
\tau=1.00013\pm 5\times 10^{-5} \quad\Big(&\mbox{error from deviation from} \\&\mbox{flatness and differences between two tails}\Big)
\end{split}
\end{equation}

\begin{figure}
	\begin{subfigure}[t]{0.5\textwidth}
   \centering
  \includegraphics[width=1\linewidth]{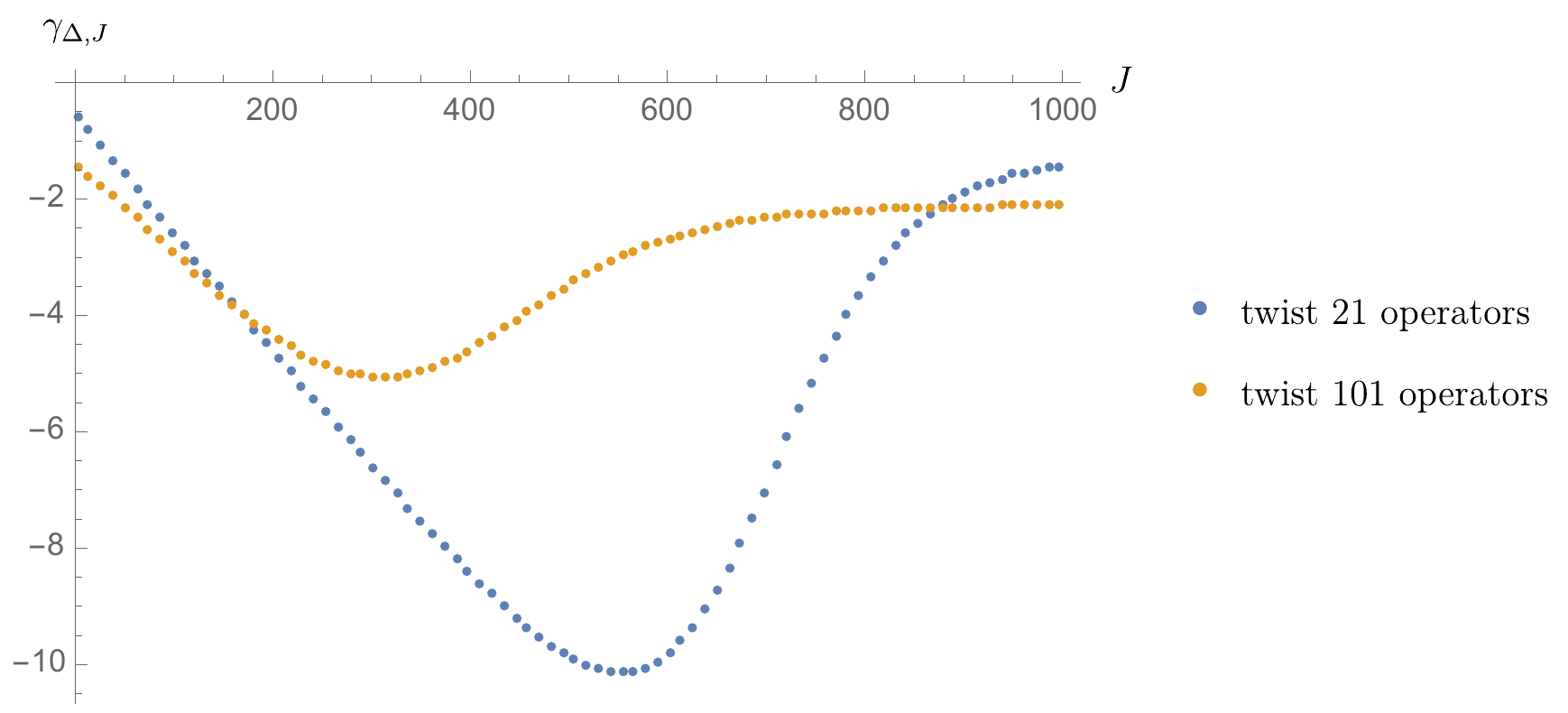}
  \caption{Evaluated at $z=10^{-4}$, .}
    \end{subfigure}
    \begin{subfigure}[t]{0.5\textwidth}
        \centering
        \includegraphics[width=1\linewidth]{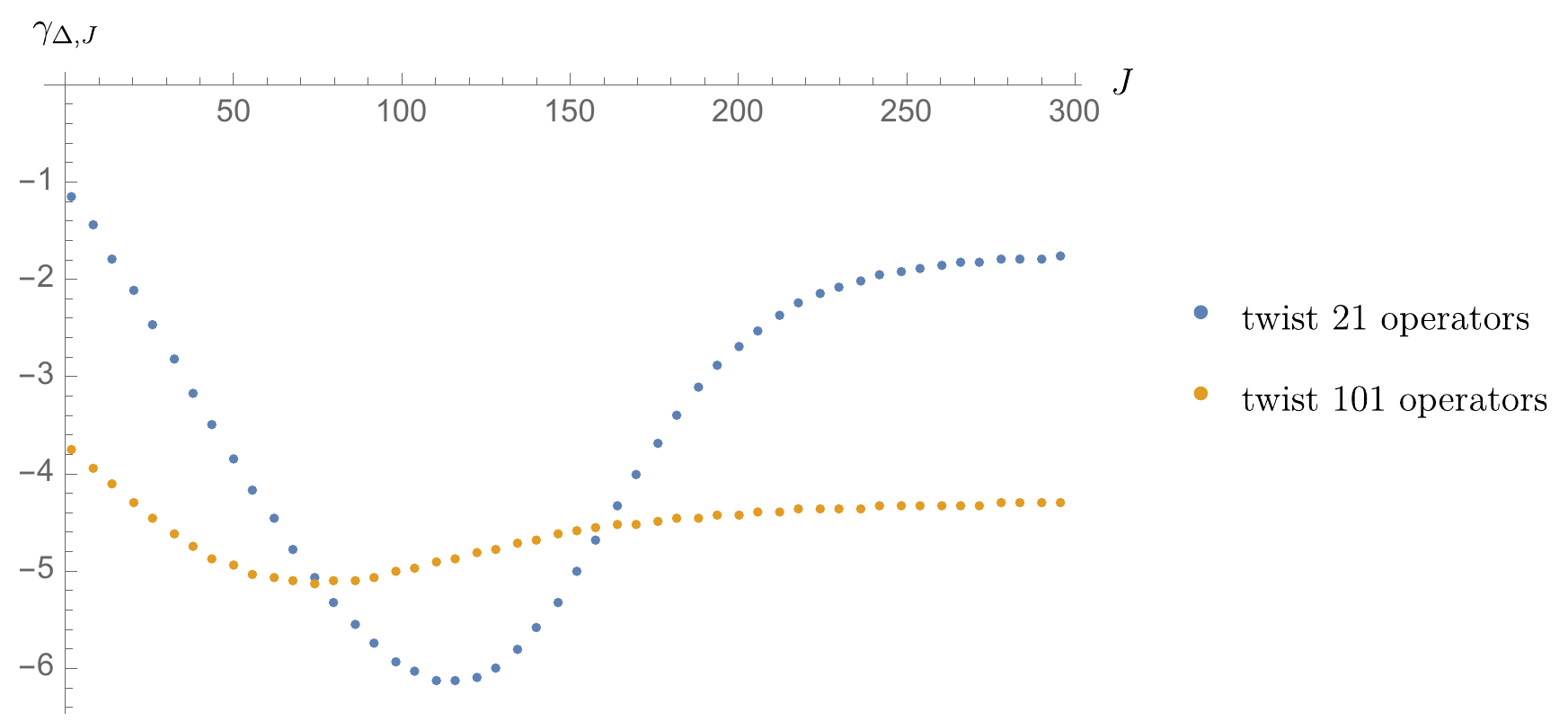}
      \caption{Evaluated at $z=10^{-3}$.}
    \end{subfigure}
      \caption{Individual high twist operators $2\frac{z\partial_z C_{\tau',j'}(z,\beta)}{C_{\tau',j'}(z,\beta)}-2\Delta\sigma$ push the curve of $\gamma_{\beta=5}$ down.}
    \label{fig:high-twist_st}
\end{figure}

It is crucial to address the question of removing the residual gap between this result and the actual twist of the stress-tensor. We argue that this can be done by including higher twist family in the OPE. To understand whether this is the right resolution, the first thing to check would be whether the high twist operators push the curve down or up. We do that by looking at $2\text{lim}_{z\rightarrow 0}\frac{z\partial_z C_{\tau',J'}(z,\beta)}{C_{\tau',J'}(z,\beta)}-2\Delta_{\sigma}$ for individual operators in fig.~\ref{fig:high-twist_st} and checking that the contribution of each is well below the average derived in eq.~\ref{eq:twist-stress-tensor}.

\begin{figure}[!htb]
  \centering
  \includegraphics[width=0.75\linewidth]{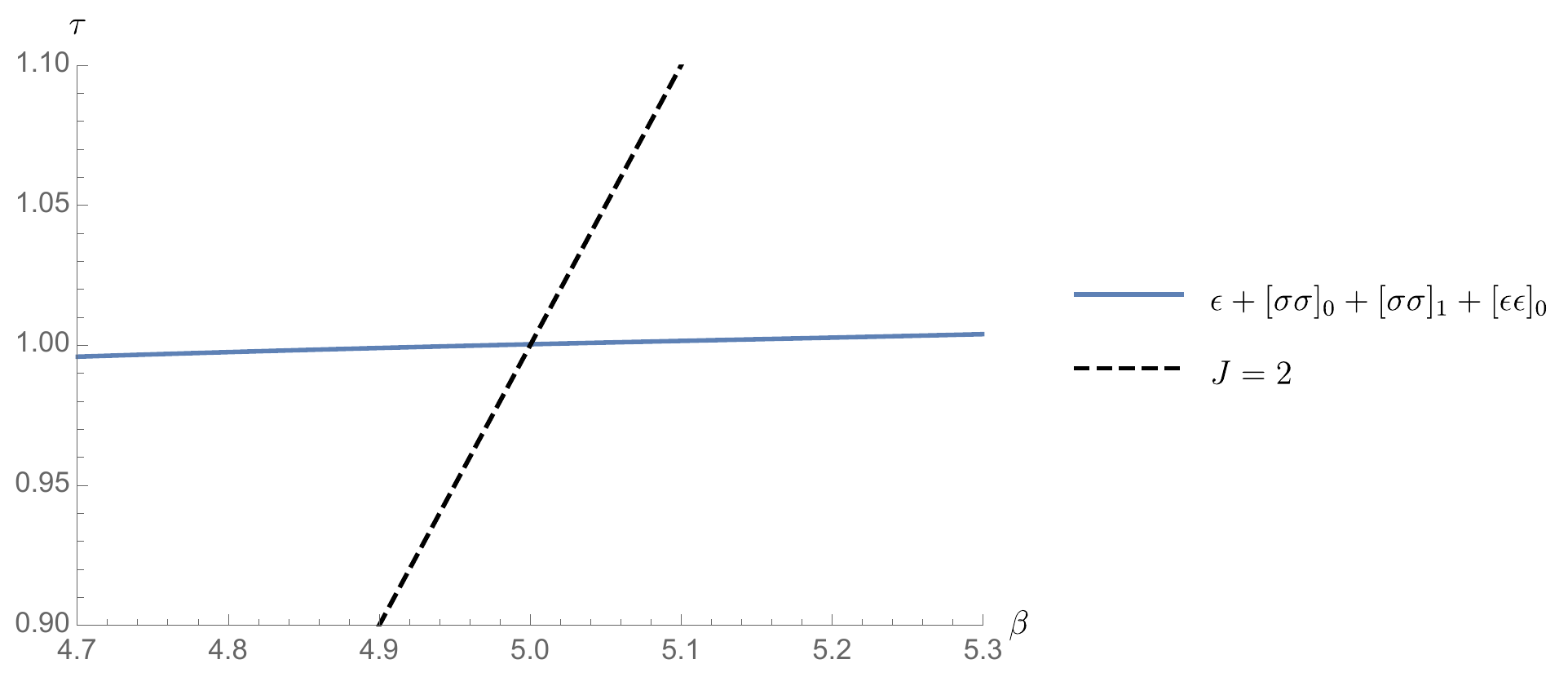}
  \caption{$\tau(\beta)$ in the vicinity of $\beta=5$. The point that the line $J=2$ crosses $c(\beta)$ gives the location of the stress-tensor. }
  \label{fig:twist-stress-tensor-interpol}
\end{figure}

Now that we have a reliable way of computing the twist for the $\beta$ as small as $\beta$ of the stress-tensor, we can calculate it for other points in the vicinity of stress-tensor by repeating this procedure. We can then use these point and get the function $\tau(\beta)$ by interpolation. The function we get is demonstrated in fig.~\ref{fig:twist-stress-tensor-interpol}.
 
 Now we know by eq.~\eqref{eq:jacobian} that the squared of the OPE coefficient is related to $ C(\beta=5)$ with a Jacobian factor which we can then calculate with the function $\tau(\beta)$ derived above. Thus we arrive at the value of the OPE coefficient. 

Again confining ourselves to the decade $\log_{10} z\in [-3.8,-2.8]$,  we get
\begin{equation}
f_{\sigma\sigma T}=0.326077\pm 12\times 10^{-6}
\end{equation}
We summarize the result for the stress-tensor in the table.~\ref{table:results-ssss}. Remarkably we are able to obtain the twist and OPE coefficient of the stress-tensor with accuracy $10^{-4}$!
\begin{table}[!htb]
\centering
\begin{tabular}{|l|l|l|l|l|}
\hline
 & $\Delta_{T}$ & $f_{\sigma\sigma T}$ \\
\hline
Inversion Formula (separate fit) & 1.00013(5) & 0.326077(12)  \\
Numerical Result & 1 & 0.32613776(45) \\
\hline
\end{tabular}
\caption{Twist and OPE of the stress-tensor, the spin 2 operator in $[\sigma\sigma]_0$, derived from the inversion integral (all the 3 families and the high spin tail from the fit is included) compared with the value derived from numerical bootstrap.}
\label{table:results-ssss}
\end{table}

 \subsubsection*{Comparison between $z\rightarrow 0$ and finite $z$}

\begin{figure}[!htb]
  \centering
  \includegraphics[width=1\linewidth]{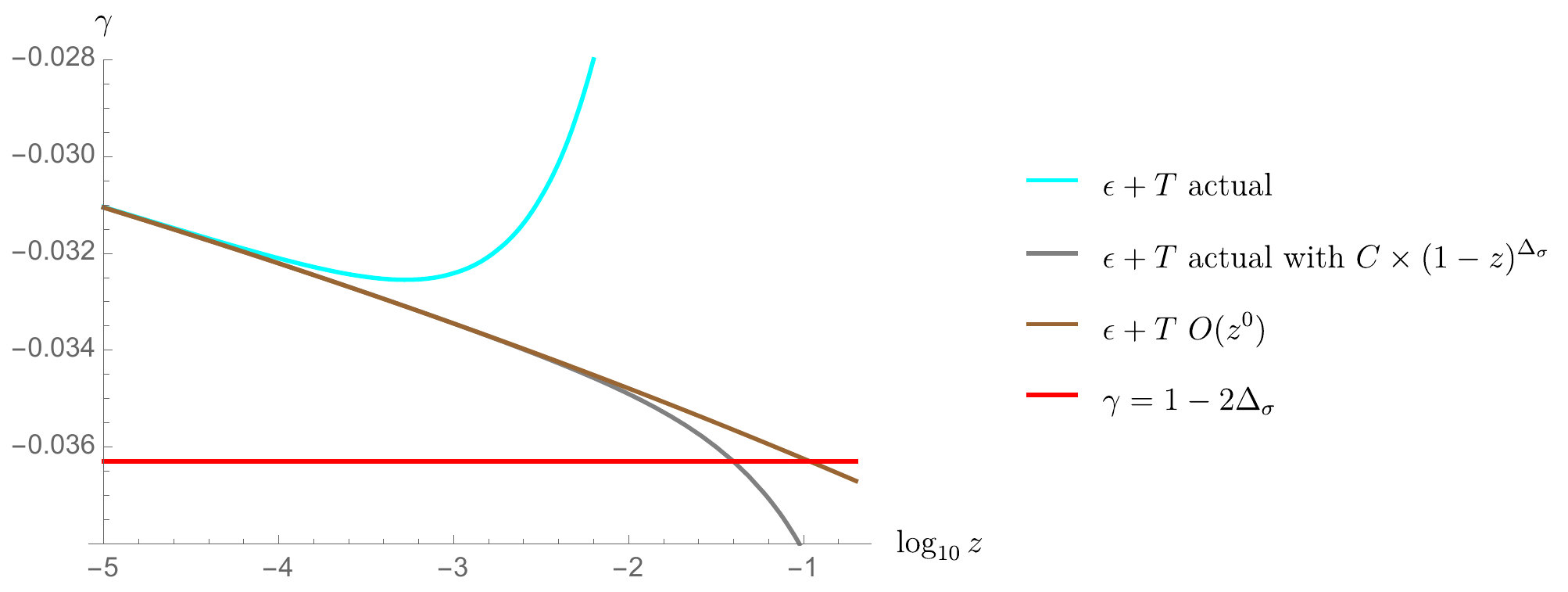}
  \caption{
  Comparison between the twist derived by evaluating exactly individual blocks (using 2d expansion)
 and their small-$z$ limit.
The agreement extends to larger values of $z$ when the former is multiplied by $(1-z)^{\Delta_\sigma}$.}
  \label{fig:zsmall-T}
\end{figure}

 As mentioned in previous sections, when we are interested in the exchange of the first few leading twist blocks in the cross-channel, we have the luxury of taking $z\rightarrow 0$ limit in eq.~\ref{eq:generating_int} since there is no infinite sum involved. In this limit the inversion formula simplifies significantly. The closed form of the inversion formula is known and is calculated in \cite{Cardona:2018dov} which is a ${}_7F_6$ function.
However, one need to take into account the errors introduced both by truncation in the cross-channel OPE expansion of the correlator in dDisc as well as error introduced by higher order terms in the $z$ expansion. As an illustration, in fig.~\ref{fig:zsmall-T}, we compare the anomalous dimension of the stress-tensor derived by using the $z\rightarrow 0$ expansion of eq.~\eqref{eq:generating_int} when only $\epsilon$ and T are exchanged in the cross-channel with the one derived with  by keeping the full $z$ dependence (plotted in fig.~\ref{fig:st-twist1} in cyan color) as well as the subtracted full  $z$ dependant one (plotted in fig.~\ref{fig:st-twist1} in gray color) . We emphasize again that one is allowed to do that because there is no infinite sum involved. 

  From fig.~\ref{fig:zsmall-T} one can observe that the $z\rightarrow 0$ limit matches with the subtracted full $z$ dependant one up to $z\sim 10^{-2}$. However the twist obtained by the unsubtracted integral starts to differ at $z=10^{-3.5}$. We can then conclude that $z\rightarrow 0$ expansion can be safely used as long as we confine ourselves to $z<10^{-3}$ for the exchange of $\epsilon$ and T.

\subsubsection*{Comparison between $\zb\rightarrow 1$ and finite $\zb$}

\begin{figure}[!htb]
  \centering
  \includegraphics[width=1\linewidth]{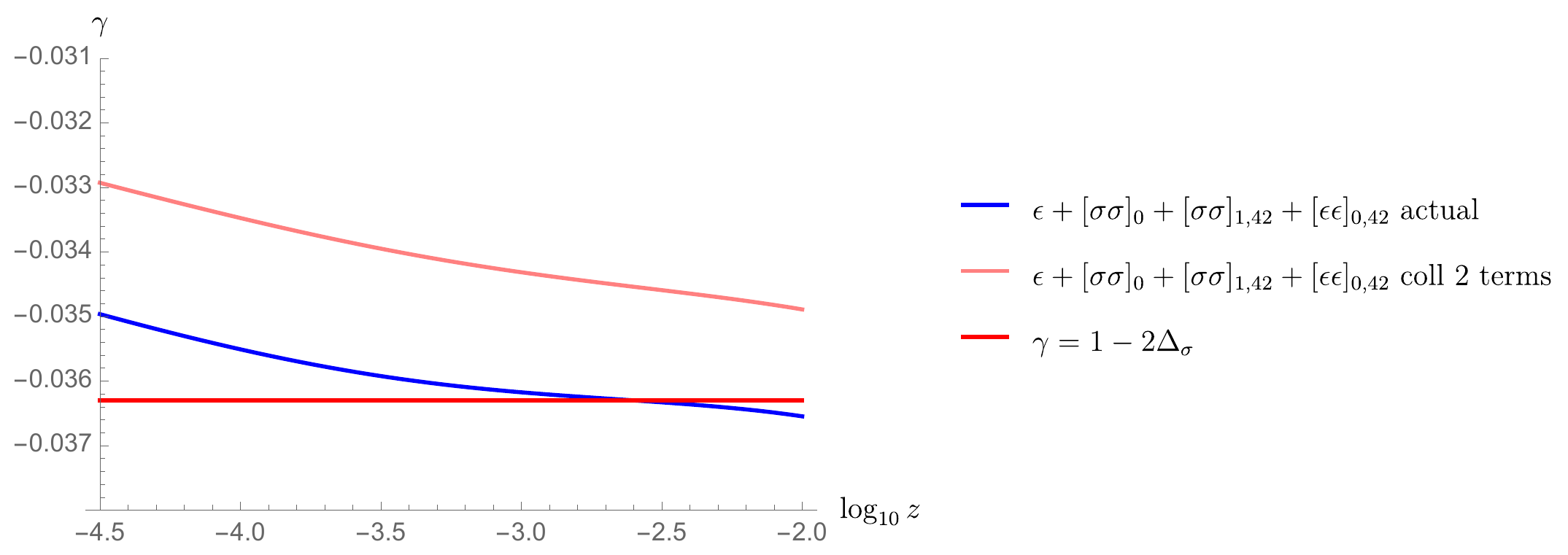}
  \caption{Comparison between the twist derived from collinear expansion versus the exact evaluation
  of individual blocks using 2d expansion.}
  \label{fig:coll-twist-T}
\end{figure}

 When $\beta$ is large enough, most of the contribution to the integral in $C(z,\beta)$ comes from $\zb\rightarrow 1$. As was shown before, in this limit, the blocks appearing in the cross channel expansion simplify a lot. In this section we compare the third order result in $1-\zb$ expansion with the result non-perturbative in $1-\zb$. In section \ref{subsec:L-S Expansion} we have already seen the error of the collinear expansion for the stress-tensor is not negligible. However, it is worthwhile to compare the final results derived with this expansion with the non-perturbative one. In figs.~\ref{fig:coll-twist-T} we compare the final answer for the twist.

One can see from this plot that the relative error on the anomalous dimension is approximately $6\%$ which is compatible
with the analysis in subsection \ref{subsec:L-S Expansion}.

\subsection{Intercept}
\label{sec:intercept}

\begin{figure}[h]
  \centering
  \includegraphics[width=0.75\linewidth]{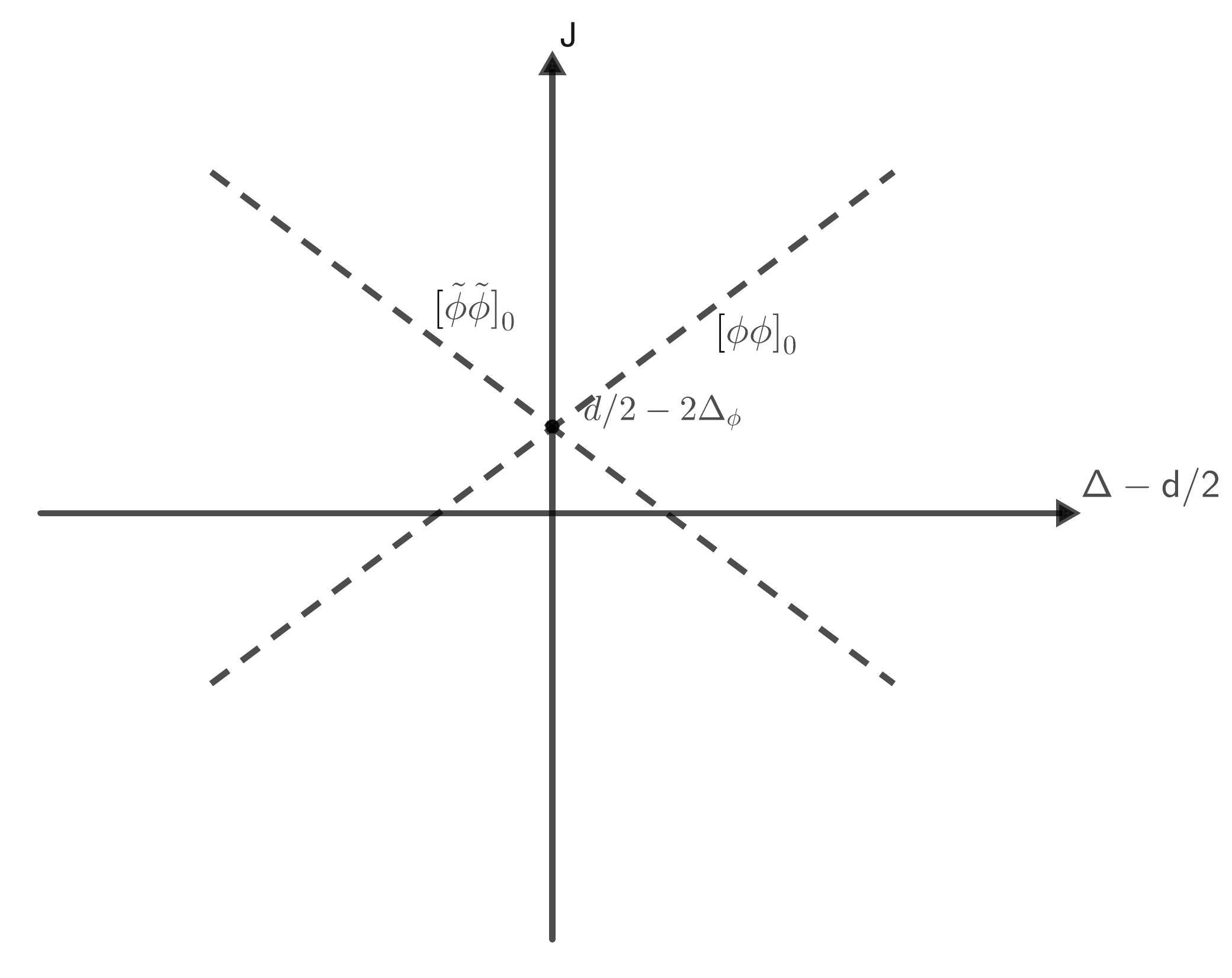}
  \caption{The Regge trajectory of $[\phi\phi]_0$ and its shadow in free field theory. The two trajectories intersect
  at the shadow symmetric point.}
  \label{fig:free_field}
\end{figure}

To understand the extent of validity of inversion formula for low spin it is crucial to study $[\sigma\sigma]_0$ regge trajectory at spin below two. Similar to stress-tensor, we extract the information from $\expval{\sigma\sigma\sigma\sigma}$ correlator. 
\\
 In free field theory, or the UV fixed point, we know that operators in $[\phi\phi]_0$ family lie on a straight line in $J-\Delta$ plane with $\Delta-J=2\Delta_{\phi}$.  In addition due to the shadow symmetry ($\Delta\leftrightarrow d-\Delta$),  we have the straight line trajectory for the shadow family as well. The trajectory and its shadow are plotted in fig.~\ref{fig:free_field}.

Note that in fig.~\ref{fig:free_field} the two curves must intersect each other at the shadow symmetric point with $\Delta=d/2$, the spin at this point is $d/2-2\Delta_{\phi}$.

Moving on the RG flow from this Gaussian fixed point to Wilson Fischer fixed point, operators acquire anomalous dimension and they move away from the straight line trajectories and lie on a smoother curve.
 Analyticity of the mentioned curve for $J\geq 2$ has been established by the proof of the Lorentzian inversion formula. However, in perturbation theory ($\varepsilon$ expansion) operators with spin smaller than two, has also been shown to be analytic in spin and lie on the Regge trajectory  (see \cite{Alday:2017zzv}).

What we are interested in this section is to extend our methods to spin smaller than 2 and capture the non-perturbative characteristic of the leading Regge trajectory, $[\sigma\sigma]_0$ and its intercept, $j_*$. 

There are subtleties associated with going to such low $\beta$. One for instance is that as discussed in section.~\ref{sect:inversion}, close to the shadow-symmetric point, $\Delta=d/2$ the mentioned decomposition of the s-channel block in eq.~\ref{eq:block-decomp} breaks down and our collinear approximation for the s-channel block is not relevant any more. This means other expansion of the s-channel block is required. We find the relevant expansion of the block is an expansion in another set of complex variables $\rho$ and $\bar{\rho}$  which is related to $z$ and $\zb$ as follows (see \cite{Hogervorst:2013sma})
 \begin{equation}
 \begin{split}
 &\rho=\frac{z}{(1+\sqrt{1-z})^2} \leftrightarrow z=\frac{4\rho}{(1+\rho)^2}\\
 &\bar{\rho}=\frac{\zb}{(1+\sqrt{1-\zb})^2} \leftrightarrow \zb=\frac{4\bar{\rho}}{(1+\bar{\rho})^2}
 \end{split}
 \end{equation}
 We expand the s-channel block in eq.~\ref{eq:inversion-integral} in $\rho, \bar{\rho} \rightarrow 0$ which captures the contribution of the block for $z\sim\zb\sim 0$ and $\Delta\sim d/2$ and then we transform back to $z$ and $\zb$. This is the appropriate range for the vicinity of the intercept.
The generating function replacing eq.~\ref{eq:generating_int} is then as follows:
\begin{equation}
\label{eq:generating_int2}
C(z,\beta)=\kappa(\beta
)\int_z^1\frac{1}{(\zb z)^2}\frac{(z - \zb)}{(z \zb)} z^{(\tau/2 + 1)}G_{\rho}^{\beta,4-\tau}(z,\zb)\text{dDisc}[\mathcal{G}(z,\zb)]
\end{equation}

\begin{figure}[!htb]
  \centering
  \includegraphics[width=1\linewidth]{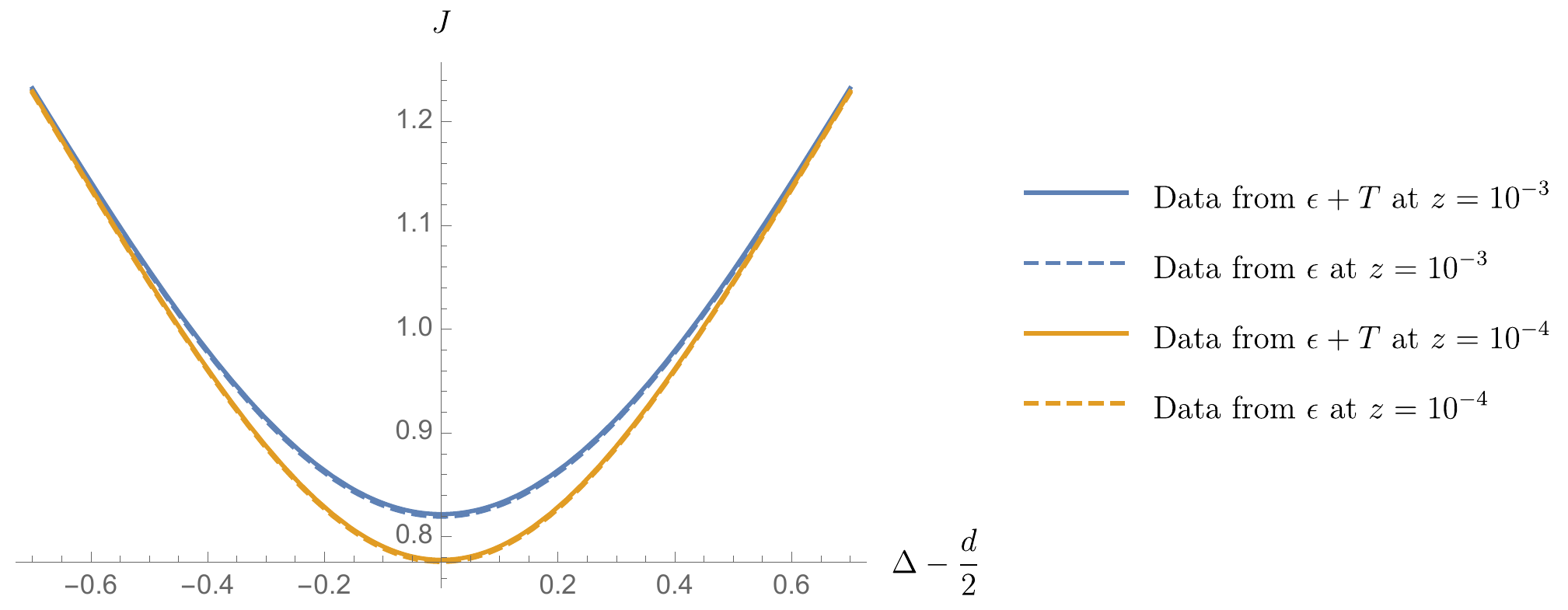}
  \caption{The function $J(\Delta)$ is plotted at low spin, for the exchange of $\epsilon$  and for the exchange of $\epsilon$ and stress-tensor at two different values of $z$. The important takeaway is the spin of intercept is below 1. The width created by these curves estimates the error of the analysis.}
  \label{fig:intercept}
\end{figure}

For calculating eq.~\ref{eq:generating_int2},  the s-channel block is expanded in $\rho$ and $\bar{\rho}$ to 6th order. In addition, in order to perform the integral, we expand the correlator in cross-channel expansion. Since we are interested in very small $\beta$, it would be beneficial to use the full cross-channel blocks. By numerically integrating the integral form of the conformal block for spin 0 which is introduced in  \citep{Ferrara:1972cq,Ferrara:1973vz,Ferrara:1974ny}, we obtain the $\epsilon$ exchange.
\begin{equation}
\label{eq:GWD}
\begin{split}
&G_{\Delta,0}(z,\bar{z})=\frac{\Gamma(\Delta)}{\Gamma(\frac{\Delta+\Delta_{34}}{2})\Gamma(\frac{\Delta-\Delta_{34}}{2})}u^{\frac{\Delta}{2}}\int^1_0d\sigma (1-(1-(1-z)(1-\zb))\sigma)^{\frac{-\Delta+\Delta_{12}}{2}}\\&\times\sigma^{\frac{\Delta+\Delta_{34}-2}{2}}(1-\sigma)^{\frac{\Delta-\Delta_{34}-2}{2}}{}_2F_1\left(\frac{\Delta+\Delta_{12}}{2},\frac{\Delta-\Delta_{12}}{2},\Delta-\frac{d-2}{2},\frac{z\zb\sigma(1-\sigma)}{1-(1-(1-z)(1-\zb))\sigma}\right)
\end{split}
\end{equation}
This integral representation is exploited in different contexts in the literature, see for instance  \cite{Hijano:2015zsa}. In addition, one can also derive similar integral representation for the exchange of conserved current:
\begin{equation}
\label{eq:GTI}
G_{j+d-2,j}(z,\zb)=\int^1_0 dt\frac{(2^{2 j} \Gamma(1 + j))}{(\sqrt{\pi}\Gamma(1/2 + j))} \frac{\sqrt{z \zb}}{(\sqrt{1 - t} \sqrt{t} \sqrt{
     1 - t z - \zb + t \zb})} \left(\frac{1 - \sqrt{1 - t z - \zb + t \zb}}{1 + \sqrt{1 - t z - \zb + t \zb]}}\right)^j
\end{equation}
specifying to spin 2 gives us the exchange of stress tensor. Now we have all the ingredient to perform the inversion formula for the exchange of $\epsilon$ and $T$ which are the leading twist operators.  Once the integral of the blocks are done, one can perform the inversion integral in eq.~\ref{eq:generating_int2} numerically as well to obtain the generating function. We can then use this generating function to obtain the twist at different values of conformal spin using eq.~\ref{eq:procedure}. Now trivially the function $\tau(\beta)$, gives us the function $J(\Delta)$, from which we can read of the intercept as can be seen in fig.~\ref{fig:intercept}.

In our analysis, we find that the smallest value of $\beta$ for which eq.~\eqref{eq:generating_int} agrees with eq.~\eqref{eq:generating_int2} is $\beta_{\text{min}}\sim 3$. For $\beta$s smaller than this value we must use the latter. 

However, note that since we are not exchanging the twist families, precision of our result will be moderate. To quantify our error, we compare the twist derived at different values of $z$ with exchange of only $\epsilon$ as well as exchange of both $\epsilon$ and T. We see that our results does not change drastically in any of the mentioned cases. This comparison is plotted in fig.~\ref{fig:intercept}
Remarkably one can see in fig.~\ref{fig:intercept}, that the intercept of the leading Regge trajectory, $[\sigma\sigma]_0$ is below one, $j_*\approx 0.8$, which conclusively shows that 3D Ising theory is transparent at high energies. We also estimate the (shadow symmetrical) residue at the intercept to be:
\begin{equation}
\Res\limits_{J=j_*}\frac{c^t(\Delta, J)}{K(\Delta,J)}\Bigg |_{\Delta=\frac{3}{2}}\approx  0.02.
\end{equation}

Recently a similar estimate $j_*^{O(2)}\approx 0.82$ was obtained in \cite{Liu:2020tpf} for the $O(2)$ model using a related method. It would be interesting to compare the details. See section \ref{sect:CF} for a detailed discussion on the implication of this result.

\subsection{Analytic continuation to spin 0: looking for $\epsilon$}
\label{sec:Epsilon}

It was shown in perturbation theory (see \cite{Alday:2019clp} ), i.e., $\varepsilon$-expansion, one can obtain an analytic curve for the leading Regge trajectory and analytically continue it to recover $\epsilon$ and its shadow on the continued curve. 

\begin{figure}[!htb]
  \centering
  \includegraphics[width=1\linewidth]{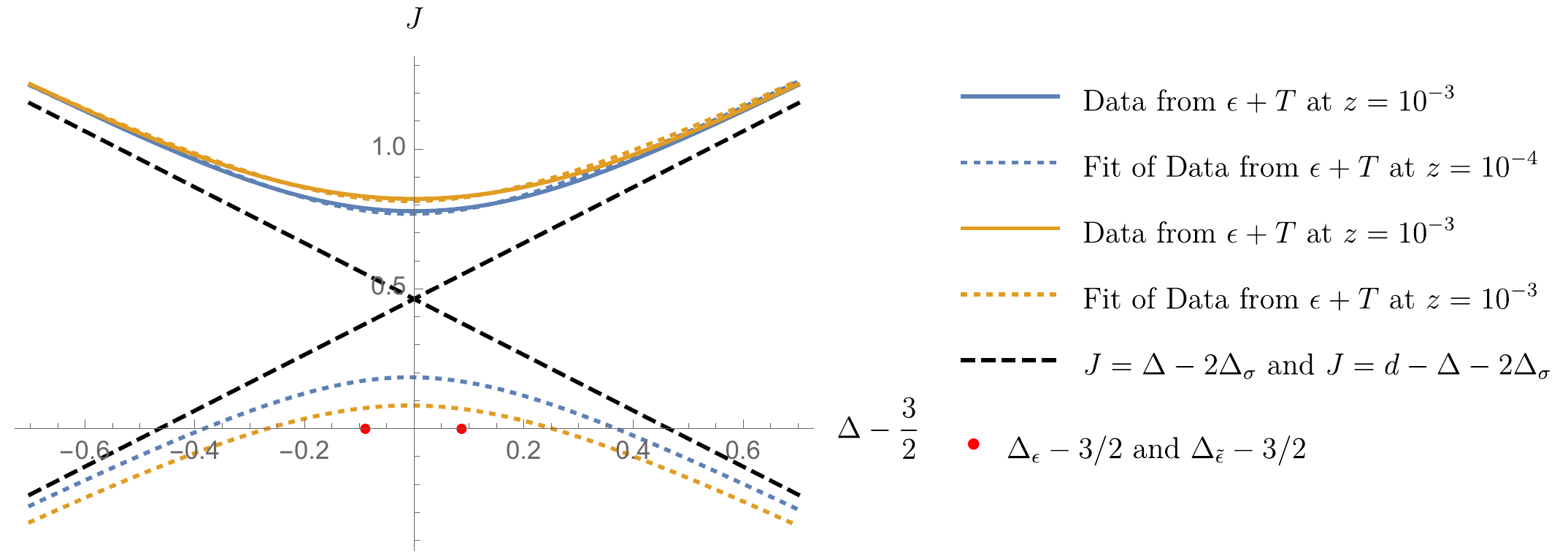}
  \caption{The interpolation of data points and their fit for two sets of points considered in fig.~\ref{fig:intercept} is given. The analytic continuation is obtained once the fit is found. The $\epsilon$ and its shadow are represented by the two red dots. We see the prediction for $\Delta_{\epsilon}$ by looking at the intersection of the curves and the $\Delta$ axis. }
  \label{fig:intercept2}
\end{figure}

In this subsection we study the possibility of finding $\epsilon$ operator on the curve obtained by the analytic continuation of the Regge trajectory to  spin below intercept in the full non-perturbative 3D Ising CFT. In order to perform the analytic continuation, we need an ansatz for the leading trajectory near the intercept which reproduces the data obtained from inversion formula properly. The ansatz we use for the function $J(\Delta)$ needs to have two important characteristic. First, it should be symmetric under the shadow transform ($\Delta\leftrightarrow d-\Delta$). Second, asymptotically it must approach the lines $J=\Delta-2\Delta_{\sigma}$ and $J=-\Delta+d-2\Delta_{\sigma}$. One simple ansatz that satisfies both of these condition is as follows.
\begin{equation}
(J-d/2+2\Delta_{\sigma})^2=(\Delta-d/2)^2+A
\end{equation}
This corresponds to the following $\tau(\beta)$ function:
\begin{equation}
\tau(\beta)=\frac{A}{\beta-\beta_0}+B-\beta_0
\end{equation}
It follows that we must have $\beta_0=d-2\Delta_{\sigma}$ and $B=d-\beta_0$, however, we keep their values unidentified. The data points are fitted with this ansatz for each of the different cases considered above to obtain the value for $A$, $\beta_0$ and $B$ in each case. As an example the values for the fit of data obtained from $\epsilon+T$ at $z=10^{-4}$ is given (blue lines in fig.~\ref{fig:intercept2}):
\begin{equation}
A=-0.085593 \qquad \beta_0=1.96857 \qquad B=2.98551
\end{equation} 

Having the ansatz, the analytical continuation of the Regge trajectory to $j<j_*$ is straightforward (see fig.~\ref{fig:intercept2} for the trajectory and its analytic continuation). 
Evaluating the function at $J=0$ gives us an estimation of the conformal dimension of $\epsilon$ operator and its shadow.  As an example  the fit of data obtained from $\epsilon+T$ at $z=10^{-4}$ gives the following estimates :
\begin{equation}
\Delta_{\epsilon}=1.11752 \qquad \Delta_{\tilde{\epsilon}}=1.868
\end{equation}
These values are not close to $\epsilon$ operator. However the analysis predict their existence as was seen before in perturbation theory. Note that a slight shift in the vertical axis of this curve can land us on a curve which includes $\epsilon$. However it should be obvious that using this method to determine quantitative properties of $\epsilon$
would be numerically unstable.

\section{Leading $Z_2$-odd twist family}\label{sect:ising-unequal}

In this section we study the low spin operators in $[\sigma\epsilon]_0$ family by analysing the correlator $\expval{\sigma\epsilon\epsilon\sigma}$, which in our notation corresponds to $\Delta_1=\Delta_{\sigma}$ and $\Delta_2=\Delta_{\epsilon}$. According to our previous discussion this correlator leads to data about the $[\sigma\epsilon]_0$ family. The operators exchanged in the t-channel, where we fuse $\sigma$ with $\sigma$ and $\epsilon$ with $\epsilon$ are $\epsilon$ and the families $[\sigma\sigma]_0$, $[\sigma\sigma]_1$, $[\epsilon\epsilon]_0$. The u-channel involves the same fusion as the s-channel, $\sigma$ with $\epsilon$ so $\sigma$ and the $[\sigma\epsilon]_0$ family are exchanged. We will use data provided in \cite{Simmons-Duffin:2016wlq} for the dimensions and OPE coefficients.

There are many interesting physical facts about 3D Ising model that one can understand by studying this family of operators. For instance, the absence of a global conserved current with spin 1. According to the unitarity bound in  eq.~\ref{eq:unitarity}, this spin one operator, if existed, must have scaling dimension 2 and thus, must belong to $[\sigma\epsilon]_0$ family. Our goal is to verify this fact analytically by extending the inversion formula to conformal dimension of order $\sim 3$ and observe its prediction.

 Another compelling question is whether this family of operators contains a spin 0 operator and if it does what is that operator. We try to answer this question by extending $C_{\text{even}}(z,\beta)$ to low spin. We show that indeed this family contains a spin 0 operator and the operator is compatible with being the shadow of $\sigma$ operator. This is showed by comparing the scaling dimension and the OPE coefficient and showing that they are in the right neighbourhood.

As a warm up example, we start by studying the spin 2 operator, for which we have controlled error and accurate data. Again, we apply the procedure explained in section \ref{sect:procedure} to derive the twist and the OPE coefficient. To control the error of the result same as section \ref{sect:ising-equal}, we illustrate the importance of including the subleading twist families to obtain a stable answer with controlled error. In addition, we see the range of $z$ in which the twist expansion is consistent, the sum over families are convergent and no resummation is required. However, the overall accuracy of the result for the spin 2 operator in $[\sigma\epsilon]_0$ is less compared to the spin 2 operators in $[\sigma\sigma]_0$ and this is due to the fact that there are  families of higher twist that need to be exchanged in the cross-channel (for instance $[\sigma\epsilon]_1$, $[\epsilon\epsilon]_1$ and $[\sigma\sigma]_2$) to get less error and a more stable result. 

Once we have familiarized ourselves with the procedure, we move on to the spin 1 and spin 0 operators by following the same steps. As expected, the result will be less stable and we have less control over the errors.



\subsection{Benchmark case: the spin 2 operator}
\label{sec:spin2}

\begin{figure}
	\begin{subfigure}[t]{0.5\textwidth}
   \centering
  \includegraphics[width=1\linewidth]{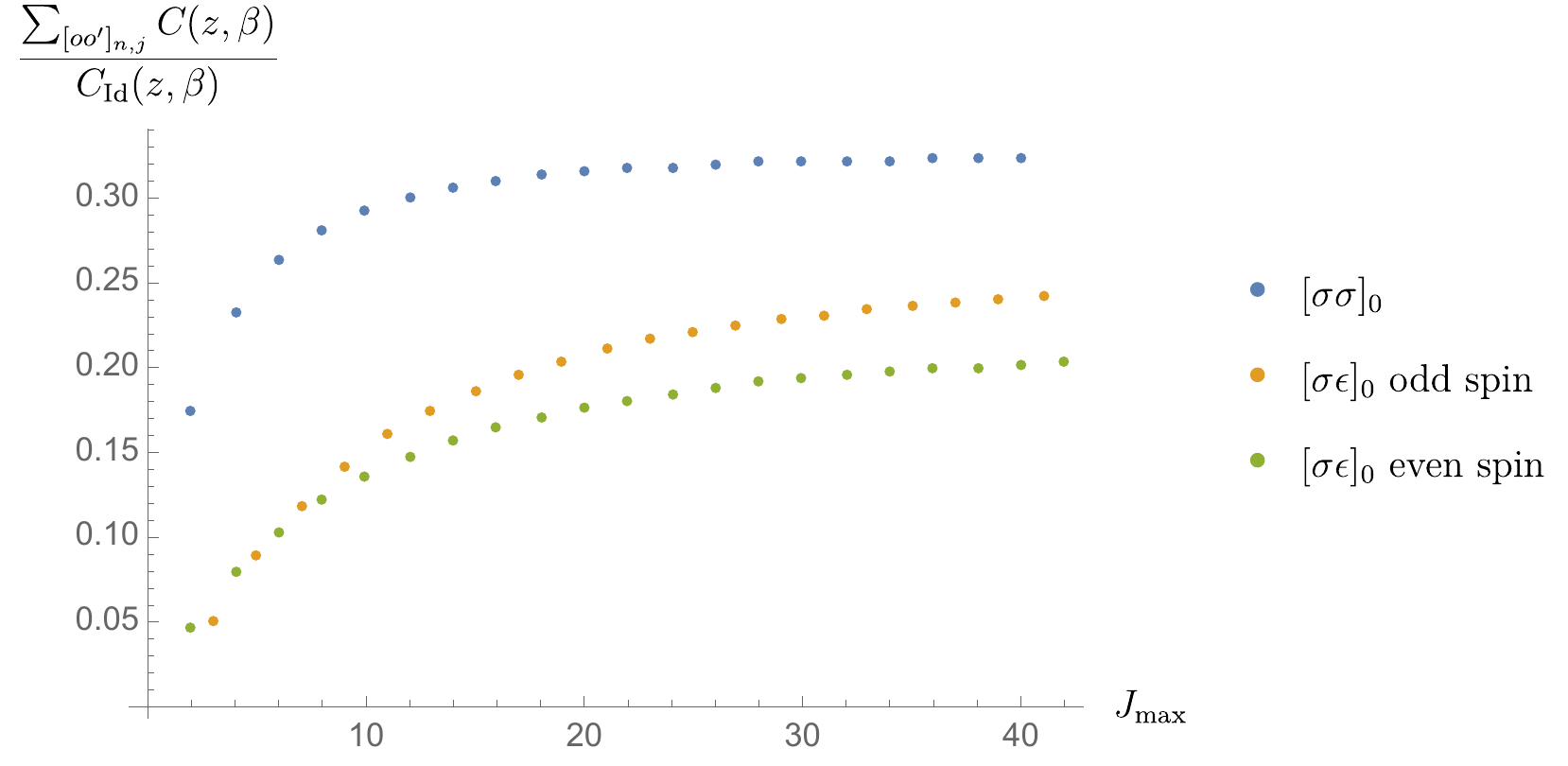}
  \caption{For the value of $z=10^{-3}$.}
    \end{subfigure}
    \begin{subfigure}[t]{0.5\textwidth}
        \centering
        \includegraphics[width=1\linewidth]{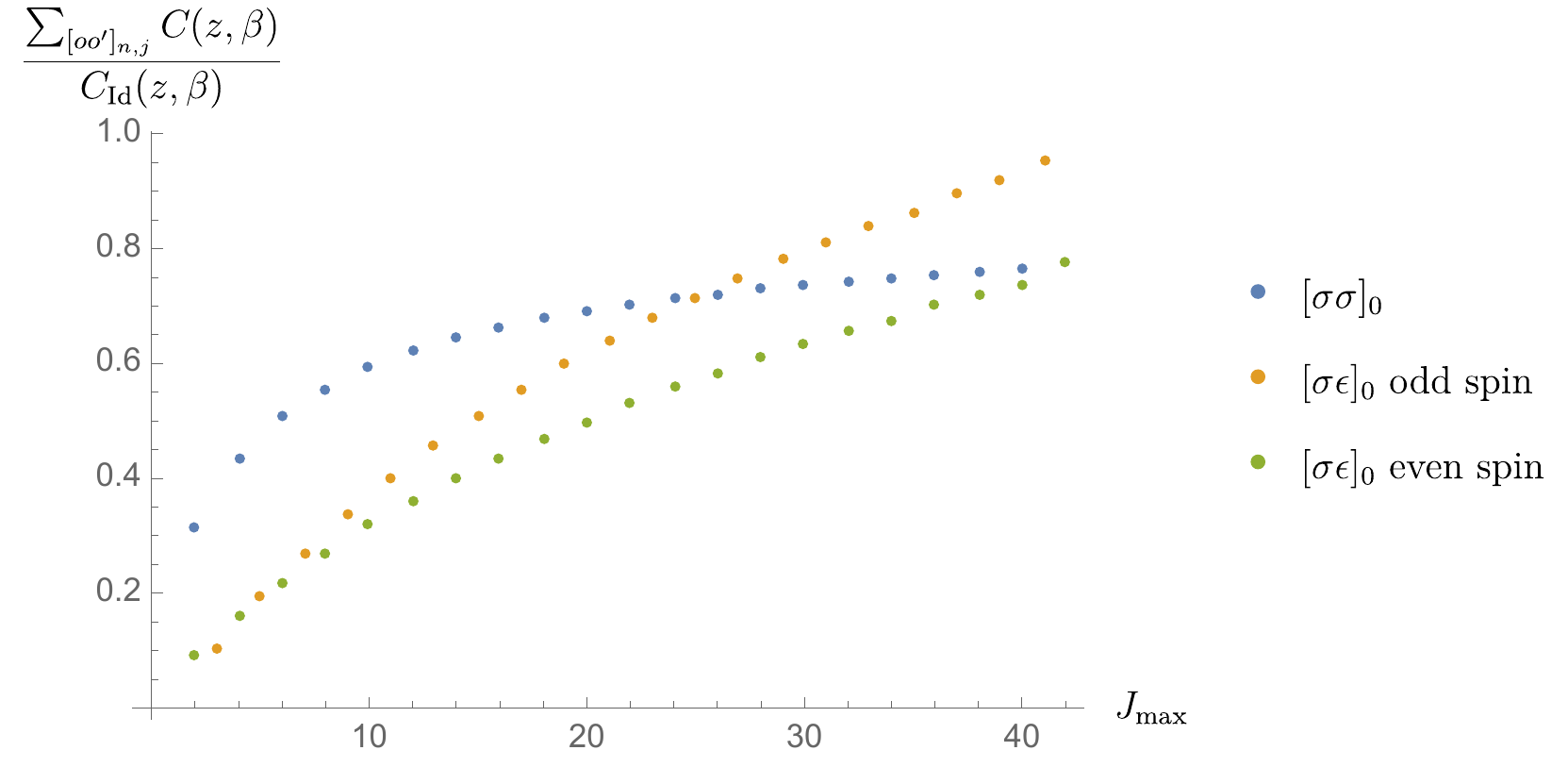}
        \caption{For the value of $z=10^{-4.5}$}
    \end{subfigure}
       \caption{Partial sum over the contributions of different families for the spin 2 operator with $\beta=6.18$ .}
    \label{fig:partial_s2}
\end{figure}

In this section we verify with what accuracy the anomalous dimension and OPE coefficient, $f_{\sigma\epsilon O_2}$, of spin 2  operator in $[\sigma\epsilon]_0$ family, $O_2$, can be derived from inversion formula with various truncation in the t-channel expansion. The procedure and the steps taken here are the same as the calculation for stress-tensor. However, the details of the calculation are of course different. Again since we are dealing with low spin, d-1 expansion of the block is the correct tool to use. But before we proceed to that, we need to find the relevant range of $z$ in which eq.~\ref{eq:procedure} can be applied. This would be the range in which the twist expansion of the argument of the dDisc breaks down. One can indeed confirm that indeed convergence of the twist expansion breaks in relatively large $z$, which is shown in fig.~\ref{fig:partial_s2}.

With the same analysis performed in fig.~\ref{fig:partial_s2}, we can conclude that the range of $z$ in which we have a valid twist expansion of the dDisc argument begins at $10^{-3}$. Thus this will be the lowest $z$ for evaluation of the twist and the OPE coefficient.

\begin{figure}[!htb]
  \centering
  \includegraphics[width=1\linewidth]{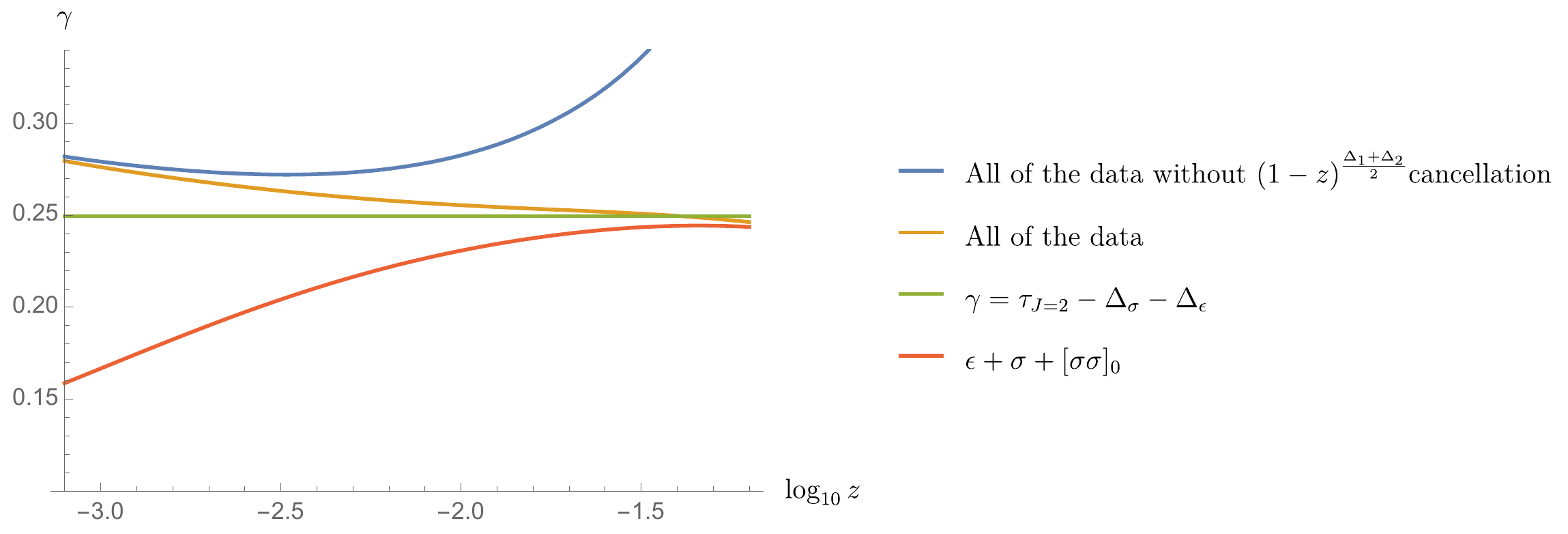}
  \caption{$\gamma(\beta=6.18)$ is given for comparison between different truncation in the cross-channel expansion of the dDisc. We see the importance of including the subleading families to get a plateau. In addition, multiplying $C_{\text{even}}(z,\beta)$ with $(1-z)^{\frac{\Delta_1+\Delta_2}{2}}$ has been shown to extend the range of $z$ in which we have a flat curve (see the paragraph below fig.~\ref{fig:st-twist1} for explanation).}
  \label{fig:s21}
\end{figure}

In fig.~\ref{fig:s21} we show the result for the anomalous dimension of the spin 2 operator.  The correlator in the argument of the dDisc in $C^t$ is built by exchanging $\epsilon$, $[\sigma\sigma]_0$, $[\sigma\sigma]_1$ and $[\epsilon\epsilon]_0$ in the t-channel and the argument for $C^u$ is built by exchanging  $\sigma$ and $[\sigma\epsilon]_0$ for the u-channel. We sum the contribution of the t-channel and u-channel since we are interested in the spin 2 operator in the s-channel (see eq.~\ref{t-pm-u}). We will call this function  $C_{\text{even}}(z,\beta)$ (in section \ref{sec:spin1}, where we are interested in odd spin operators of $[\sigma\epsilon]_0$, we subtract the u-channel contribution from the t-channel and subsequently the function derived this way will be called  $C_{\text{odd}}(z,\beta)$.

In order to understand the importance of the subleading families we compare the result with when only  $\epsilon$, $[\sigma\sigma]_0$ and $\sigma$ are exchanged in fig.~\ref{fig:s21}.

\begin{figure}[!htb]
  \centering
  \includegraphics[width=0.75\linewidth]{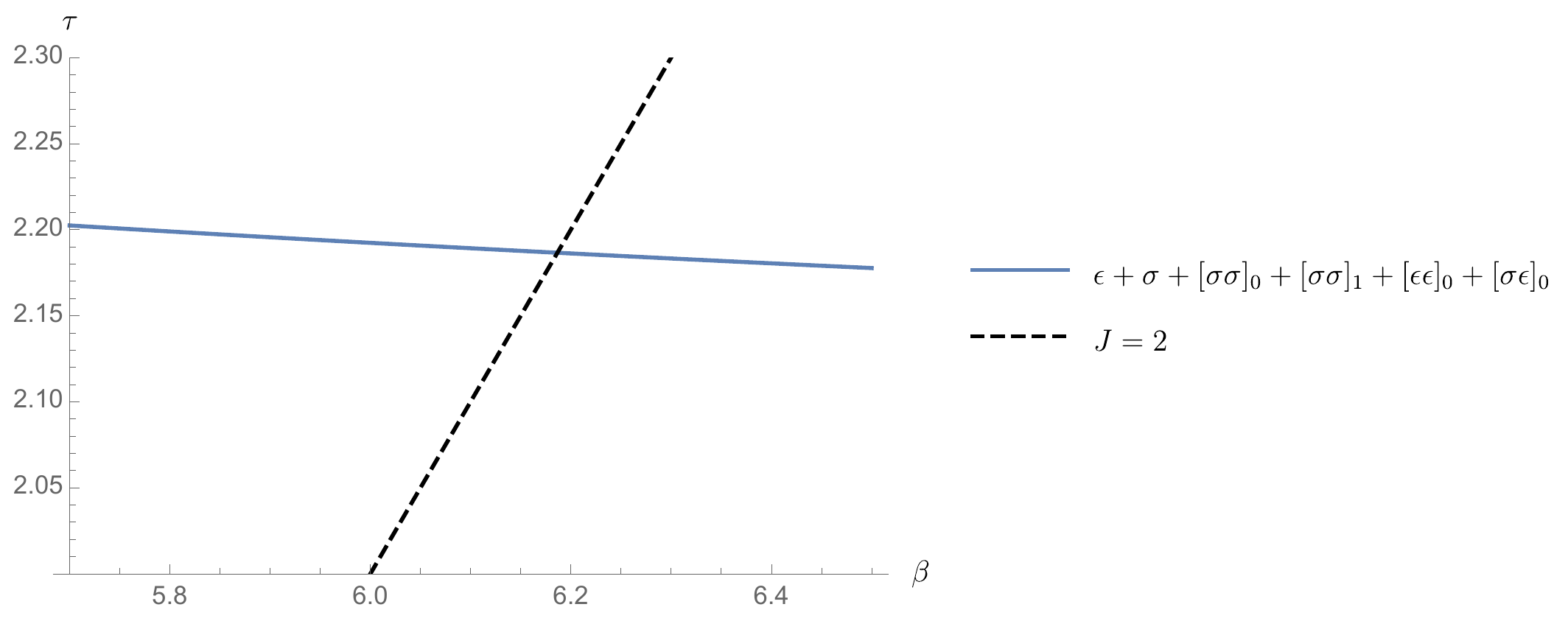}
  \caption{The intersection point of the line $J=2$ and the curve $\tau(\beta)$ is where the spin 2 operator is located.}
  \label{fig:twist-s2-interpol}
\end{figure}

Following the same procedure as the one for stress-tensor, we choose the decade in which we have the most stable result (the smallest standard deviation) which is $\log_{10} z \in[-2.3,-1.3]$, then by averaging over $z$ in this region, we get the following result for the twist of the spin 2 operator of $[\sigma\epsilon]_0$:
\begin{equation}
\tau=2.1845\pm 0.0035.
\end{equation}
We can calculate $\tau(\beta)$ for a number of operators in the vicinity of the spin 2 operator, i.e., $\beta=6.18$ for a fixed $z$. This will help us derive the function $\tau(\beta)$ by interpolation. As another method of getting the anomalous dimension of the spin 2 operator, we again intersect the spin 2 line with $\tau(\beta)$ as done in figure.~\ref{fig:twist-s2-interpol}. The point of intersection is where the spin 2 operator is located.
 
Same as $[\sigma\sigma]_0$, once we have the function $\tau(\beta)$ , we can compute the OPE coefficient by calculating $C(\beta)$ according to eq.~\ref{eq:procedure} and multiplying it by the relevant Jacobian factor in eq.~\ref{eq:jacobian} to get: 
\begin{equation}
f_{\sigma\epsilon O_2}=0.3907\pm 0.0014.
\end{equation}
In the table.~\ref{table:results-sees}, we summarised our analytical result for the scaling dimension and OPE coefficient of spin 2 operator along with values predicted by numerical bootstrap (see \cite{Simmons-Duffin:2016wlq}).
\begin{table}[!htb]
\centering
\begin{tabular}{|l|l|l|l|l|}
\hline
 & $\Delta_{O_2}$ & $f_{\sigma\epsilon O
_2}$ \\
\hline
Inversion Formula & 2.1845(35) & 0.3907(14)\\
Numerical Result & 2.180305(18) & 0.38915941(81) \\
\hline
\end{tabular}
\caption{Twist and OPE of spin 2 operator in $[\sigma\epsilon]$ derived analytically compared with the value derived from numerical bootstrap.}
\label{table:results-sees}
\end{table}
\subsection{Absence of spin 1 operator in the $[\sigma\epsilon]_0$ family}
\label{sec:spin1}

Now that we have familiarized ourselves with the basic process for extracting the low spin in $[\sigma\epsilon]_0$ family, we try to extend the analysis to study the absence of spin 1 conserved current in the odd spin sector of this family. Note that throughout this subsection all of the operators discussed in section \ref{sec:spin2} is exchanged in the cross-channel expansion of the double discontinuity.

From the unitarity bound in eq.~\ref{eq:unitarity} we know that the conserved spin 1 operator, if existed, has dimension 2. Thus we start the analysis by examining the function $C_{\text{odd}}(z,\beta)$ in the vicinity of $\beta=3$. 
 
The first step is to realize the range of $z$ in which we can trust the twist expansion following the procedure illustrated for $O_2$ operator in fig.~\ref{fig:partial_s2}. We find that the inversion formula is safe to use for $z>10^{-2.5}$. Thus all of the analysis in this section is done with 3 different values of $z$ in this range, $z=10^{-2.5}$, $z= 10^{-2}$ and $z= 10^{-1.5}$, to evaluate the stability.

\begin{figure}[!htb]
  \centering
  \includegraphics[width=1\linewidth]{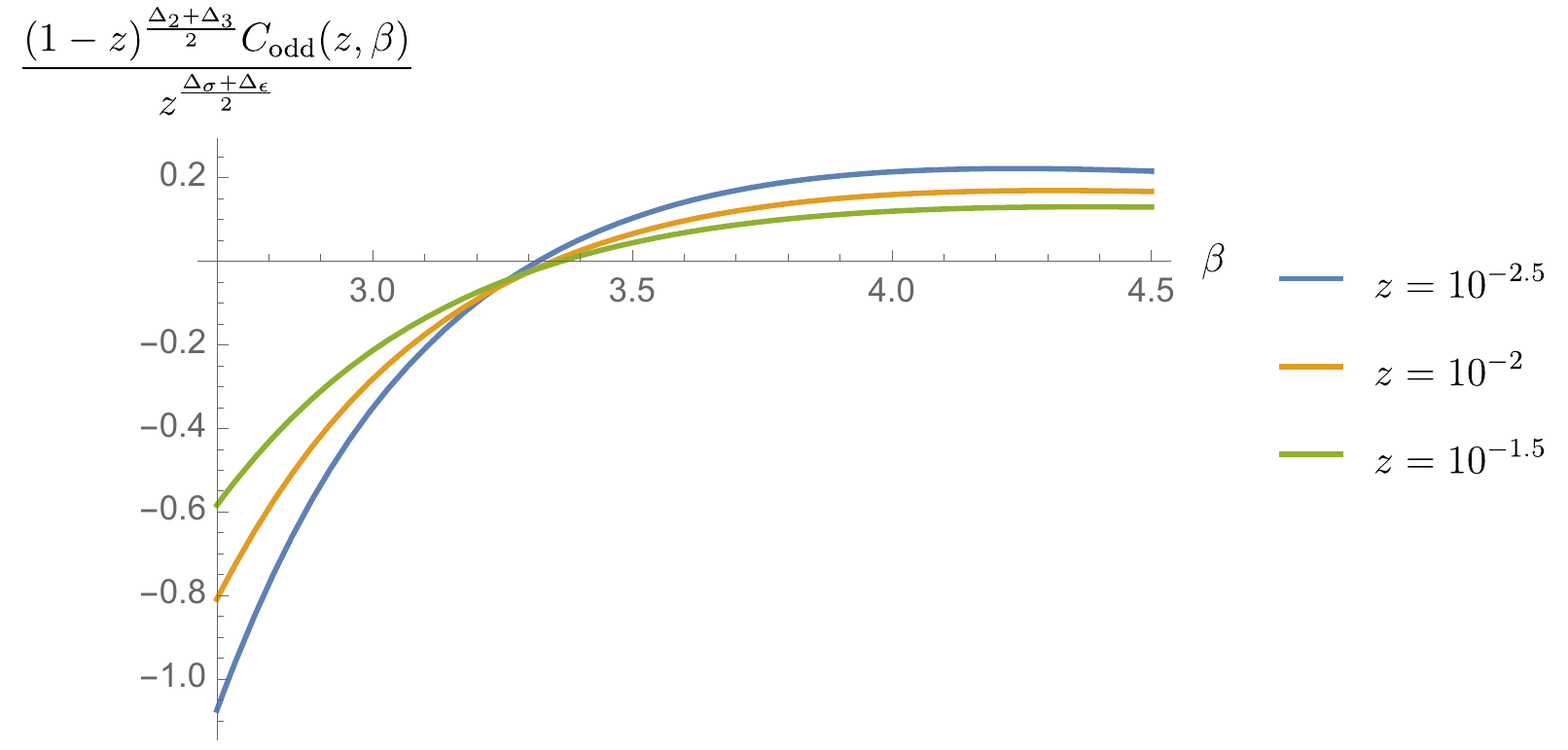}
  \caption{$(1-z)^{\frac{\Delta_2+\Delta_3}{2}}C_{\text{odd}}(z,\beta)/z^{\frac{\Delta_{\sigma}+\Delta_{\epsilon}}{2}}$ for $\beta$ in the vicinity of spin 1 operator for three different values of $z$. One can observe the vanishing of the OPE coefficient for $\beta\sim 3.3$ }
  \label{fig:codd-s12}
\end{figure}

 In fig.~\ref{fig:codd-s12}  $C_{\text{odd}}(z,\beta)$ multiplied by the factor $(1-z)^{\frac{\Delta_2+\Delta_3}{2}}/z^{\frac{\Delta_{\sigma}+\Delta_{\epsilon}}{2}}$ is depicted. The $(1-z)^{\frac{\Delta_2+\Delta_3}{2}}$  factor subtract the collinear descendants and higher twist contamination (see the paragraph below fig.~\ref{fig:st-twist1} for explanation) and division by $z^{\frac{\Delta_{\sigma}+\Delta_{\epsilon}}{2}}$ reduces the $z$-dependence of $C_{\text{odd}}(z,\beta)$ to $z^{\gamma_{\beta}/2}$.
 
A few notes must follow: we emphasize much alike section \ref{sec:intercept} the stability in $z$ is moderate. The difference in the curve is suggestive of what the error should be.

We see that all three curves cross zero at $\beta \sim 3.3$. This implies the vanishing of the OPE coefficient for that conformal spin. 

\begin{figure}[!htb]
  \centering
  \includegraphics[width=1\linewidth]{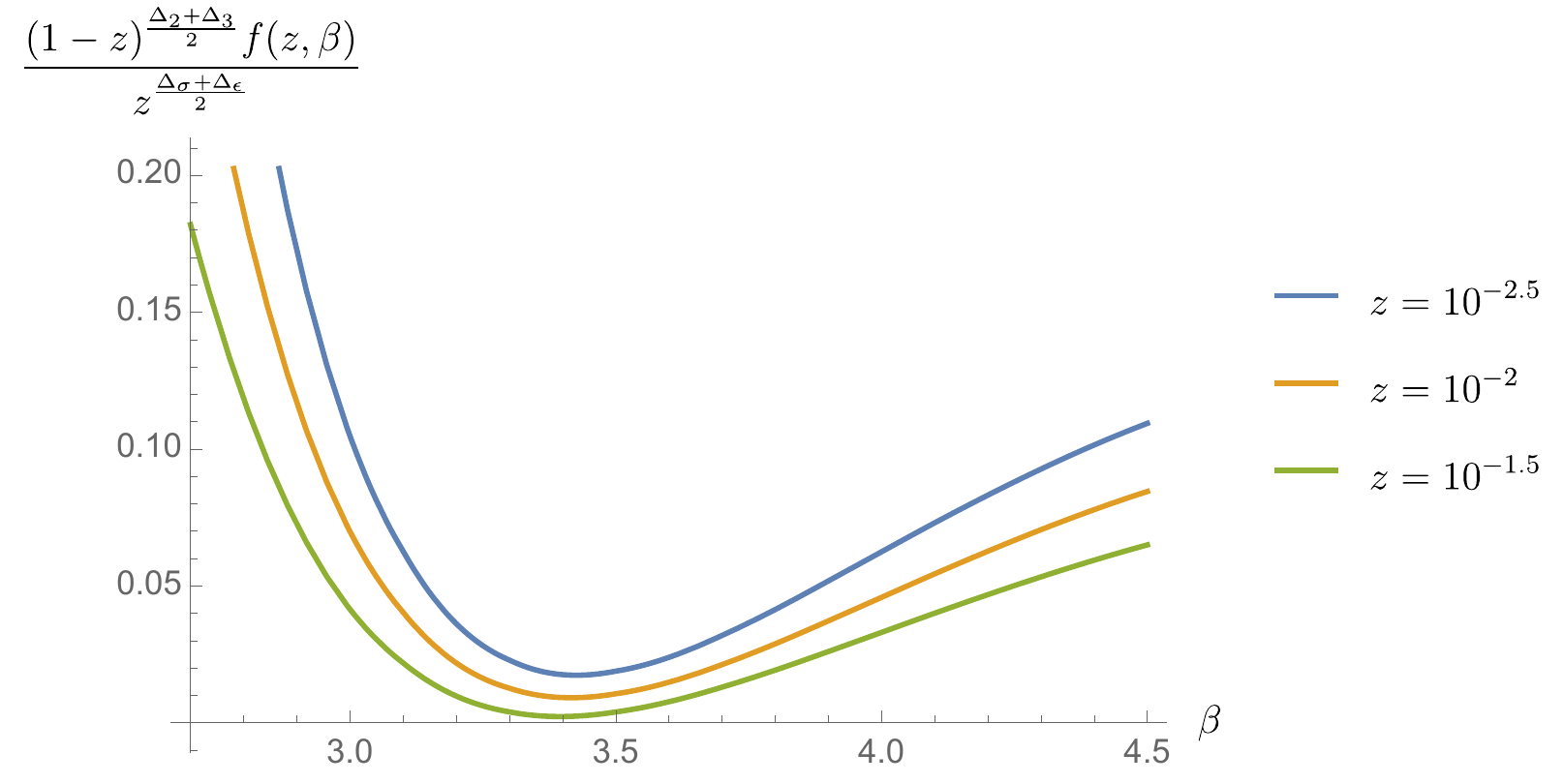}
  \caption{$f(z,\beta)/z^{\frac{\Delta_{\sigma}+\Delta_{\epsilon}}{2}}=(J-1)C_{\text{odd}}(z,\beta)/z^{\frac{\Delta_{\sigma}+\Delta_{\epsilon}}{2}}$ for $\beta$ close to $\Delta_{\sigma}+\Delta_{\epsilon}+2$ for three different values of $z$.
Within errors, this function seems compatible with having a double zero touching the real axis.}
  \label{fig:codds1-2}
\end{figure}

Theoretically, the $J=1$ operator is absent from the spectrum if and only if the vanishing of $C$ occurs
precisely when the trajectory crosses $J=1$.  Because of $C$ vanishing, the numerical evaluation
of the twist using $z\partial_z\log C$ is however unstable.  To assess whether the vanishing of $C$
and $J=1$ occur at the same point, we consider the following combination:
\begin{equation}
f(z,\beta)=\big(\beta/2-1-z\partial_z\big)C_{\text{odd}}(z,\beta).
\end{equation}
Using eq.~(\ref{eq:procedure}) it can also be written as:
\begin{equation}
f(z,\beta)=(J-1)C_{\text{odd}}(z,\beta).
\end{equation}
Note that this function vanishes when the trajectory contains an operator with spin 1 or when $C_{\text{odd}}(z,\beta)$ is zero. If the residue at the spin 1 point vanishes (which in turn implies the absence of spin 1 operator), then the two zeros in $f(z,\beta)$ must be at the same place and we expect to find a curve tangent to the $x$-axis.
By examining the function $f(z,\beta)$ in fig.~\ref{fig:codds1-2}, we see that the curves are almost tangential but not completely! We speculate this to be caused by the truncation of the t-channel OPE.

\begin{figure}[!htb]
  \centering
  \includegraphics[width=1\linewidth]{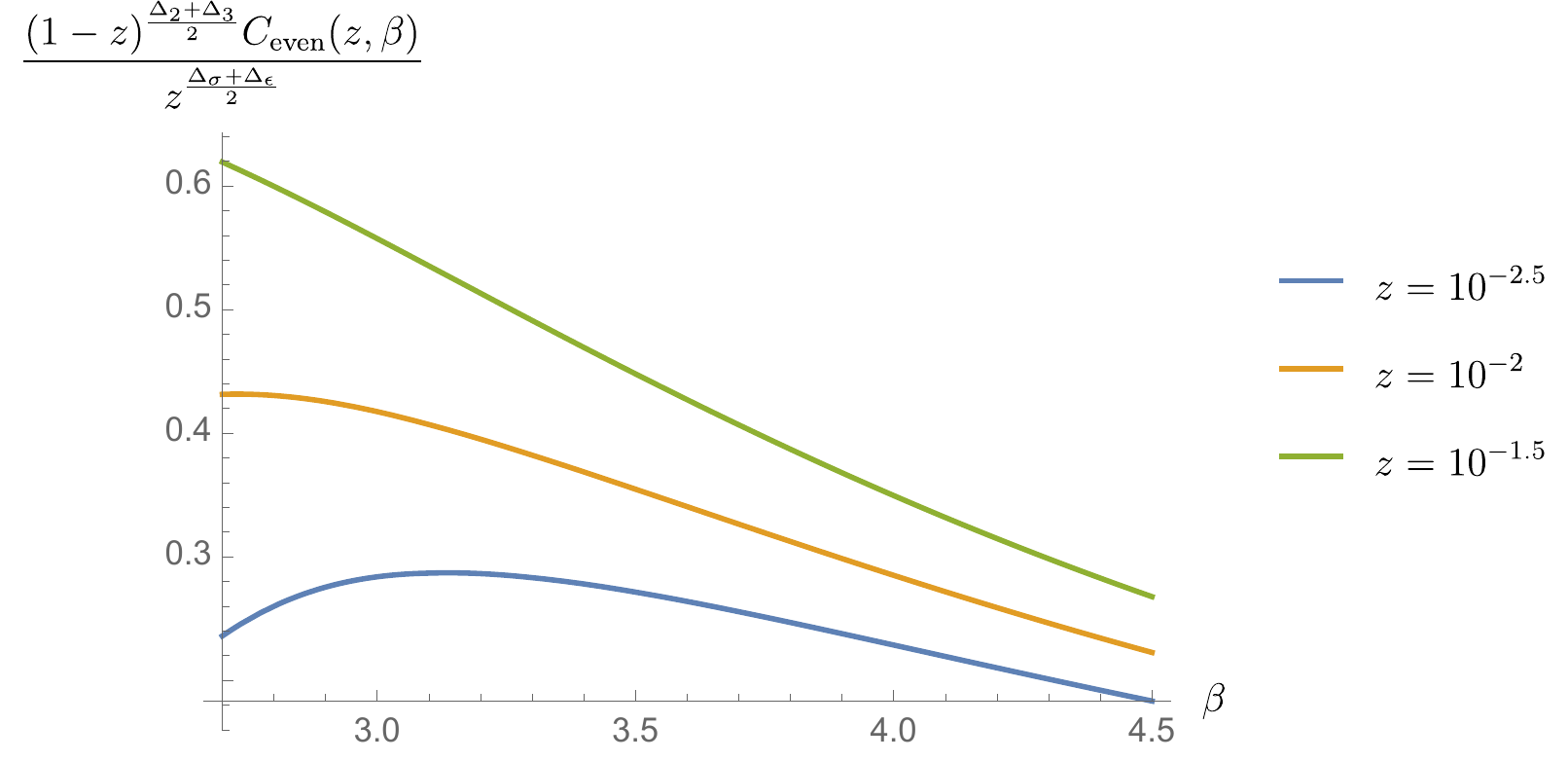}
  \caption{$(1-z)^{\frac{\Delta_2+\Delta_3}{2}}C_{\text{even}}(z,\beta)/z^{\frac{\Delta_{\sigma}+\Delta_{\epsilon}}{2}}$ for $\beta$ close to $\Delta_{\sigma}+\Delta_{
  \epsilon}+2$ for three different values of $z$. One can observe that the  function does not vanish for any $\beta$. This is in contrast with $(1-z)^{\frac{\Delta_2+\Delta_3}{2}}C_{\text{odd}}(z,\beta)/z^{\frac{\Delta_{\sigma}+\Delta_{\epsilon}}{2}}$  in fig.~\ref{fig:codd-s12} for the same range of $\beta$. }
  \label{fig:c0-2}
\end{figure}

Lastly, to recognize whether vanishing of $C_{\text{odd}}(z,\beta)$ is due to a subtle cancellation between the t-channel and u-channel, we study $(1-z)^{\frac{\Delta_2+\Delta_3}{2}}C_{\text{even}}(z,\beta)/z^{\frac{\Delta_{\sigma}+\Delta_{\epsilon}}{2}}$ for the same values of $\beta$ in fig.~\ref{fig:c0-2}. 

In fig.~\ref{fig:c0-2}, we can explicitly observe that indeed when the t-channel and the u-channel are added instead of subtracted as it is done in $C_{\text{even}}(z,\beta)$, there is no vanishing of the OPE coefficient.

To sum up, we studied the odd-sector of $[\sigma\epsilon]_0$ family through constructing $C_{\text{odd}}(z,\beta)$ and its derivative. We showed that the analytic calculation is consistent with the absence of global conserved current with spin 1 operator and this absence results from a very interesting conspiracy between the u-channel and t-channel terms.

 \subsection{Continuing to spin 0}

In this section, we try to push the analysis to see whether $[\sigma\epsilon]_0$ trajectory can contain a spin 0 operator. Again the accuracy of our analysis is moderate since the range of $z$ accessible to us is small (by an analysis similar to what has been done in section \ref{sec:spin2} we realize that $z<10^{-2}$ cannot be used)  and even in this region the result varies quite a bit since the expansion of the dDisc converges more slowly for such a small value of $\beta$ and more subleading twist operators need to be exchanged as the contribution of each operator falls with its twist as $1/\beta^{\tau'}$.

\begin{figure}[!htb]
  \centering
  \includegraphics[width=1\linewidth]{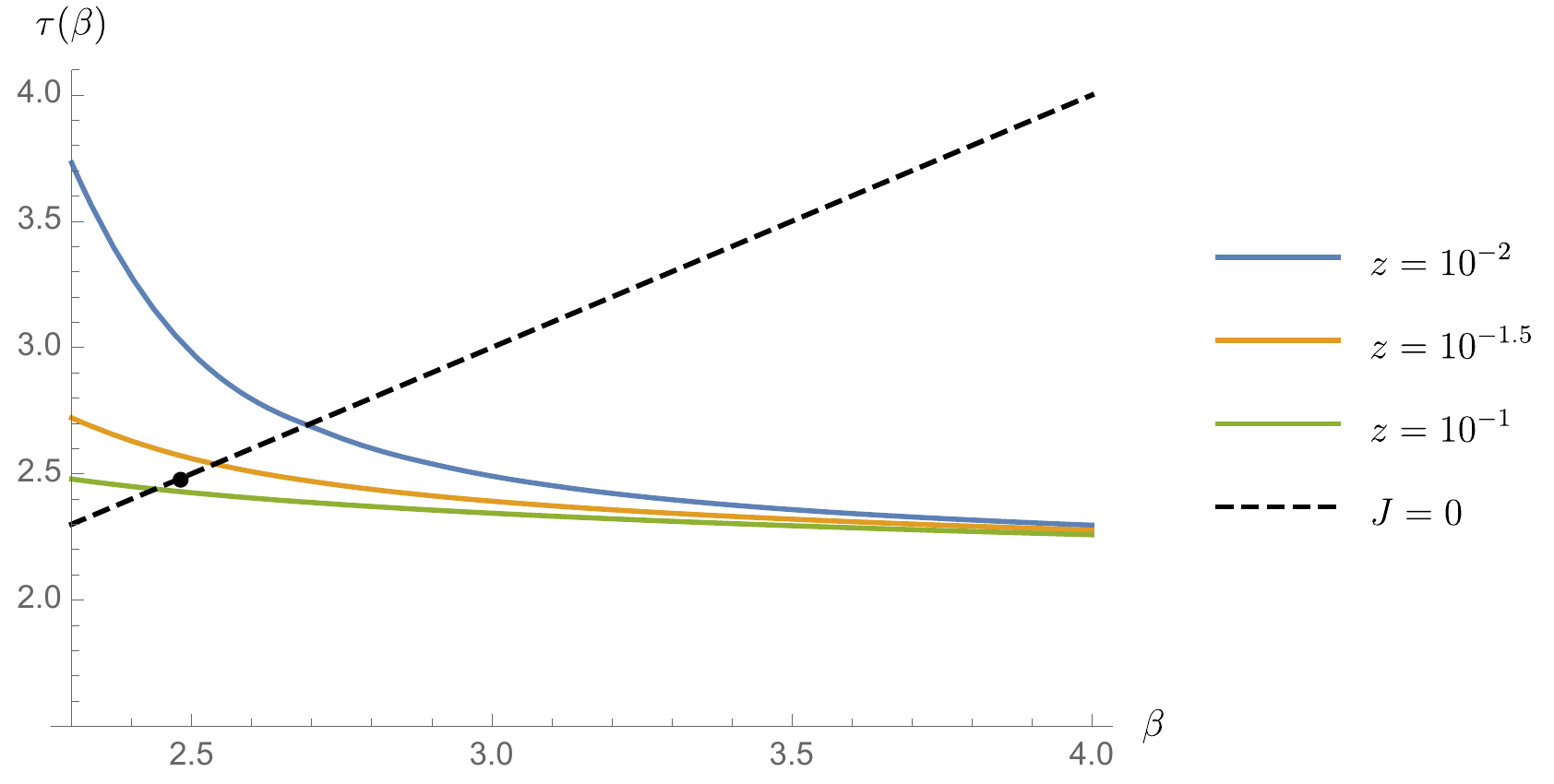}
  \caption{The intersection between the three curves $\tau(\beta)$ and the line $J=0$ is the predicted location of the spin 0 operator. The black dot on the $J=0$ line is the actual location of shadow of $\sigma$ operator.}
  \label{fig:c0-1}
\end{figure}

That being said, we can still proceed with extracting the twist of such low $\beta$ and intersect it with $J=0$ line to find out if the trajectory admits a spin 0 operator and if it does, what the twist of such spin 0 operator is. Once again, the difference of the result for different value of $z$, gives us an estimation of the error. This analysis is done in fig.~\ref{fig:c0-1}

We see in fig.~\ref{fig:c0-1} that the error of the analysis is indeed not negligible meaning that we cannot pin down the operator with great accuracy. However what is completely manifest from this analysis is that $[\sigma\epsilon]_0$ trajectory does include a spin 0 operator with dimension in the neighbourhood of $\Delta \sim 2.5$. If one calculates the squared of OPE coefficient of this operator, one gets $f_{\sigma\epsilon\tilde{\sigma}}^2\sim 1.15$  We make the conjecture that this operator is indeed shadow of $\sigma$ operator. As a support for this conjecture we remind the reader of the scaling dimension of shadow of sigma, $\tilde{\sigma}$ which is $\Delta_{\tilde{\sigma}}=3-\Delta_{\sigma}\simeq 2.48$ and its OPE coefficient, $f_{\sigma\epsilon\tilde{\sigma}}^2\simeq1.44$ (this is calculated using eqref.~\ref{eq:ope-shadow} which relates $c(\Delta,J)$ and $c(d-\Delta,J)$). For convenience, our results along with what is expected from numerical bootstrap is summarised in the table.~\ref{tab:shadow-sigma}.
\begin{table}[!htb]
\centering
\begin{tabular}{|l|l|l|l|l|}
\hline
 & $\Delta_{\tilde{\sigma}}$ & $f_{\sigma\epsilon \tilde{\sigma}}$ \\
\hline
Inversion Formula at $z=10^{-2}$ & 2.69 & 1.16\\
Inversion Formula at $z=10^{-1.5}$ & 2.53 & 1.10\\
Inversion Formula at $z=10^{-1}$ & 2.44 &1.15\\
Numerical Result & 2.48185 & 1.4393\\
\hline
\end{tabular}
\caption{Twist and OPE of spin 2 operator in $[\sigma\epsilon]$ derived analytically compared with the value derived from numerical bootstrap}
\label{tab:shadow-sigma}
\end{table}

Our analysis predicts that the spin 0 operator of $[\sigma\epsilon]_0$ is in the vicinity of shadow of $\sigma$ operator. However, the gap between the  OPE coefficient of shadow of $\sigma$ and the result obtained from the inversion formula indicates that even though our analysis is compatible with shadow of sigma belonging to $[\sigma\epsilon]_0$ trajectory,
using this method to predict quantitatively properties of the operator would not be numerically very effective.
This is similar to what we observed for $\epsilon$ operator in section.~\ref{sec:Epsilon}.

We also get the Chew-Frautschi plot for this analytic trajectory $[\sigma\epsilon]_{0,\text{even}}$ (as we did for $[\sigma\sigma]_0$ family in fig.~\ref{fig:intercept}) in fig.~\ref{fig:c0-2} which has been placed in section \ref{Leading $O(N)$-Fundamental Twist Family} for compariosn with similar trajectories in $O(N)$ model.


\section{Extended discussion}
\label{sect:discussion}

To shed light on the results presented in this paper, 
we give an extended discussion on the following aspects.
First, we discuss the qualitative distinctions between the Regge trajectories of transparent and opaque theories,
we compare 3D Ising with the critical $O(N)$ model at large $N$ (which is in the transparent class),
and we work out a novel formula showing that transparency implies regularity of the heavy spectrum.

\subsection{Transparent versus opaque theories}
\label{sect:CF}

When do we expect the spectrum to be analytic down to $J=0$?
Here we argue that, in many situations, this is closely related to asymptotic transparency.%
\footnote{A discussion along these lines was first presented by one of the authors
at the 2018 Azores workshop on the analytic bootstrap.}

\begin{figure}[!htb]
	\begin{subfigure}[t]{0.5\textwidth}
  \centering
  \includegraphics[width=1\linewidth]{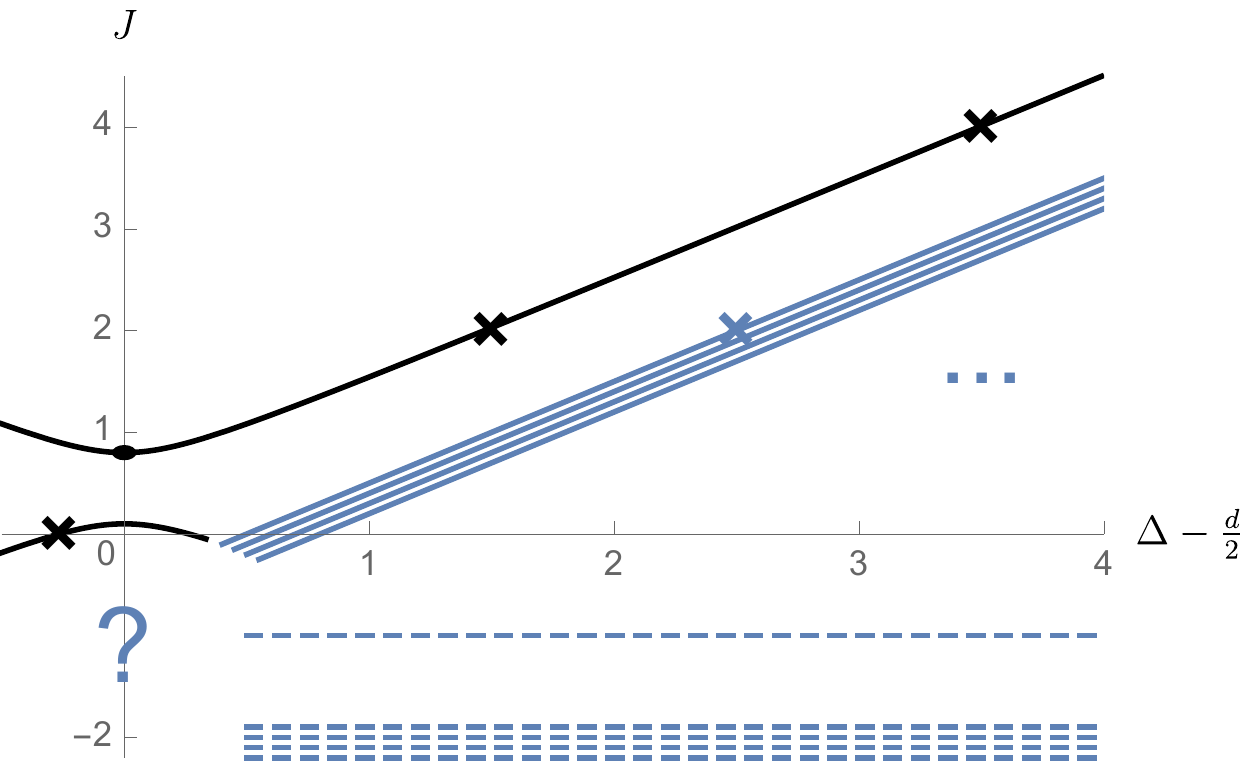}
  \caption{Asymptotically transparent}
  \end{subfigure} 
\quad  \begin{subfigure}[t]{0.5\textwidth}
  \centering
  \includegraphics[width=1\linewidth]{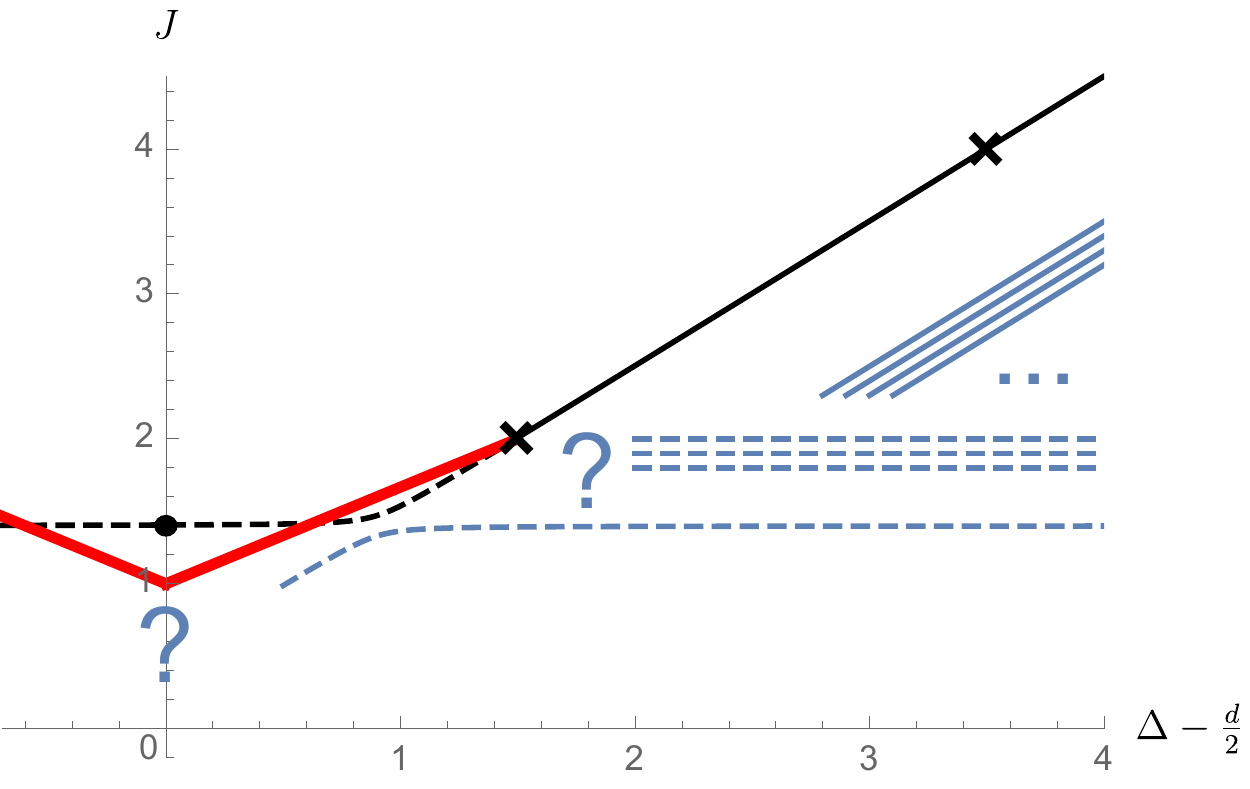}
  \caption{Asymptotically opaque}  \end{subfigure} 
  \caption{ \label{fig:CF}
Chew-Frautschi sketches for transparent and opaque theories.
Solid lines indicate multi-twist trajectories,
and dashed lines show possible BFKL-like horizontal trajectories.
Complicated behavior could occur where accumulation points of trajectories intersect.
 (a) In scalar-like theories, most serious complications would seem restricted to $J<0$ making
 analytic continuation to $J=0$ plausible.
 (b) In nonabelian gauge theories, here at weak coupling, the perturbative leading trajectory
has intercept $j_*^{\rm transient}>1$ (solid disk).
At higher orders in perturbation theory,
near-horizontal composites with ever-increasing spin must exist.
Singularities above the thick red lines must disappear nonperturbatively, by convexity.
}
\end{figure}

Let us try to sketch, more generally, what singularities we expect in the complex $(\Delta,J)$-plane.
We begin with the region of large spin and dimension.
There we certainly find double-twist trajectories, which have approximately constant twist
$\tau\approx \Delta_i+\Delta_j+2n$ and lie near to 45$^\circ$ in the figure.
More generally we also expect multi-twist operators,
built of products of many primaries and derivatives, and it is interesting to try and track their trajectories.
Since the number of local operators grows with spin, we expect the number of trajectories
to be infinite, likely accumulating at discrete twist values (with, presumably,
only a finite number of them having a nonzero OPE coefficient at a given integer spin).
These are the solid lines shown in fig.~\ref{fig:CF}.  
As explained in ref.~\cite{Kravchuk:2018htv},
Regge trajectories represent non-local operators, which reduce
to line integrals of local operators at the position of the crosses. 

Although general classification of nonlocal operators is still lacking, we also expect near-horizontal trajectories.
In a weakly coupled gauge theory these are well-known to arise as color-singlet combinations
of null-infinite Wilson lines $U(x_\perp)\propto \mathcal{P}e^{i\int_{-\infty}^{\infty} dx^+A_+(x^+,0^-,x_\perp)}$
where $x^\pm=t\pm x$ are lightcone coordinates.
The simplest such trajectory, the BFKL Pomeron, is labelled by the positions of two Wilson lines, where the
quantum number $\Delta$ is conjugate to their transverse separation (see
\cite{Lipatov:1996ts,Brower:2006ea,Cornalba:2008qf,Caron-Huot:2013fea} for various distinct perspectives).
Notice that since the gauge fields have spin 1 the integral $\int dx^+ A_+$ is formally boost-invariant
(momentarily neglecting the need to introduce rapidity cutoff); products of multiple Wilson lines
thus have the same spin (boost) quantum number.
More generally, non-local operators satisfy the standard addition law from Regge theory: 
\be
 J_{[\mathcal{O}_1\mathcal{O}_2]} \approx J_{\mathcal{O}_1} +J_{\mathcal{O}_2}-1\,, \label{spin law}
\ee
where the offset is due to the mismatching number of $dx^+$ on both sides.

The sharp difference between Lagrangian theories which contain vector bosons, and those which do not (``matter-like" theories) is where these near-horizontal trajectories lie.
A nonlocal composite of two scalars would give a single trajectory near $J\approx -1$, and the first accumulation point of
trajectories is delayed to $J\approx -2$; a composite of two fermions may produce a single trajectory near $J\approx 0$,
but it is still effectively isolated from more complicated composites.
In contrast, in gauge theories one immediately runs into infinitely many trajectories that mix with each other.
One reason BFKL were able to make progress is that mixing between states of different number of elementary Reggeized gluons is suppressed by two effects: by weak coupling and/or the planar limit, see \cite{Caron-Huot:2013fea}. 
Quantum corrections move the two-Reggeon intercept above 1 in both limits: the intercept is $j_*^{\rm transient}=1+\mathcal{O}(\alpha_s)$ at weak coupling, and at strong coupling $j_*^{\rm transient}\approx 2$ in holographic theories,
leading to interactions which grow with energies (opacity).

Now if exchange of one object grows like some power $e^{\eta(j_*^{\rm transient}-1)}$ at large boost,
one naturally expects double exchange to grow
twice as fast, giving an effective excitation of spin $2j_*^{\rm transient}-1>j_*^{\rm transient}$.  This argument seems rather unavoidable due to cluster decomposition in spacetime dimensions $d>2$, since excitations can be widely separated in the transverse plane.
It is not possible to have just a single trajectory with $j_*>1$, there must be an infinite tower!
The growth of course can only be transient because the correlator is bounded;
it is generally expected that the higher trajectories stop the growth rather than speed it up,
in the same way that the higher-order Taylor coefficients of the function $(1-e^{-x})$ limit its initial linear growth.
See \cite{Brower:2006ea} for further discussions; we do not have anything to add here about how saturation works,
if only to note that convexity requires that all singularities cancel below the red line in fig.~\ref{fig:CF}.

An important lesson from this discussion is that while in asymtptotically transparent theories it seems perfectly reasonable,
if numerically challenging, to analytically continue trajectories to $J=0$, in opaque theories there may be much more serious obstructions to crossing $J=1$.

\subsection{Analyticity to spin $J=0$ in the large-$N$ $O(N)$ model}\label{sect:critical}
 
 $O(N)$ models at large $N$ is a theory for which we have analytical control by using $1/N$ expansion. This theory is a specifically suitable theory for testing the ideas put forward in this paper as its operator's contents resembles the one in 3d Ising. This is because 3d Ising is given by $O(N)$ model at $N=1$. The leading $O(N)$-Bilinear twist families of the $O(N)$ model has been studied at large $N$ to order $1/N^2$ in \cite{Alday:2019clp}. We will reproduce their result of $O(N)$-Bilinear twist family with slightly different approach up to order $1/N$ and compare with our results in section \ref{sect:ising-equal}. In addition, we study the $O(N)$-Fundamental twist family up to order $1/N$ to study our conjecture in this model and compare with the results obtained in 3d Ising model in section \ref{sect:ising-unequal}.

$O(N)$ model is a theory of $N$ scalar fields $\phi_i$ that transform in the fundamental representation of $O(N)$. The OPE of these fields can be separated into three different tensor structures:
\begin{equation}
 \phi_i\times\phi_j = \sum_S \delta_{ij}\mathcal{O} + \sum_T \mathcal{O}_{(ij)} + \sum_A \mathcal{O}_{[ij]},
 \end{equation}
where $S$ stands for singlet of even spin, $T$ stands for symmetric traceless of even spin and $A$ stands for anti-symmetric tensors of odd spin. Similar to previous sections, we want to derive the CFT data (OPE coefficient and anomalous dimension) for low spin operators in the spectrum. 

\subsubsection*{Leading $O(N)$-Bilinear Twist Family}
First we review how this works for the low spin operators in $[\phi_i\phi_j]$ double-twist families, which are the leading twist families of the $O(N)$ theory.  This discussion will follow closely \cite{Alday:2019clp}.
The data for the spectrum will be derived in the limit that $N$ is large and is thus given as an analytic expansion in $1/N$. We look at a 4-point function of these scalar fields, which can be again separated in three independent tensor structures:
\be\ba
 &\hspace{20mm} x_{12}^{2\DeltaP}x_{34}^{2\DeltaP}\langle\phi_i(x_1)\phi_j(x_2)\phi_k(x_3)\phi_l(x_4)\rangle \\
 &= \delta_{ij}\delta_{kl}\,\mathcal{G}_S(u,v) 
 + \left(\delta_{il}\delta_{jk}+\delta_{ik}\delta_{jl}-\frac{2}{N}\delta_{ij}\delta_{kl}\right)\mathcal{G}_T(u,v) 
+  \Big(\delta_{il}\delta_{jk}-\delta_{ik}\delta_{jl}\Big)\mathcal{G}_A(u,v) \,.\ea\ee
The functions that appear are just the usual conformal block expansion but with the sums only over the operator in the given sector. We can also expand this correlator in the t and u channels to obtain the following crossing symmetry equation: 
\begin{equation}\label{crossing}\begin{split} f_S(u,v)=&
\frac{1}{N}f_S(v,u)+\frac{N^2+N-2}{2N^2}f_T(v,u)+\frac{1-N}{2N}f_{A}(v,u)
\\ & \rightarrow\frac{1}{N}f_S(v,u)+\frac{1}{2}f_T(v,u)+\frac{1}{2}f_{A}(v,u)\, ,
\\ f_T(u,v)=&f_S(v,u) + \frac{N-2}{2N}f_T(v,u)+\frac{1}{2}f_A(v,u)\ , 
\\ f_A(u,v)=&-f_S(v,u)+\frac{2+N}{2N}f_T(v,u)+\frac{1}{2}f_A(v,u)\ .
\end{split}\end{equation}
The leading equations at large $N$ are the main tool of this section. The crossing equations into the u-channel are essentially the same but with minus signs everywhere in the A sector. The functions that appear are defined by $f(u,v)=u^{-\DeltaP}\mathcal{G}(u,v)$ to incorporate the factors coming from crossing. The full correlation function can be expanded in $1/N$ and the explicit expression at $N=\infty$ follows from Wick contraction and is as follows :
\begin{equation}
\mathcal{G}_{ijkl}^0(u,v)=\delta_{ij}\delta_{kl}+u^{\frac{d-2}{2}}\delta_{ik}\delta_{jl}+\left(\frac{u}{v}\delta_{il}\right)^{\frac{d-2}{2}}\delta_{jk}.
\end{equation}
This equations means that we have the following expansion for each decomposition:
\begin{equation}
\label{expansion 1/n}
\begin{split}
\mathcal{G}_S(z,\zb)&=1+\frac{1}{N}\mathcal{G}_S^{(1)}(u,v)+\ldots\\
\mathcal{G}_T(z,\zb)&=u^{\frac{d-2}{2}}\left(1+\frac{1}{v^{\frac{d-2}{2}}}\right)+\frac{1}{N}\mathcal{G}_T^{(1)}(u,v)+\ldots\\
\mathcal{G}_A(z,\zb)&=u^{\frac{d-2}{2}}\left(1-\frac{1}{v^{\frac{d-2}{2}}}\right)+\frac{1}{N}\mathcal{G}_A^{(1)}(u,v)+\ldots
\end{split}
\end{equation}
Now we are equipped to look at the N scaling of different terms in the crossing. First we see that for T and A operators, there is a whole tower of double twist operators exchanged at $N\rightarrow \infty$ limit, so their OPE coefficients are of order 1 and are the ones from generalized free fields. The double twist operators in the S sector do not appear at this order so their OPE coefficients must have term that scale as a negative power of N ( $1/N^{1/2}$). From dimensional analysis we can find that the leading scaling dimension of $\phi$ is $1/2$. This means that double twist operators of spin $J$ have a leading dimension of $1+J$. There is however still the possibility that the scalars that appear in the OPE are shadows of the double twists, similar for the $\sigma$ operator in the $[\sigma\epsilon]_0$ family of the 3d Ising model. This possibility will be incompatible with the leading $N$ behaviour of the correlator for the T sector. However, for the S sector, if the shadow does not appear we indeed run into trouble, as the crossing would imply that  T and A operator do not have anomalous dimension of order $1/N$. We will then call the operator appearing in the OPE $S$ and it has a leading dimension of $d-\Delta_{[\phi\phi]_{S,0}}=2$.

The idea for studying this theory using the inversion formula is to use the crossing symmetry equations (\ref{crossing}) to understand how to combine the different elements that appear in the generating function $C(z,\beta)$. Crossing then dictates what combination of $f_R(v,u)$ appears in t-channel correlator of the inversion formula for $[\phi\phi]_{R}$ double twist operators in each sector.


One can see from from crossing eqs.~\eqref{crossing}, the leading behaviour of the OPE coefficients is then given by the identity contribution. This can be found by evaluating \eqref{eq:identity-contribution} at $\beta=1+2J$ and multiplying by 2 because the identity appears both in the t and u channels. The result for $S$ can be deduced from evaluating the answer at spin 0 and transforming to the shadow with (\ref{eq:ope-shadow}). We find:
\be\ba & f^2_{\phi\phi S}=\frac{4}{\pi^2N}+O\left(\frac{1}{N^2}\right), \\ & f^2_{\phi\phi[\phi\phi]_{T/A}}(J)=N f^2_{\phi\phi[\phi\phi]_S}(J)=\frac{2\Gamma(J)\Gamma\left(J+\frac{1}{2}\right)^2}{\pi\Gamma(2J)\Gamma(J+1)}+O\left(\frac{1}{N}\right) \,.\ea\ee

The next step is to calculate anomalous dimensions. One important point to emphasize with this setup is that whether we are summing over the whole twist family or considering only the exchange of a single operator, taking $z\rightarrow 0$ in the first step does not create any problem. This is because the analyticity of the data in $1/N$ prevents problems caused by loss of the log$^2$ terms when the  wrong order of limits are taken and this problem is simply resolved because $\log^2$ terms  appear only in next order in $1/N$. 

The simplest anomalous dimensions to first calculate are those in the T and A families because they receive contribution only from $S$ at order $1/N$.  This is done using the inversion formula at $z\rightarrow 0$ in eq.~\ref{eq:7F6} together with $f_{\phi\phi\phi_S}$ calculated above and it leads to
\be \gamma_{[\phi\phi]_{T/A}}(\beta)=-\frac{8}{\pi^2N(2J+1)(2J-1)} \,.\ee
One nice use for this result comes from the fact that we should recover that the spin 1 operator in the A sector is a conserved current with dimension 2. This can be used to fix the correction to the dimension of $\phi$ since the dimension of the spin 1 operator is given by $2\Delta_{\phi}+1+\gamma_{[\phi\phi]_A}(1)$. The result is
\be \Delta_{\phi}=\frac{1}{2}+\frac{4}{3\pi^2N} \,,\ee
which is consistent with the previous calculations summarized in \cite{Kos:2013tga, Alday:2019clp, Vasiliev:1982dc}.

We can continue by calculating the anomalous dimensions of the singlet double twists. Here the double twist operators in the other sectors contribute. We could again use eq.~\ref{eq:7F6}  but it gets complicated for general spin. We instead lean on the fact that conformal blocks for conserved currents, which the double twists are at leading order, are very simple. Using the method described in Appendix \ref{app:collinear} we find that for conserved currents the coefficient of the log is
\be \eval{\sqrt{\frac{\zb}{1-\zb}}G_{\Delta,J}(1-z,1-\zb)}_{\log}=-\frac{2\Gamma(2J+1)}{\Gamma\left(J+\frac{1}{2}\right)} \,.\ee
This can be used directly in (\ref{eq:generating_int}) along with the data already found for the T and A double twist operators. The contributions are exactly the same for both sectors, except that they contribute with even and odd spins. The result for the sum of the contributions from $\phi_S$ (same as for T and A sectors) and the double twists is
\be \gamma_{[\phi\phi]_S}(J)=-\frac{8}{\pi^2N(2J-1)} \,.\ee
We have now come to a point where an consistency check is possible. Indeed the spin 2 operator in this family should be the stress tensor and it should have a dimension of 3, which we find to be the case.

\begin{figure}[!htb]
  \centering
  \includegraphics[width=1\linewidth]{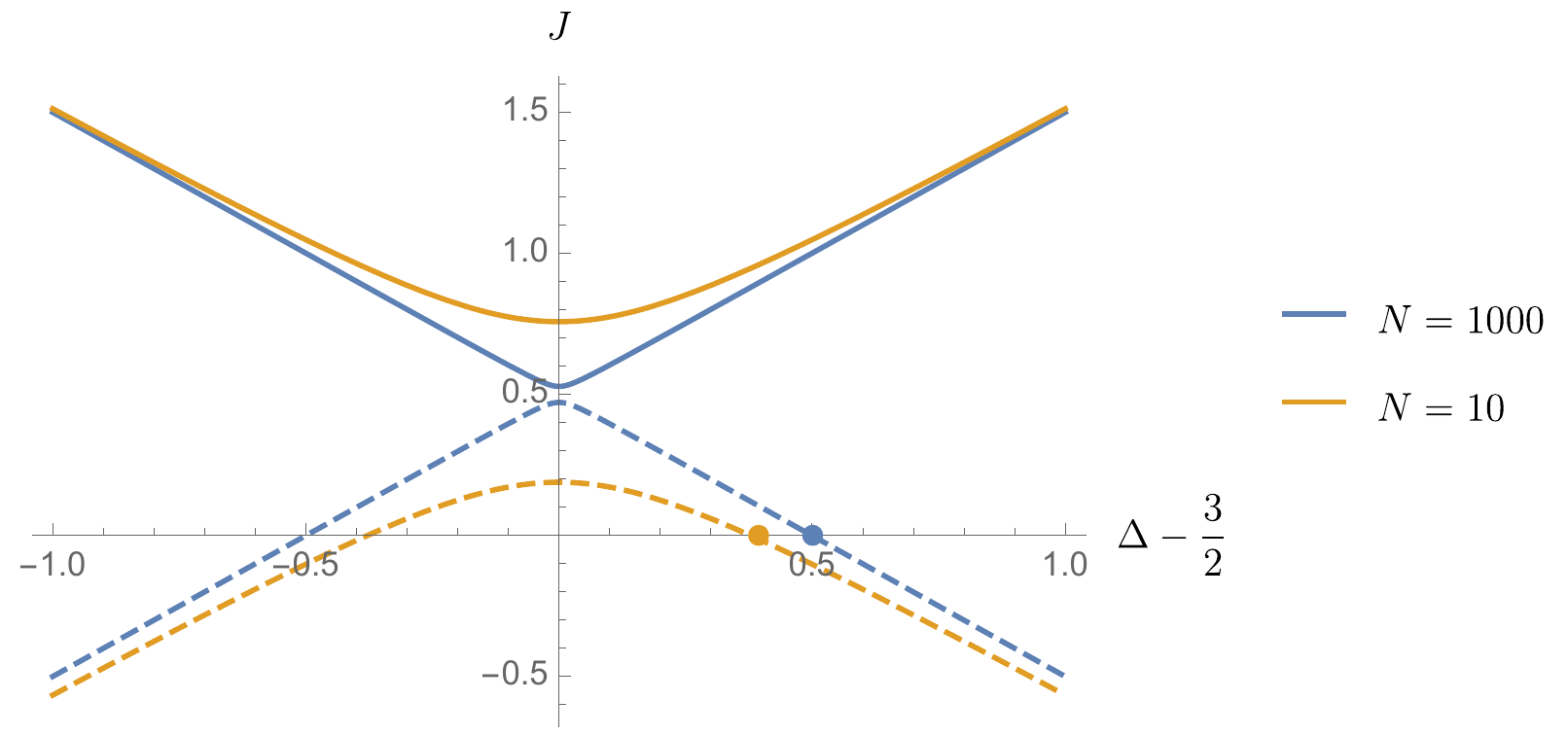}
  \caption{The Chew-Frautschi plot for $[\phi\phi]_S$ and its analytic continuation (dashed lines) of the $[\phi\phi]_{0,S}$ is plotted for $N=1000$ and $N=10$. We see that as $N$ increase the intercept approaches $1/2$ from above.
  Note that the curves plotted here are valid up to order $1/N$. Dots indicate the spin-0 $S$ operator.}
  \label{fig:ON-intercept}
\end{figure}

As a concluding remark we plot the Chew-Frautschi plot of $[\phi\phi]_{0,S}$ at $N=1000$ in fig.~\ref{fig:ON-intercept}. The  analogous plot for $[\sigma\sigma]_0$ is fig.~\ref{fig:intercept}. At $N\rightarrow \infty$ the intercept approaches $1/2$. As N decreases we see that the value of the intercept increases:
\begin{equation}
\label{eq:intercept-ON}
j_*=\frac{1}{2} - \frac{8}{3 \pi^2 N} + \sqrt{\frac{8}{\pi^2 N}}
\end{equation}
This result can be compared with our result for the intercept in section \ref{sec:intercept}. We emphasize that for small $N$ we do not expect this formula to capture the true physics. As mentioned above the intercept for $O(2)$ model is recently calculated to be $\sim 0.82$ in \cite{Liu:2020tpf}, where eq.~\ref{eq:intercept-ON} would predict 1.00152.
The difference for 3D Ising is of course is more drastic; eq.~\ref{eq:intercept-ON} for $N=1$ predicts the intercept to be 1.13013, we see indeed in section \ref{sec:intercept} that it is $\sim 0.8$.
However, the formula should be reliable for sufficiently large $N$.

Another remarkable fact is that if we analytically continue the Chew-Frautschi plot to $J$ smaller than the intercept, we can recover the spin 0 operator of $[\phi\phi]_0$ and its shadow (S) on the continued curve (dashed line in fig.~\ref{fig:ON-intercept}). 
\footnote{Analytic continuation of the leading $Z_2$ even trajectory to $J<J_0$ has also been studied in $\epsilon$-expansion in \cite{Alday:2017zzv}. There it was shown that one can discover $\epsilon$ operator on the analytically curve as the shadow of spin 0 operator in $[\sigma\sigma]_0$ family. One might hope to apply the same procedure to 3d Ising and analytically continue the trajectory in fig.~\ref{fig:intercept} to find $\epsilon$ operator, however since we do not have an analytical expression for the trajectory, this continuation cannot be done in a numerically controlled convincing way.}

\subsubsection*{Leading $O(N)$-fundamental twist family}
\label{Leading $O(N)$-Fundamental Twist Family}
 Now we turn to the leading $O(N)$-Fundamental twist family, $[\phi_i S]_0$ . We study this family by considering the 4-point function $\langle \phi_i S S \phi_j\rangle$. The index structure of crossing equations is trivial for this correlator as there is one possible option, $\delta_{ij}$. This means that we can discard the indices from crossing and consequently from inversion formula and use its stripped version as:
 \begin{equation}
\mathcal{G}(u,v)=\frac{u^{\frac{\Delta_1+\Delta_2}{2}}}{v^{\frac{\Delta_2+\Delta_3}{2}}}\mathcal{G}(v,u)
 \end{equation}
 To obtain information about $[\phi_i S]_0$ by using the inversion formula, we first need to calculate the t-channel and u-channel dDisc of the correlator. For the $1/N$ expansion $\text{dDisc}_t$ we have the identity operator which is the only operator exchanged at $O(N^0)$, then at the next order we have the exchange of operator S which is a single twist operator and its $1/N$ suppression comes from its OPE coefficient, $f_{\phi\phi S}f_{SSS}$ (In fact there is additional $1/N$ suppression because of  $\sin (\Delta_S-2\Delta_S)/2$, however this is cancelled with the factor of $\Gamma(1-\Delta_S/2)$ from the inversion formula as in eq.~\ref{eq:gamma}).  Contribution of all the double twist operators are additionally suppressed with a factor of $1/N$ due to the sine factors of dDisc (see eq.~\ref{eq:dDisc}). To calculate the u-channel discontinuity we exchange operators 1 and 2. Then the leading contribution to this expansion comes from the exchange of $\phi$. The large $N$ scaling of this contribution can be assumed to firstly comes from the OPE coefficient $f_{\phi\phi S}^2$. In addition, there is an additional $1/N$ suppression because $S$ has integer scaling dimension: the sine factors which can be calculated to be
 \begin{equation}
\sin [\pi\frac{\Delta_{\phi}-\Delta_S-\Delta_{\phi}}{2}]^2,
 \end{equation}
 scales as $1/N^2$ . However, only one of these will contribute as the other one gets cancelled with a factor of $\Gamma(1-\Delta_S/2)$ in the inversion integral using the following identity:
\begin{equation}
 \label{eq:gamma}
\Gamma(z)\gamma(1-z)=\frac{\pi}{\sin(\pi z)}
\end{equation}
 At the end we observe that the exchange of $\phi$ is not suppressed by $1/N$ as might have first guessed but it is suppressed with a factor of $1/N^2$. Now, we have enough information to obtain the leading order OPE coefficient and anomalous dimension of $[\phi_i S]_0$ operator. For our purpose which is to compare with section \ref{sect:ising-unequal}, we would like obtain the data for spin 0 operator of this family and verify whether this operator can be the shadow of $\phi$. This would be analogues to our conjecture that shadow of $\sigma$ belongs to $[\sigma\epsilon]_0$ family. The leading contribution to the OPE coefficient of $[\phi S]$ family comes from the identity using eq.\eqref{eq:identity-contribution}:
 \begin{equation}
 f_{\phi S [\phi S]_{0,j}}^2=\mathcal{I}^{(\Delta_{\phi},\Delta_S)}(2j+\Delta_{\phi}+\Delta_S)
 \end{equation}
 If one uses this formula for the spin 0 operator,  i.e., $\beta=5/2$  one obtains the OPE coefficient at order $O(N^0)$ to be 1. Now we can use the shadow transform given in eq.~\eqref{eq:ope-shadow} to obtain the $f_{\phi S \tilde{[\phi S]}_{0,0}}$ to be :
 \begin{equation}
 \label{eq:ope-ON-2}
 f_{\phi S \tilde{[\phi S]}_{0,0}}^2=0+\frac{4}{\pi^2N}
 \end{equation}
 Notice that $f_{\phi S \tilde{[\phi S]}_{0,0}}^2$ vanishes at order $N^0$, this is because the shadow transform gives zero at leading order for operator with dimension $\frac{d-2}{2}$ as can be easily seen from eq.~\eqref{eq:ope-shadow}. The fact that $f_{\phi S \tilde{[\phi S]}_{0,0}}^2$ matches with $f_{\phi \phi S}^2$ up to order $1/N$ is an evidence for our conjecture that $\phi_i$ operator is the shadow of the spin 0 operator in the $[\phi_i S]_0$ family.

 Interestingly, once we assume analyticity at spin 0, we can have a prediction for the OPE coefficient $f_{SSS}$ at leading order which is otherwise not trivial to obtain. This can be done by imposing that the leading order anomalous dimension of spin 0 operator in $[\phi_i S]_{0,0}$ is such that 
\begin{equation}
d-\Delta^{(0)}_{[\phi S]_{0,0}}-\gamma_{[\phi S]_{0,0}}^{(1)}=\Delta^{(0)}_{\phi}+\gamma_{\phi}^{(1)}+\Delta^{( 0)}_{S}+\gamma_{S}^{(1)}
\end{equation} 
 
\begin{figure}[!htb]
	\begin{subfigure}[t]{0.5\textwidth}
  \centering
  \includegraphics[width=1\linewidth]{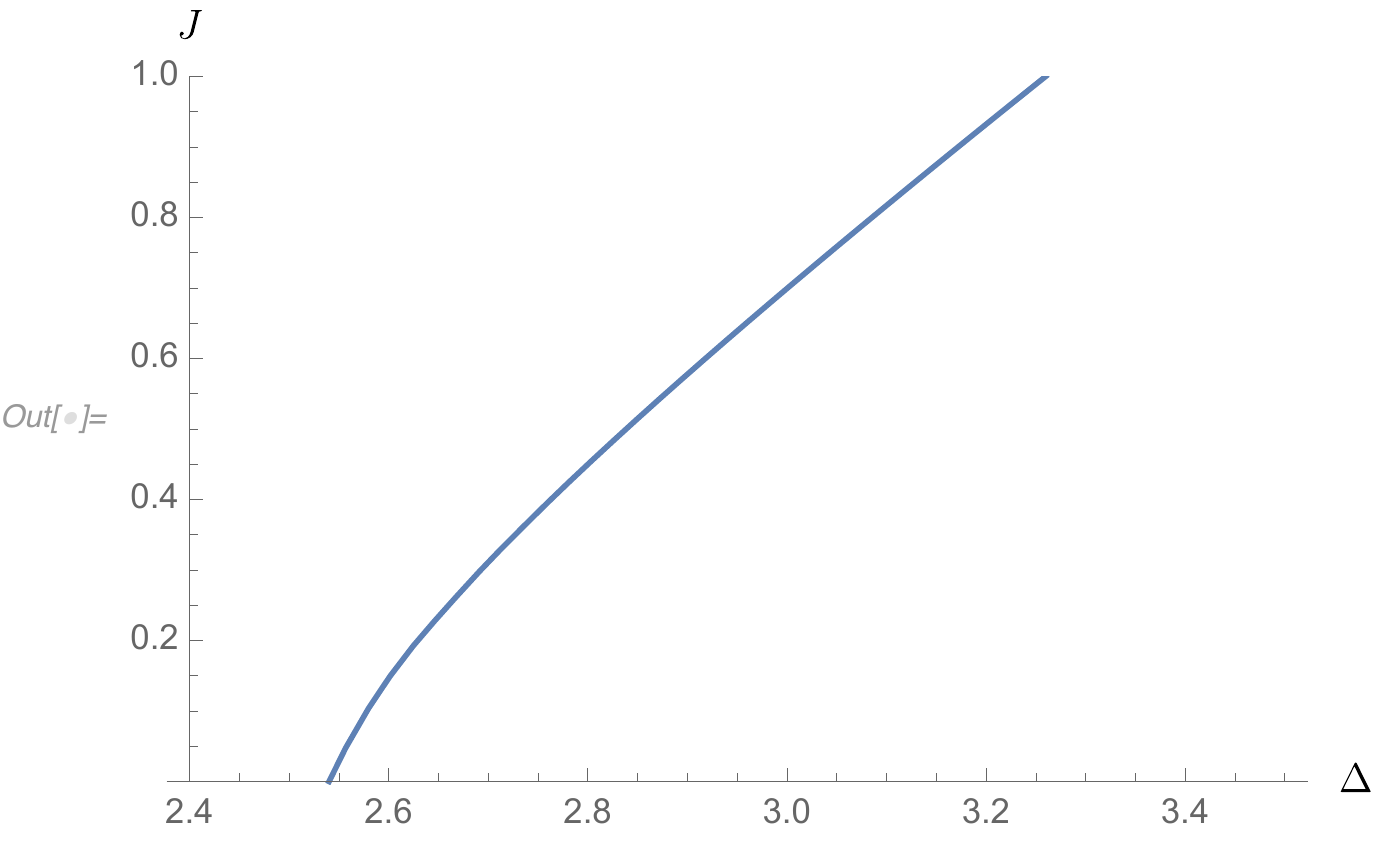}
  \caption{ $[\sigma\epsilon]_{0,\text{even}}$ family obtained at $z=10^{-1.5}$.}
  \label{fig:c0-ising}
  \end{subfigure} 
  \begin{subfigure}[t]{0.5\textwidth}
  \centering
  \includegraphics[width=1\linewidth]{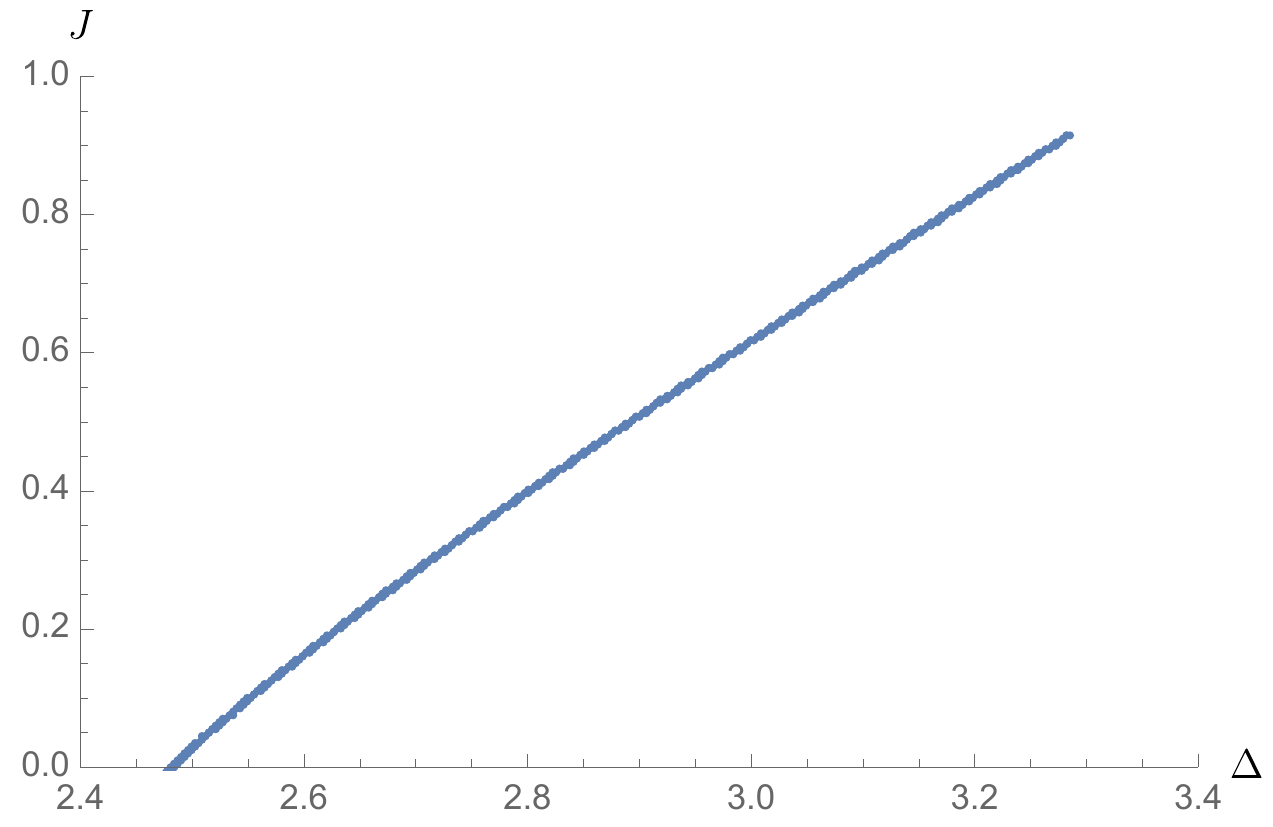}
  \caption{ $[\phi_i S]_{0}$ family for $N=6$  including correction of order $1/N$.}
  \label{fig:c0-ON}
  \end{subfigure} 
  \caption{Chew-Frautschi plot for the tow families mentioned above. See fig.~\ref{fig:intercept} and fig.~\ref{fig:ON-intercept} for similar plots of the leading trajectory in 3D Ising and $O(N)$ model respectively.}
\end{figure}

$\gamma_{[\phi S]_{0,0}}^{(1)}$ comes from the exchange of S in the t-channel and is proportional $f_{\phi\phi S}f_{SSS}$. This allows us to predict the value of $f_{SSS}$ in terms of the other know variables to be:
\begin{equation}
f_{SSS}=0+\frac{2}{\pi \sqrt{N}}
\end{equation}
The Chew-Frautschi plots of $[\phi_i S]_0$ as well as $[\sigma\epsilon]_0$ are plotted in fig.~\ref{fig:c0-ising} and ~\ref{fig:c0-ON} for comparison of their similarity.

\subsection{Implications for anomalous dimensions heavy operators?}
\label{sect:predictions}

What does transparency imply for the spectrum of heavy operators?
Naively one may expect that it implies a certain \emph{regularity} in the spectrum,
to prevent different components of the wavefunctions from arriving with random phases.
This subsection summarizes our attempt to derive a quantitative version of this statement.
The argument is similar to that used already in \cite{Cornalba:2006xm}
to show that, in holographic CFTs, anomalous dimensions of large-$(\Delta',J')$
OPE data in the $t$-channel grows with energy $\Delta'$ in a way controlled by exchange
of a $s$-channel Reggeized graviton. 
Our proposed formula, eq.~(\ref{sine squared average}), conversely shows
that an intercept $j_*<1$ implies decaying anomalous dimensions: a \emph{regular spectrum}.

The results in ref.~\cite{Cornalba:2006xm} were obtained pre-Conformal Regge theory using the
so-called impact parameter representation, and were derived in the context the language of holography.
We try to give them a fresh look in light of ref.~\cite{Costa:2012cb},
which allows us to strip the formula from its holographic context.
We proceed in two steps: first we work out implications for the Regge limit of the correlator in $(z,\zb)$-space,
then we convert those to heavy cross-channel operators.  The second step will rely on
an unproven identity about blocks in eq.~(\ref{relative Regge}), but otherwise we believe that all steps are rigorous.
The first step is achieved by the Watson-Sommerfeld resummation of our eq.~(\ref{eq:ope-principal-series}),
as given in eqs.~(5.21) and (5.22) of ref.~\cite{Kravchuk:2018htv}
(restricting the blocks to the leading power given in eq.~(\ref{limit block}); see also \cite{Costa:2017twz}):\footnote{
In our conventions, eq.~(5.22) of ref.~\cite{Kravchuk:2018htv} reads, to leading power:
\be
e^{i\pi(a+b)}F_{\Delta,J}(z,\zb)^{\circlearrowleft}-F_{\Delta,J}(z,\zb) \to
 C_{1-\Delta}\left(\frac{z+\zb}{2\sqrt{z\zb}}\right)\frac{1}{2\pi i \kappa(\Delta+J)} (z\zb)^{\frac{1-J}{2}}\,.
\ee
}
\be
 e^{i\pi(a+b)}\mathcal{G}(z,\zb)^{\circlearrowleft}-\mathcal{G}(z,\zb) \to\!\!
 \int\limits_{\frac{d}{2}-i\infty}^{\frac{d}{2}+i\infty} \frac{d\Delta}{2\pi i}\  C_{1-\Delta}\left(\frac{z+\zb}{2\sqrt{z\zb}}\right)
  \Res\limits_{J=j_*(\Delta)} \left[
  \frac{c^t(\Delta,J)+e^{-i\pi J}c^u(\Delta,J)}{\kappa(\Delta+J)(e^{-2i \pi J}-1)} (z\zb)^{\frac{1-J}{2}}\right] \label{G regge}
\ee
The formula simplifies significantly when considering the dDisc in eq.~(\ref{eq:dDisc}). 
We need to add the complex conjugate conjugation path,
which gives the same thing with just $i\pi\mapsto -i\pi$ inside the square bracket,
and eq.~(\ref{G regge}) reduces to:
\be
\lim_{z,\zb\to 0} {\rm dDisc}\ \mathcal{G}(z,\zb) =\int\limits_{\frac{d}{2}-i\infty}^{\frac{d}{2}+i\infty} \frac{d\Delta}{2\pi i}\
 C_{1-\Delta}\left(\frac{z+\zb}{2\sqrt{z\zb}}\right)
 \Res\limits_{J=j_*(\Delta)} \left[ \frac{c^t(\Delta,J)}{2\kappa(\Delta+J)}(z\zb)^{\frac{1-J}{2}}\right]\,.
\label{dDisc regge}
\ee
To our knowledge, this formula has not appeared in print before.
Notice that the $u$-channel coefficient has canceled, as well as the integer-spin poles:
the $t$-channel dDisc is directly related to the $t$-channel contribution to the Lorentzian inversion formula.
This is perhaps not too surprising given the form of the
Lorentzian inversion formula in eq.~(\ref{eq:inversion-integral}):
as a consistency check, we tried inserting eq.~(\ref{dDisc regge}) back into the latter,
and we indeed recover $c^t(\Delta,J)$ using the orthogonality relation between $C_{1-\Delta}$ along the principal series.
In other words, to leading power, eq.~(\ref{dDisc regge}) is just the inverse of the inversion formula.

Because of this interpretation, we can assume that eq.~(\ref{dDisc regge}) is valid even for correlators which do
not grow in the Regge limit, even though the validity of eq.~(\ref{G regge}) in this case does not strictly follow
from the works~\cite{Costa:2012cb,Costa:2017twz,Kravchuk:2018htv} and may require further discussion \cite{Sandor:2020}.

What does eq.~(\ref{dDisc regge}) imply for heavy $t$-channel operators?
We follow the logic of refs.~\cite{Cornalba:2006xm,Mukhametzhanov:2018zja}, where the $z\sim \zb\to 0$ limit
of the correlator is related to $t$-channel operators with large dimension and spin $\Delta'\sim J' \sim 1/\sqrt{z}$.
We review the Euclidean case \cite{Mukhametzhanov:2018zja}.
The starting point is the fact that the unit operator in the $s$-channel is reproduced by an infinite sum over $t$-channel operators:
\be
 1= \sum_{n,J'=0}^\infty (1+(-1)^J) P^{(0)}(\Delta',J')
\underbrace{\left(\frac{z\zb}{(1{-}z)(1{-}\zb)}\right)^{\Delta_\sigma} G_{\Delta',J'}(1{-}z,1{-}\zb)}_{G^{(t)}_{\Delta',J'}(z,\zb)}
\ee
where $\Delta'=2\Delta_\sigma+2n$ and the average spectral density
$P^{(0)}(\Delta',J')\equiv P^{\Delta_\sigma}_{\Delta'-J',J'}$ is defined in eq.~(\ref{MFT}).
The term with $(-1)^J$ is regular in the $z,\zb\to 0$ limit and can be ignored for the present discussion.
More generally, to study the limit we rewrite the OPE sum (exactly) as an integral
\be
\label{asympt OPE}
\mathcal{G}(z,\zb) = \int \frac{d\Delta' dJ'}{2}\ c^{(0)}(\Delta',J') G_{\Delta',J'}^{(t)}(z,\zb) \times \langle C(\Delta',J')\rangle\,,
\ee
where the bracket is a sum over $\delta$-function at each local operators, divided by the mean free spectral density:
\be
 \langle C(\Delta',J') \rangle \equiv \sum_{\Delta'} \frac{f^2_{\sigma\sigma \mathcal{O}'}}{C^{(0)}(\Delta_{\mathcal{O}},J')}\delta(\Delta'-\Delta_{\mathcal{O}})\,.
\ee
Note that our normalization of the spectral density is slightly different from \cite{Mukhametzhanov:2018zja}.
As discussed there, the fact that the $z,\zb\to 0$ limit is dominated by identity implies that
$\langle C(\Delta',J') \rangle$, after suitably smearing out in $\Delta'$ and $J'$, goes to 1 asymptotically,
with computable corrections. Namely,
exchange of a $s$-channel scalar operator of dimension $\Delta$ produces a correction suppressed by a relative
$\Delta'^{-2\Delta}$:
\be \label{relative Eucl}
 (z\zb)^{\frac{\Delta}{2}} = \int \frac{d\Delta' dJ'}{2}\ c^{(0)}(\Delta',J') G_{\Delta',J'}^{(t)}(z,\zb) \times 
 \frac{\gamma(0)\gamma(d-2)}{\gamma(\Delta)\gamma(\Delta+d-2)}(h'\hb')^{-\Delta}  + \mbox{subleading}\,,
\ee
where
\be
 \gamma(x) = \Gamma\left(\frac{\Delta_1+\Delta_2-x}{2}\right)\Gamma\left(\frac{\Delta_3+\Delta_4-x}{2}\right)
\ee
is a combination which will re-occur often, and
\be \label{h}
 h'=\Delta'+J'-1, \quad \hb'=\Delta'-J'-d+1
\ee
are combinations which transform simply ($h'\mapsto \pm h'$ or $\pm \hb'$)
under all SO$(d,2)$ Weyl reflections ($\Delta\leftrightarrow d-\Delta$, $\Delta\leftrightarrow 1-J$ and $j\leftrightarrow 2-d-J$).
Although in eq.~(\ref{relative Eucl}) we focus on the leading term at large-$\Delta'$ and $J'$,
we find that using  the Weyl-friendly form of $h$ and $\hb$ makes
subleading terms smaller (suppressed by a relative $1/\Delta'^2$).

What about $s$-channel operators with spin?
Using the Casimir recursion in Dolan-Osborn coordinates (see section 2 of \cite{Hogervorst:2013sma})
we could compute exactly the OPE coefficient
dual to a power of $\frac{z \zb}{(1-z)(1-\zb)}$ times a Gegenbauer polynomial.
This will be detailed elsewhere \cite{Yue-Zhou} and here we simply record a compelling formula
that we observed for the leading behavior at large $\Delta'$ and $J'$:
\be\begin{aligned} \label{relative Eucl 2}
 (z\zb)^{\frac{\Delta}{2}}C_J\left(\frac{z+\zb}{2\sqrt{z\zb}}\right)
 &= \int \frac{d\Delta' dJ'}{2}\ c^{(0)}(\Delta',J') G_{\Delta',J'}^{(t)}(z,\zb)
  \times 
\frac{\gamma(0)\gamma(d-2)}{\gamma(\Delta-J)\gamma(\Delta+J+d-2)}
 \\ &\hspace{20mm} 
\times(h'\hb')^{-\Delta} C_J\left(\frac{h'^2+\hb'^2}{2h'\hb'}\right)  + \mbox{subleading}\,.
\end{aligned}\ee
Notice the parallel between $z$ and $h^{-2}$ on the two sides of the formula, with Gegenbauers turning onto Gegenbauers.
This is the key observation made long ago in ref.~\cite{Cornalba:2006xm}
using an auxiliary impact parameter representation, which allowed them to generalize
the statement that the Fourier transform of a Gegenbauer is a Gegenbauer.
Here we sidestepped the auxiliary space and we are simply making a statement about conformal blocks.

Comparing eqs.~(\ref{asympt OPE}) and (\ref{relative Eucl 2}) and summing over the $s$-channel OPE gives
a formal series expansion for the asymptotic spectral density:
\be \label{Euclidean average}
\langle C(\Delta',J') \rangle = \sum_{\Delta,J} f_{12\mathcal{O}}f_{34\mathcal{O}}
\frac{\gamma(0)\gamma(d-2)(h'\hb')^{-\Delta}}{\gamma(\Delta-J)\gamma(\Delta+J+d-2)}
C_J\left(\frac{h'^2+\hb'^2}{2h'\hb'}\right)+\ldots
\ee
where the dots stand for the omitted subleading terms in eq.~(\ref{relative Eucl 2}).
The formula shows that the presence of an operator $(\Delta,J)$ on the $s$-channel OPE implies
$(h'\hb')^{-\Delta}$ corrections to the large-dimension spectrum in the cross-channel with the same Gegenbauer
angular dependence.
The factors $1/\gamma(\Delta-J)$ produce a double-zero when $(\Delta,J)$ is a double-twist operator:
this was expected since such exponents can be generated by individual blocks and do not affect the heavy spectrum.
This factor grows at large $\Delta$ and likely causes eq.~(\ref{Euclidean average}) to be an asymptotics series in $1/h$.
Eq.~(\ref{Euclidean average}) represents
a technical extension of ref.~\cite{Mukhametzhanov:2018zja} to account for spinning $s$-channel operators.  As explained there, a minimal but rigorous ``smearing" can be provided via Cauchy moments.


We now apply the same logic to the Regge limit (\ref{dDisc regge}) of the double discontinuity.
From the $t$-channel perspective, the double discontinuity simply multiplies the average in
eq.~(\ref{asympt OPE}) by two sines:
\be
\label{asympt dDisc OPE}
{\rm dDisc}\ \mathcal{G}(z,\zb) = \int \frac{d\Delta' dJ'}{2}\ c^{(0)}(\Delta',J') G_{\Delta',J'}^{(t)}(z,\zb) \times \langle
2\sin(\tfrac{\Delta'-J'-\Delta_2-\Delta_3}{2})\sin(\tfrac{\Delta'-J'-\Delta_1-\Delta_4}{2})
C(\Delta',J')\rangle\,.
\ee
We need to find the average which reproduce the conformal Regge prediction (\ref{dDisc regge}).
This requires a generalization of eq.~(\ref{relative Eucl 2}) where the Gegenbauer function is no longer a polynomial,
and for which the Casimir recursion mentioned above eq.~(\ref{relative Eucl 2}) does not terminate.
However, we find the form of eq.~(\ref{relative Eucl 2})
compelling enough to conjecture that it is valid in general:
\be\begin{aligned} \label{relative Regge}
 (z\zb)^{\frac{1-J}{2}}C_{1-\Delta}\left(\frac{z+\zb}{2\sqrt{z\zb}}\right)
 &\stackrel{?}{=} \int\frac{d\Delta' dJ'}{2}\ c^{(0)}(\Delta',J') G_{\Delta',J'}^{(t)}(z,\zb) \times 
 \frac{\gamma(0)\gamma(d-2)}{\gamma(J-\Delta)\gamma(J-d+\Delta)}
\\&\hspace{20mm}\times (h'\hb')^{J-1}C_{1-\Delta}\left(\frac{h'^2+\hb'^2}{2h'\hb'}\right)  + \mbox{subleading}\,.
\end{aligned}\ee
We do not have a proof of eq.~(\ref{relative Regge}), but we give indirect evidence for it below.
Plugging eq.~(\ref{relative Regge}) into eq.~(\ref{dDisc regge}) and comparing with
eq.~(\ref{asympt dDisc OPE}), we obtain the following formula for the asymptotics of the spectral density:
\be\label{sine squared average}
\boxed{
 \frac{\langle C(\Delta',J') 2\sin^2(\cdots) \rangle}{\gamma(0)\gamma(d-2)} \simeq\!\!
 \int\limits_{\frac{d}{2}-i\infty}^{\frac{d}{2}+i\infty} \frac{d\Delta}{2\pi i}\
 \frac{\Gamma(\Delta-1)}{\Gamma\big(\Delta-\tfrac{d}{2}\big)}C_{1-\Delta}\!\left(\eta_{h}\right)
 \Res\limits_{J=j_*(\Delta)} \tfrac12 b^t(\Delta,J)
 (h'\hb')^{J-1}
}
\ee
where $\eta_h=\frac{h'^2+\hb'^2}{2h'\hb'}$ with $h',\hb'$ defined in eq.~(\ref{h}) and 
\be
 b^t(\Delta,J) = \frac{c^t(\Delta,J)}{K(\Delta,J)\gamma(\Delta-J)\gamma(d-\Delta-J)} \ . \label{bt}
\ee
Eq.~(\ref{sine squared average}) is the main result of this subsection.
It shows that the heavy spectrum must be regular in theories that are asymptotically transparent,
at least in so far as probed by the four-point function.
That is, if $j_*<1$ the right-hand-side vanishes like $\Delta'^{2(j_*-1)}$ implying that
$2\sin^2$ averages to zero.
Operators whose dimensions differ appreciably from double-twists, $\Delta'\approx \Delta_2+\Delta_3+2n+J'$, if they exist,
must thus have small coefficients.

At large dimension with $h'\approx \hb'$, we expect the integral (\ref{sine squared average})
to be dominated near the intercept, which was studied in section \ref{sec:intercept}.
It would be interesting to confront the prediction of this formula with the heavy spectrum of the 3d Ising model.

Conversely, for theories that are asymptotically opaque where ${\rm dDisc}\to 1$,
we would naively expect $\Delta$ to be uniformly distributed modulo 2 so that $2\sin^2$ averages to 1, {\it ie.} the phase of
$e^{i\pi\Delta}$ must be random. The condition that ${\rm dDisc}\to 1$ is however an additional physical assumption
which does not follow from eq.~(\ref{sine squared average}) alone.

An important comment on the regime of validity of eq.~(\ref{sine squared average}).
We know of no guarantee that the function $b^t$ is power-behaved at large imaginary dimensions,
since the $\gamma$-factors in eq.~(\ref{bt}) imply that $b^t\propto c^t e^{\pi |{\rm Im} \Delta|}$, and
all we know from the Lorentzian inversion formula is that $c^t$ is power-behaved.
This means that the integral in eq.~(\ref{sine squared average}) may not converge pointwise for a given $h',\hb'$.
This is not a fundamental problem and simply means that we need some smearing in $\Delta'$:
eq.~(\ref{sine squared average}) is an expression for the \emph{average} spectral density.
Just how much smearing is needed is a question we leave to future work.


\subsubsection{Comments on conformal Regge theory}

As mentioned, a relation between Regge trajectories and the heavy spectrum
is not new and was discussed in the holographic context in \cite{Cornalba:2006xm}.
To our knowledge, however, this was not discussed in the more general context.
Let us thus make contact with the conventions and results of the
conformal Regge theory paper \cite{Costa:2012cb}.  We begin by rewriting
our Gegenbauer-like function $C_{1-\Delta}$ in eq.~(\ref{limit block})
in terms of the harmonic function $\Omega$ defined there, which are
the same up to a proportionality factor:
\be
 4\pi^{\frac{d}{2}}\Omega_{i\nu}(\eta) \equiv \frac{\Gamma(\Delta-1)}{\Gamma\big(\Delta-\tfrac{d}{2}\big)}C_{1-\Delta}(\eta)\Big|_{\Delta=\frac{d}{2}+i\nu}\,.
\ee
We note that this is shadow-symmetric: $\Omega_{i\nu}=\Omega_{-i\nu}$.
We may then rewrite our eq.~(\ref{G regge}) for the Regge limit of the correlator as
\be
 \lim_{z,\zb\to 0}\mathcal{A}(z,\zb) = \int_{-\infty}^{+\infty} \frac{d\nu}{2\pi}
 \left[4\pi^{\frac{d}{2}}\Omega_{i\nu}\left(\frac{z+\zb}{2\sqrt{z\zb}}\right)\right]
 \Res\limits_{J=j_*(\Delta)}\left[\gamma(\Delta{-}J)\gamma(d{-}\Delta{-}J)
 \frac{b^t+b^u e^{-2\pi i J}}{e^{-2\pi i J}-1}(z\zb)^{\frac{1-J}{2}}\right] \label{regge 2}
\ee
where $\mathcal{A}\equiv e^{i\pi(a+b)}\mathcal{G}(z,\zb)^{\circlearrowleft}-\mathcal{G}(z,\zb)$,
$b^u(\Delta,J)$ is defined similarly to eq.~(\ref{bt}), and $\Delta=\tfrac{d}{2}+i\nu$ where it appears.
This form is in precise agreement with eq.~(56) of \cite{Costa:2012cb}.\footnote{
We used that $K(\Delta,J)^{\rm there} =
\frac{1}{K(\Delta,J)^{\rm here}} \frac{(2\Delta-d)4^{-J}}{\pi^2\gamma(\Delta-J)\gamma(d-\Delta-J)}$.
Furthermore, comparing eqs.~(28) and (43) of \cite{Costa:2012cb} with our eq.~(\ref{f from poles}) we find
$b_J^{+}(\nu)^{\rm there}= \frac{2^{-J}}{\pi^2}(b^t+b^u)^{\rm here}$,
and, from eq.~(54) there: $\beta(\nu)^{\rm there} = -\frac{\pi}{2}\Res\limits_{J=j_*(\Delta)} b_J^{+}(\nu)^{\rm there}$.}
For the double discontinuity we get the same formula with $\frac{b^t+b^u e^{-2\pi i J}}{e^{-2\pi i J}-1}\mapsto \frac{b^t}{2}$.

We see that the only differences between the Lorentzian inversion and Mellin-space formalisms is how
the double-twist poles $\gamma\gamma$ are treated.
What comes out of the Lorentzian inversion formula is $c\propto \gamma\gamma b$ and from this perspective
it seems like an arbitrary choice to explicit
this factor in eq.~(\ref{regge 2}).  In holographic CFTs, however,
the function $b$ turns out to be simple rational function (at tree-level)
which makes the writing in eq.~(\ref{regge 2}) natural.
In the Mellin space approach of ref.~\cite{Costa:2012cb}, the double-trace poles are built-in. 

Comparing eqs.~(\ref{regge 2}) and (\ref{sine squared average}), we see that the principal change
in going from $(z,\zb)$ space to $(h',\hb')$ space is the disappearance of the $\gamma\gamma$ double-twist poles.
This was already observed using the \emph{impact parameter representation}
in ref.~\cite{Cornalba:2006xm}, understood there to be directly related to
the $(h',\hb')$ spectrum (called $(h,\hb)$ in section 3.2 there);
it was later explained that going from the $(z,\zb)$ Regge limit to the impact parameter representation
simply cancels factors of $\gamma\gamma$ (see eq.~(2.31) of \cite{Costa:2017twz}).
Here our starting point was simply an observed
identity regarding the asymptotics of blocks, eq.~(\ref{relative Regge}), generalized from integer spins,
and we view the results of \cite{Cornalba:2006xm,Costa:2017twz} as further supporting that identity.

To our knowledge, it is an open question whether the function $b$ is power-behaved or not,
or equivalently, whether the integral (\ref{sine squared average}) (or equivalently (2.31) of \cite{Costa:2017twz}) converges pointwise.
This is not implied by the Lorentzian inversion formula, but it is known to be true perturbatively in holographic theories.
If this were to hold nonperturbatively, one could imagine a version of eq.~(\ref{sine squared average}) (including subleading corrections)
that represents the \emph{exact} spectral function, that is a sum over discrete $\delta$-function, as opposed
to just smeared averages as considered here. We leave this to future investigation.




\section{Concluding remarks}
\label{sect:conclusions}

In this paper we study the spectrum of the 3D Ising model at low spin, combining the
Lorentzian inversion formula developed in \cite{Caron-Huot:2017vep} with the numerical
data from \cite{Simmons-Duffin:2016wlq}.
Two leading twist families are our main focus; $[\sigma\sigma]_0$ which is the leading $Z_2$-Even twist family and $[\sigma\epsilon]_0$ which is the leading $Z_2$-Odd twist family. Two compelling questions are studied in this work. First, can these trajectories which are proven to exist for $J\geq 2$ be extended to spin 0 and include $\epsilon$ and shadow of $\sigma$? Second, what is the intercept of the leading Regge trajectory, $[\sigma\sigma]_0$?

We started by studying a benchmark case, the stress tensor in section \ref{sec:Benchmark: Stress-Tensor}. This is the spin 2 operator in $[\sigma\sigma]_0$ family. To evaluate the inversion integral to high accuracy we used the method of dimensional reduction to  express 3 dimensional as sums of 2d ones with practically neglibigle error (see \cite{Hogervorst:2016hal} and Appendix.~\ref{appen:dimensional-reduction}). We then summed the conformal blocks over the known (truncated) spectrum determined in ref.~\cite{Simmons-Duffin:2016wlq}, i.e., $\epsilon$ and operators belonging to $[\sigma\sigma]_0$, $[\sigma\sigma]_1$ and  $[\epsilon\epsilon]_0$ family up to spin 40 are included. We also added the high-spin tails to these families, which are under analytic control. In fig.~\ref{fig:st-twist1} we compare the stress-tensor anomalous dimension obtained from different truncation in cross-channel expansion in terms of stability and overall error. We obtain a stable result for twist and OPE coefficient with a controlled error at $10^{-4}$ levels (see table.~\ref{table:results-ssss}). 

We then proceeded to continuously lower the value of $\Delta$ (and spin) to reach the intercept.
The intercept answers the question of whether high-energy scattering in the 2+1-dimensional version of the model
is transparent or opaque. However, the method is very different from studying such scattering processes directly
using the OPE, since the integrals computed in section \ref{sec:intercept} are dominated by a region of large impact parameter, where OPE convergence is improved. Such shuffling around of information is familiar from dispersion relations.
In the vicinity of the intercept we use accurate integral representations
of the $\epsilon$ and $T$ cross-channel blocks (see eq.~\ref{eq:GWD} and eq.~\ref{eq:GTI}), 
as well as suitable approximations ($\rho$-expansion) for the direct channel blocks.
The truncation of spectrum is reflected in the $z$-dependence of the result as depicted in fig.~\ref{fig:intercept}, and we obtain the value of the intercept to be $\sim 0.8$ with uncertainty less than 1 in the last digit.

By analytically continuing the leading trajectory fitted to a hyperbola (see fig.~\ref{fig:intercept2}) we find a consistent picture where the $\epsilon$ operator and its shadow lie on a different branch of the leading trajectory, although we were not able  to use this method to compute the properties of $\epsilon$ numerically stable way.
In the $Z_2$-odd sector we find similar conclusions, with a compelling picture of a odd-spin $[\sigma\epsilon]_0$
trajectory having a zero at the location of an (absent) spin-1 current, and where the even-spin is compatible with passing through the shadow of $\sigma$.  This qualitatively picture arises rigorously in the large-$N$ $O(N)$ model,
as shown in section \ref{sect:critical}, where the $[\phi_i\phi_j]_0$ (discussed previously in \citep{Alday:2017zzv,Alday:2019clp}) and $[\phi_iS]_0$ trajectories corresponding to $[\sigma\sigma]_0$ and $[\sigma\epsilon]_0$ (see fig.~\ref{fig:intercept2}).

The finding that the intercept is below unity, $j_*<1$, indicates transparency in the high-energy scattering of lumps
or equivalently a negative Lyapunov exponent and absence of chaos when the theory is placed in Rindler space.
A specific prediction is regularity of the heavy spectrum in eq.~(\ref{sine squared average}).
While not chaotic, the 3D Ising CFT is certainly not integrable,
and a useful analogy may be the KAM theorem in classical mechanics,
which states (very roughly) that certain small enough deformations of an integrable system are
not chaotic.\footnote{We thank Alex Maloney for this analogy.}.  The approximately conserved quantities of the Ising CFT
are likely higher-spin currents \cite{Maldacena:2012sf,Alday:2015ota} and transparency
suggests that they become increasingly powerful for heavy operators with increasing dimensions.

\acknowledgments

We thank Anh-Khoi Trinh, David Simmons-Duffin, David Meltzer, Petr Kravchuk for discussions, and Alex Maloney
for initial collaboration.
Zahra Zahraee acknowledges support from the Schulich fellowship.
Work of SCH is supported by the National Science and Engineering Council of Canada, the Canada Research Chair program,
the Fonds de Recherche du Qu\'ebec - Nature et Technologies, and the Simons Collaboration on the Nonperturbative Bootstrap.  


\appendix

\section{Inversion Integrals}
\label{app:integrals}

In this appendix we discuss various inversion integrals employed in the body of the paper.
We first describe the dimensional reduction of 3-dimensional blocks over 2d ones, which relies itself on a recursion
for the series expansion of blocks; we present (for the first time) a closed formula for the latter.
Then we give analytic formulas for the inversion integral of 2d block.  Combined, these results provide an accurate
way to compute the contribution of a single cross-channel block.
Lastly, we consider the collinear approximation to the exchange of a single cross-channel block,
which has been used to make comparison plots such as in section \ref{subsec:L-S Expansion}.

\subsection{Dimensional reduction for cross-channel blocks}
\label{appen:dimensional-reduction}

In this section, we briefly review of the dimensional reduction method introduced in \cite{Hogervorst:2016hal} and employed in the body of the paper. The idea is to break the (Euclidean) $d$-dimensional conformal group, SO($d{+}1,1$), to its subgroup SO($d$,1). This will help us to write different representations of the latter group in terms of the former. A primary operator in $d$-dimensions is a sum of infinitely many primaries in $(d-1)$-dimensions.
This is because a state corresponding to a primary operator must be annihilated by all the generators of special conformal translations, $K_{\mu}$. Primaries of both of both groups are annihilated by $K_1, ..,K_{d-1}$ but only SO($d$) primaries
are annihilated by $K_d$.  Loosely speaking, taking derivatives with $P_d$ generates new SO($d-1$) primaries.
The SO($d$) angular momentum multiplets also decompose into SO($d-1$) multiplets.
This consequently means that any $d$-dimensional conformal multiplet of of spin $J$ and dimension $\Delta$ can be decomposed in terms of infinitely many $d-1$ dimensional mutiplet with spin $0\leq \ell\leq J$ and dimensions $\Delta+m$ with $m\geq 0$.  This in turn means that conformal blocks should also follow this decomposition rule so that a $d$-dimensional conformal block can be written as follows:\footnote{Our subscripts differ from ref.~\cite{Hogervorst:2016hal}
as: $m^{\rm here}=2m^{\rm there}$,
as required for non-identical operators.}
\begin{equation}
\label{eq:2d-blocks-expansion}
G^{(a,b)}_{\Delta,J}(z,\zb;d)=\sum \mathcal{A}^{(a,b)}_{m,n}(\Delta,J)G^{(a,b)}_{\Delta+m,\ell-n}(z,\zb;d-1)  \qquad 0\leq n\leq J, \quad m=0,1,2...
\end{equation}
where again $a=\frac{\Delta_2-\Delta_1}{2}$ and $b=\frac{\Delta_3-\Delta_4}{2}$.

The coefficients with $m=0$ describe the dimensional reduction of Gegenbauer polynomials
and are given as
\be
 \mathcal{A}^{(a,b)}_{0,n}(\Delta,J)=\left\{ \begin{array}{ll}
 Z^{J}_{n/2}\ , \quad & n\ \mbox{ even} \\
 0\ ,\quad & \mbox{otherwise}\end{array}\right.
\ee
with, in our conventions,
\be
Z_t\equiv \frac{(-1)^t(\frac12)_t (-J)_{2t}}{t!\ (J-2t+\frac{d-1}{2})_t (-J-\frac{d-4}{2})_t}
\ee
with $(a)_b \equiv \frac{\Gamma(a+b)}{\Gamma(a)}$ the Pochhammer symbol.
Generally, the other coefficients vanish unless $m\equiv n$ modulo 2. 
They can be obtained recursively by comparing the radial expansion of the blocks in the two dimensions.
This recursion was given for identical operators in ref.~\cite{Hogervorst:2016hal}
and we state here the general case:
\be\label{3d 2d recursion}
\begin{aligned}
\mathcal{A}^{(a,b)}_{m,n}(\Delta,J)&=
\sum^{\frac{m+n}{2}}_{p=\max(-\frac{m-n}{2},0)} \Big(Z^{J-n+2p}_p 
a_{\frac{m+n}{2}-p,\frac{m-n}{2}+p}^{(a,b)}(\Delta,J;d)\Big)\\
&\quad -\sum_{m'=1}^m\sum_{n'} \mathcal{A}^{(a,b)}_{m-m',n-n'}(\Delta,J)a^{(a,b)}_{\frac{m'+n'}{2},\frac{m'-n'}{2}}(\Delta+m-m',J-n+n';d-1)\,.
\end{aligned}\ee
where the sum over $n'$ ranges from $\max(-m',n-J+\delta_{m'+n-\ell,\rm odd})$ to $\text{min}(m',n)$ in steps of 2.
The coefficients $a$ describe the radial expansion of blocks:
\be \label{radial}
G^{(a,b)}_{\Delta,J}(z,\zb;d) = \sum_{r,s\geq 0} a^{(a,b)}_{r,s}(\Delta,J;d)
(z\zb)^{\frac{\Delta+r+s}{2}}C_{J+s-r}\left(\frac{z+\zb}{2\sqrt{z\zb}};d\right)\,.
\ee
In the case of identical operators, a closed form solution to eq.~(\ref{3d 2d recursion}) was given in
ref.~\cite{Hogervorst:2016hal}, which we reproduced.
As shown in that reference, the expansion (\ref{eq:2d-blocks-expansion}) converges very rapidly,
always at least as fast as the $\rho$-series.

\subsection{New closed-form expression for radial expansion coefficients}

The coefficients in eq.~(\ref{radial}) are to be determined using
another recursion \cite{Hogervorst:2013sma} (see for example appendix A of ref.~\cite{Caron-Huot:2017vep} for nonidentical operators). We do not reproduce that recursion here, because, inspired by recent formulas by Li \cite{Li:2019cwm}
and a bit of guesswork,
we were able to find a closed formula!
\be\begin{split}
&a^{(a,b)}_{r,s}(\Delta,J;d)=\frac{\left(\frac{\Delta-J+2-d}{2}+a\right)_r\left(\frac{\Delta-J+2-d}{2}+b\right)_r}{
r!(\Delta{-}J{+}2{-}d)_r(-J-\frac{d-2}{2})_r(J-r+\frac{d-2}{2})_r}
\frac{\left(\frac{\Delta+J}{2}+a\right)_s\left(\frac{\Delta+J}{2}+b\right)_s}{s!(\Delta+J)_s}
 \\
 &\hspace{10mm}\times
 \sum_{p=0}^{\min(r,s)}\Bigg((-1)^p \pFq{4}{3}{{-}r{+}p\ ,{-}s{+}p\ ,p\ ,\Delta{-}1}{\Delta{+}p{-}\frac{d-2}{2}\ ,J{+}d{-}2{-}r{+}p\ ,{-}J{-}s{+}p{-}\frac{d-2}{2}}{1}\\
 &\hspace{30mm}\times \frac{(\frac{2-d}{2})_p(\frac{4-d}{2})_p(-r)_p(-s)_p(d{-}1{-}\Delta{+}J{-}r)_p(-J)_{r-p}}{
 p!(\Delta-\frac{d-2}{2})_p(-J-s-\frac{d-2}{2})_p(J{+}d{-}2)_{p-r}}\Bigg)\,.
\end{split}
\ee
Note from eq.~(\ref{radial}) that $r$ increases the twist of the descendants and $s$ the conformal spin.
The logic of this formula is that
coefficients get progressively more complicated as one goes away from the leading twist or leading conformal spin,
where only the $p=0$ term contributes in both cases;
a pattern was guessed empirically by working away from these simple limits.
The formula truncates in $d=2$ and $d=4$ due to the Pochhammers.
In fact the limit to even spacetime dimensions gives annoying $0/0$ forms,
and to evaluate eq.~(\ref{3d 2d recursion}) in $d=2$ we use the following simplified result:
\be\begin{aligned}
a^{(a,b)}_{r,s}(\Delta,J;d{=}2) &= \frac{1}{1+\delta_{J,0}}
\frac{\left(\frac{\Delta-J}{2}+a\right)_r\left(\frac{\Delta-J}{2}+b\right)_r}{r!(\Delta-J)_r}
\frac{\left(\frac{\Delta+J}{2}+a\right)_s\left(\frac{\Delta+J}{2}+b\right)_s}{s!(\Delta+J)_s}
\\ &\quad\times \left(1+\frac{(-r)_J(\Delta+s)_J}{(1-\Delta-r)_J(1+s)_J}\right)
\end{aligned}\ee
which is valid for $j+s-r\geq 0$ and should be set to 0 otherwise.

\subsection{Lorentzian inversion in 2d}
\label{app:Lorentzian inversion in 2d}
The 3d to 2d series (\ref{eq:2d-blocks-expansion}) is useful for this work
 because of exact results for inversion integrals that exist in $d=2$.
Let us denote as $c^{\Delta_1,\Delta_2,\Delta_3,\Delta_4}_{\Delta',J'}(\beta,z;d{=}2)$ the 
contribution eq.~(\ref{eq:generating_int}) coming from a 2d $t$-channel block of $(\Delta',J')$,
as defined in eq.~(\ref{eq:2d-block}).
Using eq.~(3.38) in \cite{Liu:2018jhs} (see also \cite{Cardona:2018qrt,Albayrak:2019gnz}),
the result (integrated from $\zb\geq 0$ instead of $\zb\geq z$!) is written as:
\begin{equation}
\label{eq:2d-inversion}
c^{\Delta_1,\Delta_2,\Delta_3,\Delta_4}_{\Delta',J'}(\beta,z;d{=}2)
=\frac{z^{\frac{\Delta_1+\Delta_2}{2}}}{(1-z)^{\frac{\Delta_2+\Delta_3}{2}}}\frac{I_{\frac{\Delta'+J'}{2}}^{\frac{\Delta_1}{2}\cdots\frac{\Delta_4}{2}}\left(\frac{\beta}{2}\right)\ k^{(a',b')}_{\Delta'-J'}(1-z)+(J'\mapsto -J')}{1+\delta_{J',0}},
\end{equation}
where $a'=\frac{\Delta_2-\Delta_3}{2}$, $b'=\frac{\Delta_1-\Delta_4}{2}$ and $I_{h'}^{h_1\cdots h_4}(h)$ is the one-dimensional inversion given as:
\be\label{eq:7F6}\begin{aligned}
I_{h'}^{h_1\cdots h_4}(h) &= \frac{\Gamma(h+h_{21})\Gamma(h+h_{43})}{\Gamma(h_2+h_3-h')\Gamma(h_1+h_4-h')\Gamma(2h-1)} \frac{\Gamma(h-h'+h_1+h_3-1)}{\Gamma(h+h'-h_1-h_3+1)} \\
&\qquad \times \pFq{4}{3}{h'{+}h_{23}\ ,h'{+}h_{41}\ ,h'{-}h_1{-}h_4{+}1\ , h'{-}h_2{-}h_3{+}1}{2h'\ ,h{+}h'{-}h_1{-}h_3\ ,h'{-}h{-}h_1{-}h_3{+}2}{1}
\\ &\phantom{=} +
2\sin(\pi(h'{-}h_2{-}h_3))\sin(\pi(h'{-}h_1{-}h_4))\kappa^{(h_{12},h_{34})}_{2h}\\
&\qquad\times
\pFq{4}{3}{h{+}h_{34}\ ,h{+}h_{12}\ ,h{+}h_1{+}h_2{-}1\ ,h{+}h_3{+}h_4{-}1}{2h\ ,h{+}h'{+}h_1{+}h_3{-}1\ ,h{-}h'{+}h_1{+}h_3}{1}
\end{aligned}\ee
with $h_{ij}=h_i-h_j$.
This result, inserted in the 3d to 2d expansion (\ref{eq:2d-blocks-expansion}), gives the formula (\ref{eq:2dto3d-inversion}) which is used repeatedly in this paper.

We stress that the analytic result (\ref{eq:7F6}) is only valid when integrating over the complete range
$0\leq\zb\leq 1$ in the inversion integral (\ref{eq:generating_int}), whereas the formula instructs us to invegrate
only over $z\leq \zb\leq 1$. Since in practice we work at small $z$,
we can correct for this discrepancy by subtracting from eq.~(\ref{eq:7F6}) the integral of the
first few terms in the $\zb\to 0$ Taylor series of the integrand.
For very small $z$ this is completely negligible, but for moderate values like $z\sim 10^{-2}$
this is important for our precision study.

The hypergeometric series in eq.~(\ref{eq:7F6}) terminates in special cases such as $h'=h_{32}$.
These correspond to power laws in the $t$-channel. Setting $h_1+h_2=\tau$, $h_{21}=a$ and $h_{34}=b$
the formula reduces to the integral recorded in eq.~(4.7) of \cite{Caron-Huot:2017vep}
(using also $a\leftrightarrow b$ symmetry):
\be\begin{aligned} \label{Iab}
 I^{(a,b)}_{\tau}(\beta)&\equiv\int_0^1\frac{dz}{z^2}\kappa_{\beta}^{(a,b)}k_{\beta}^{(-a,-b)}(z)
 \mbox{dDisc}\left[\left(\frac{1-z}{z}\right)^{\frac{\tau}{2}-b}z^{-b}\right] \\
 &= \frac{1}{\Gamma\left(-\frac{\tau}{2}+b\right)\Gamma\left(-\frac{\tau}{2}-a\right)}\frac{\Gamma\left(\frac{\beta}{2}-a\right)\Gamma\left(\frac{\beta}{2}+b\right)}{\Gamma(\beta-1)}\frac{\Gamma\left(\frac{\beta}{2}-\frac{\tau}{2}-1\right)}{\Gamma\left(\frac{\beta}{2}+\frac{\tau}{2}+1\right)}\ .
\end{aligned}\ee
This integral allows to deal exactly with identity exchange and more generally large-spin perturbation theory.
The first hypergeometric function in eq.~(\ref{eq:7F6}) can be interpreted as summing up
the $(1-\zb)/\zb$ series according to this integral; this series is asymptotic, and the second hypergeometric
can be interpreted as a nonperturbative correction at large spin \cite{Albayrak:2019gnz}.

\subsection{Collinear expansion $\zb\rightarrow 1$ for cross-channel exchanged blocks}
\label{app:collinear}

Taking the limit $z\rightarrow 0$ of $G_{\Delta,J}(z,\zb)$ or $\zb\rightarrow 1$ of $G_{\Delta,J}(1-\zb,1-z)$ is a straightforward procedure. In this limit the quadratic Casimir equation becomes a hypergeometric equation so the leading behaviour of the conformal blocks in this limit is
\be G_{\Delta,J}\rightarrow z^{\frac{\Delta-J}{2}}k_{\Delta+J}(\zb) \,,\ee
where $k_{\beta}(z)$ is defined in (\ref{2F1}). A nice way to organize the expansion, which was discussed for example in \cite{Caron-Huot:2017vep}, is in terms of these functions since they control the $SL_2(R)$ part of the conformal group that remains after taking the limit. With a convenient factor extracted and focusing on $d=3$,
the expansion that we use is
\be \sqrt{1-z/\zb}\ G_{\Delta,J}(z,\zb)=\sum_{m=0}^{\infty}z^{\frac{\tau}{2}+m}h^{(m)}_{\Delta,J}(\zb) \,,
\label{coll exp 1}\ee
with
\be h^{(m)}_{\Delta,J}(\zb)=\sum_{n=-m}^m h^{(m,n)}_{\Delta,J}k_{\beta+2n}(\zb) \,.\ee
The quadratic Casimir equation then gives the following recursion relation in $m$ (see \cite{Caron-Huot:2017vep}):
\be\ba
&\sum_{n=-m}^m(n(n+\beta-1)+m(m+\tau-2))h^{(m,n)}_{\Delta,J}k_{\beta+2n}(\zb)
\\
&=\left(\frac{\tau-3}{2}+m+a\right) \left(\frac{\tau-3}{2}+m+b\right)h^{(m-1)}_{\Delta,J}(\zb)-\frac{1}{4}\sum_{m'=1}^m\left(\frac{2m'}{\zb^{m'}}-\frac{2m'-1}{\zb^{m'-1}}\right)h^{(m-m')}_{\Delta,J}(\zb)\,.
\ea\ee
To isolate the coefficient of $k_{\beta+2n}(\zb)$ on the right-hand-side we need to use
the shift relation
\be \frac{1}{\zb}k_{\beta}(\zb)=k_{\beta-2}(\zb)+\left(\frac{1}{2}-\frac{2ab}{\beta(\beta-2)}\right)k_{\beta}(\zb)+\frac{(\frac{\beta^2}{4}-a^2)(\frac{\beta^2}{4}-b^2)}{\beta^2(\beta^2-1)}k_{\beta+2}(\zb) \,,\ee
to eliminate all explicit appearance of $\zb$, after which we can
solve recursively for the coefficients $h^{(m,n)}_{\Delta,J}$.

The result of the recursion can finally be combined with the prefactor in eq.~(\ref{coll exp 1})
to expand the block in pure powers of $z$. The first two terms of this expansion are
\be\ba \label{eq:coll-block}
G_{\Delta,J}(z,\zb)&\approx  z^{\frac{\tau}{2}}k_{\beta}(\zb) + z^{\frac{\tau}{2}+1}\Bigg[
\frac{\beta-\tau}{2(\beta-\tau-1)}k_{\beta-2}(\zb)
+\frac{(\beta^2-4a^2)(\beta^2-4b^2)(\beta+\tau-2)}{32\beta^2(\beta^2-1)(\beta+\tau-1)}k_{\beta+2}(\zb)
\\
&\hspace{15mm} + \left(\frac{\tau+2a+2b}{4}+ \frac{a b\big( (\beta-1)^2-\tau+1\big)}{\beta(\beta-2)(\tau-1)}\right)
k_\beta(\zb)\Bigg] + \mathcal{O}(\zb^{\frac{\tau}{2}+2})\,.
\ea\ee
For blocks with $J$ non-integer, the same formulas give the expansion of
$g^{\rm pure}_{\Delta,J}(z,\zb)$.

\section{Compact approximations from large-spin perturbation theory}
\label{app:cheap}

In this appendix we record compact but surprisingly accurate approximations for the OPE
data based on large-spin perturbation theory.
Our formulas are essentially simplified versions of results from \cite{Simmons-Duffin:2016wlq}.

Starting from the Lorentzian inversion formula, the idea is to truncate the $t$-channel sum to just identity
and a small number of operators. For each operator, we keep
only the leading term at large spin or $\beta\to \infty$, which comes from $\zb\to 1$,
and we extract anomalous dimensions by looking at logarithmic terms as $z\to 0$.

\begin{figure}
    \begin{subfigure}[t]{0.5\textwidth}
        \centering
  \includegraphics[width=\linewidth]{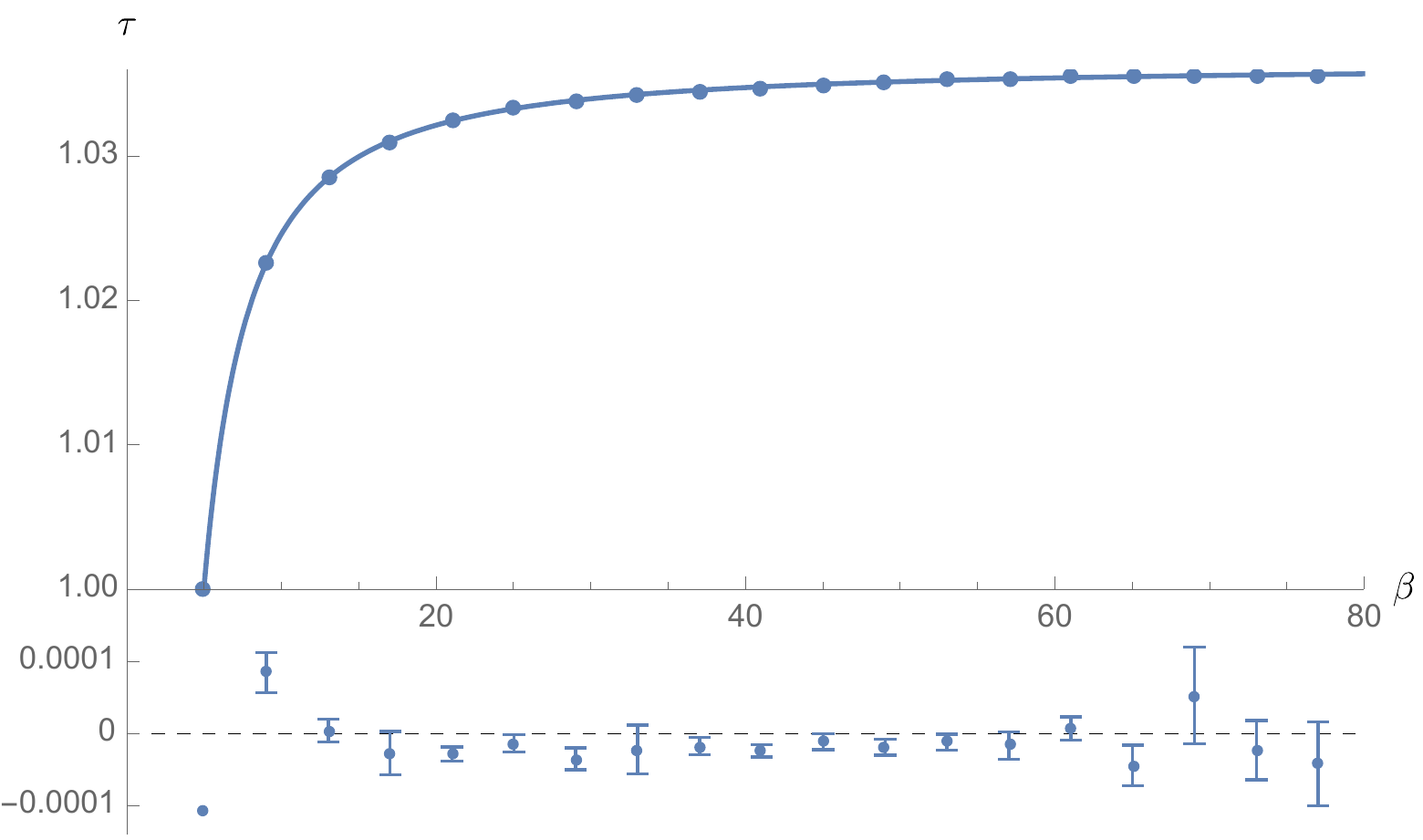}
   \caption{Twist, with
    difference (data)-(approx.) in inset.}
    \end{subfigure}
    \begin{subfigure}[t]{0.5\textwidth}
        \centering
        \includegraphics[width=1\linewidth]{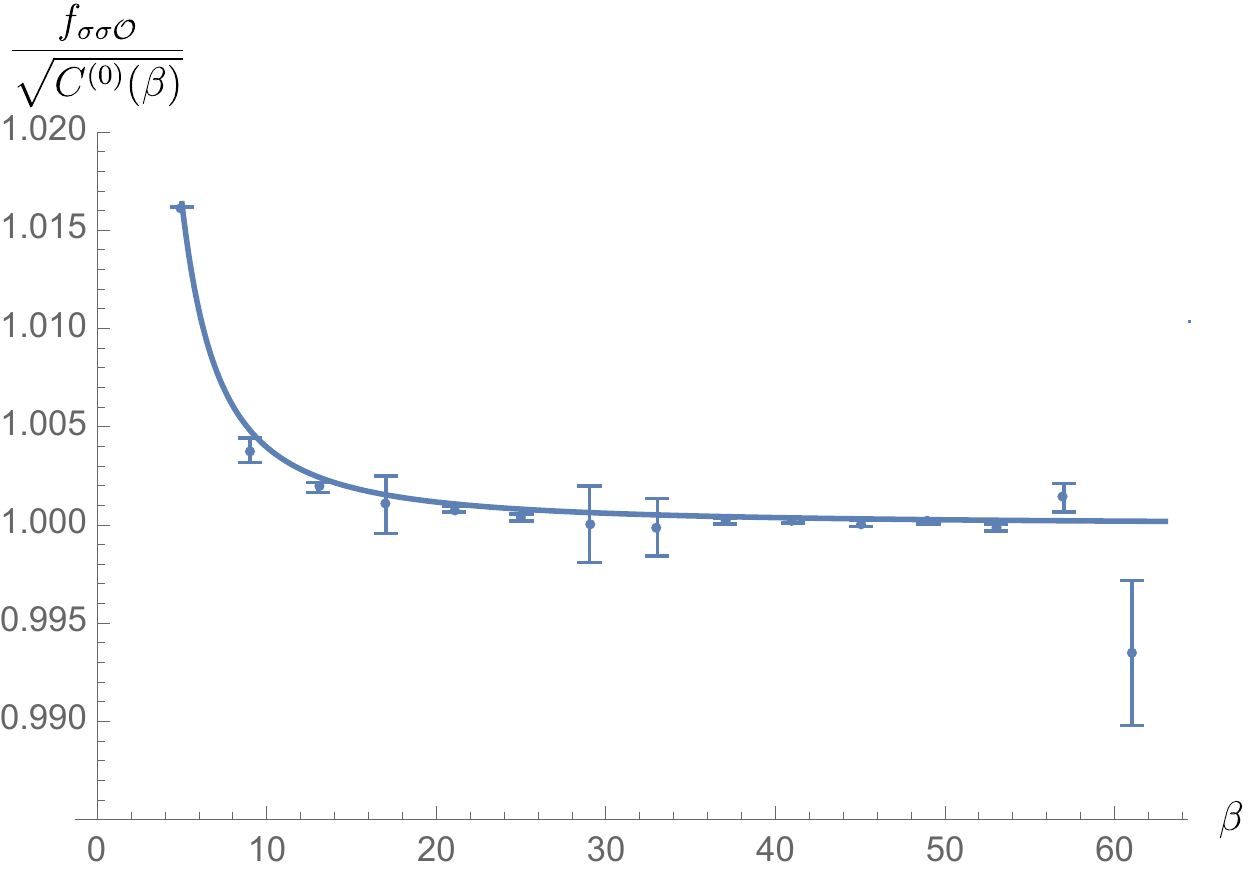}
      \caption{OPE coefficients.}
    \end{subfigure} 
        \caption{\label{fig:cheap SS0}
Comparison of the large-spin approximation (\ref{pockets}) for the $[\sigma\sigma]_0$ family
with the numerical data of \cite{Simmons-Duffin:2016wlq},
for the twist and OPE coefficients (divided by mean field theory).
Only numerical errors are shown: discrepancies on the left-hand side
of the plots should be attributed to shortcomings in the approximation not data.}
\end{figure}

Considering the exchange of an operator $\cO$ of twist $\tau_\cO=\Delta_\cO-J_\cO$ and conformal spin $\beta_\cO=\Delta_\cO+J_\cO$,
we take the double limit $(z,\zb)\to (0,1)$ (for example starting
from the $\zb\to 1$ limit recorded in eq.~(\ref{collinear limit block})):
\be \label{collinear limit block 1}
\lim_{z\to 0,\zb\to 1} G^{(0,0)}_{\Delta_\cO,J_\cO}(1-\zb,1-z)\rightarrow
-\frac{2\Gamma(\beta_\cO)}{\Gamma(\tfrac{\beta_\cO}{2}\big)^2} (1-\zb)^{\frac{\tau_\cO}{2}}
\left(
\tfrac12\log z + H(\tfrac{\beta_\cO}{2}-1)\right)
\ee
where $H(x)=\psi(x+1)-\psi(1)$ is the harmonic number.
Plugging into the inversion integral (\ref{generating_int c}) (replacing $(1-\zb)$ by $\frac{1-\zb}{\zb}$)
and expanding eq.~(\ref{Iab}) at large $\beta$, we obtain the
following approximation to the collinear generating function:
\be
 C^t(z,\beta)+C^u(z,\beta) \approx C_{[\sigma\sigma]_0}^{(0)}(\beta)z^{\Delta_\sigma}
\left[ 1-\sum_{\cO}
\frac{2f_{\sigma\sigma\cO}^2\Gamma(\beta_\cO)\Gamma(\Delta_\sigma)^2}
{\Gamma(\tfrac{\beta_\cO}{2}\big)^2\Gamma\big(\tfrac{2\Delta_\sigma-\tau_\cO}{2}\big)^2}
\frac{\frac12\log z+ H(\tfrac{\beta_\cO}{2}-1)}{[(\beta-1)/2]^{\tau_\cO}}\right]\,,
\label{approx C}
\ee
where we defined the mean-field theory coefficient on the leading trajectory:
\be
C_{[\sigma\sigma]_0}^{(0)}(\beta)\equiv 2I^{(0,0)}_{-2\Delta_\sigma} =
\frac{2\Gamma(\beta)}{\Gamma\big(\tfrac{\beta}{2}\big)^2}\frac{1}{\Gamma(\Delta_\sigma)^2}
\frac{\Gamma(\tfrac{\beta}{2}+\Delta_\sigma-1)}{\Gamma(\tfrac{\beta}{2}-\Delta_\sigma+1)}\,.
\ee
We stress that in eq.~(\ref{approx C}) only the lowest few operators should be included in the sum,
which is not a convergent sum. We include only $\epsilon$ and $T$.
As noted in the main text, we expand in $1/(\beta-1)$ because the series is even in that variable.
Taking the coefficient of $\tfrac12 \log z$, and the constant, respectively, gives the ``pocket-book" formula
recorded for the twist in eq.~(\ref{large spin gamma}), and a corresponding formula for the OPE coefficients
(including the Jacobian factor in eq.~(\ref{eq:jacobian})):
\be\begin{aligned}\label{pockets}
\tau_{[\sigma\sigma]_0} &\approx 2\Delta_\sigma - \sum_{\cO=\epsilon,T}
\frac{2 \lambda_{\sigma\sigma\cO}^2\Gamma(\Delta_\sigma)^2}{\Gamma\big(\Delta_\sigma-\tfrac{\tau_{\cO}}{2}\big)^2} \frac{\Gamma(\beta_{\cO})}{\Gamma\big(\tfrac{\beta_{\cO}}{2}\big)^2}
\left(\frac{2}{\beta-1}\right)^{\tau_{\cO}}\,, \\
f_{\sigma\sigma[\sigma\sigma]_{0}}^2&\approx\frac{C_{[\sigma\sigma]_0}^{(0)}(\beta)}{1-\frac{d\tau_{[\sigma\sigma]_0}}{d\beta}}
\left[ 1-\sum_{\cO=\epsilon,T}
\frac{2 \lambda_{\sigma\sigma\cO}^2\Gamma(\Delta_\sigma)^2}{\Gamma\big(\Delta_\sigma-\tfrac{\tau_{\cO}}{2}\big)^2} \frac{\Gamma(\beta_{\cO})}{\Gamma\big(\tfrac{\beta_{\cO}}{2}\big)^2}
H\big(\tfrac{\beta_{\cO}}{2}-1\big)\left(\frac{2}{\beta-1}\right)^{\tau_{\cO}} \right]\,.
\end{aligned}\ee
Note that $\beta=\Delta+J=\tau+2J$ enters the formula for $\tau$.
To compute the twist of an operator of given spin $J$, we first evaluate the first line with $\beta\mapsto 2\Delta_\sigma+2J$
to get a crude approximation to $\tau$; we then iterate using the improved value $\beta\mapsto \tau+2J$. The procedure converges rapidly. The resulting value of $\beta$ is then inserted in both equations.
In figure \ref{fig:cheap SS0} we compare this formula with the numerical data of ref.~\cite{Simmons-Duffin:2016wlq}.
Both plots exhibit relative accuracy better than $10^{-4}$ for all twists and the stress tensor OPE coefficients,
but the relative error is closer to $10^{-3}$ for the spin-4 OPE coefficient.

Given the error budget discussed in section \ref{sec:Benchmark: Stress-Tensor},
we believe that the remarkable accuracy of the approximation
at spin $J=2$ is a lucky accident of that particular formula.
Indeed, in the absence of accident one would expect the discrepancy to be significantly larger
for $J=2$ than for $J=4$.

\begin{figure}
    \begin{subfigure}[t]{0.49\textwidth}
        \centering
  \includegraphics[width=\linewidth]{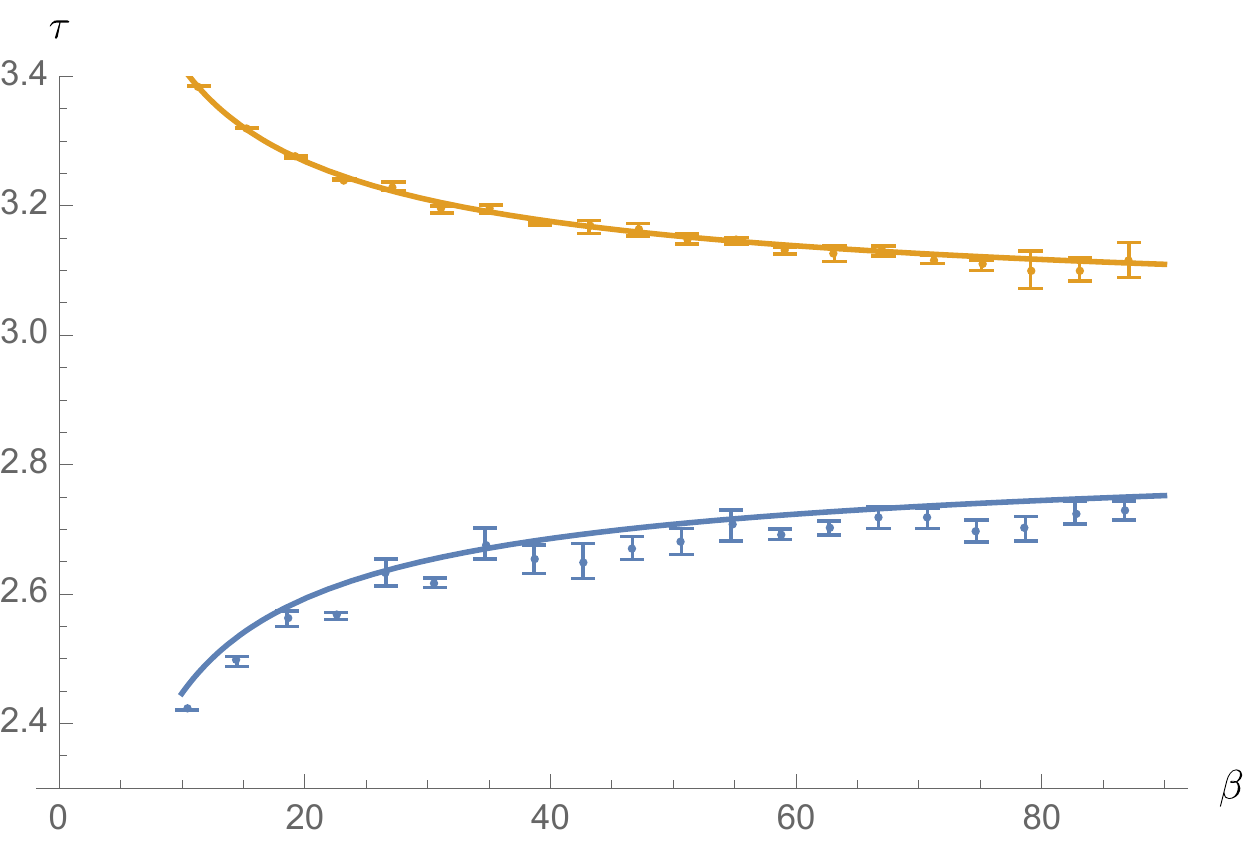}
   \caption{Twist}
    \end{subfigure}\quad
    \begin{subfigure}[t]{0.49\textwidth}
        \centering
        \includegraphics[width=1\linewidth]{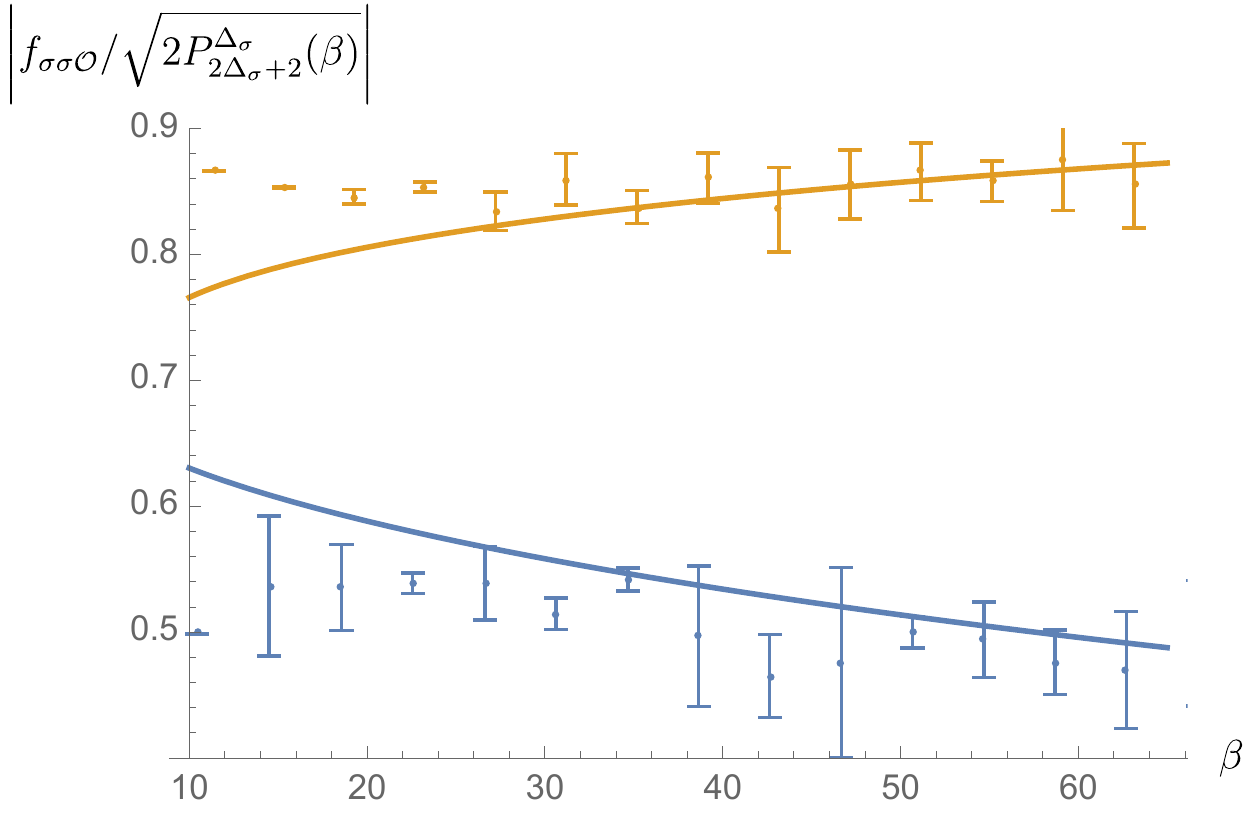}
      \caption{OPE coefficients divided by MFT (\ref{MFT}).}
    \end{subfigure} 
        \caption{\label{fig:cheap SS1}
Same as fig.~\ref{fig:cheap SS0} but for the $[\sigma\sigma]_1$ and $[\epsilon\epsilon]_0$ families (top and bottom, respectively).}
\end{figure}

For the subleading trajectories, $[\sigma\sigma]_1$ and $[\epsilon\epsilon]_0$ are near-degenerate and mix substantially,
as pointed out in \cite{Simmons-Duffin:2016wlq}. We thus need to study a $2\times2$ matrix of correlators.
We including only exchange and identity and the logarithmic term from exchange of $\sigma$.
We begin with identity exchange, first pretending that $2\Delta_\epsilon\approx 2\Delta_\sigma+2$ so as to
make the operators degenerate. We then expanding the Lorentzian inversion formula to second order in $z$ where needed
({\it ie.} to subtract descendants of $[\sigma\sigma]_0$):
\be
M^{(0)}\equiv \left(\begin{array}{cc} C_{\sigma\sigma\sigma\sigma}& C_{\sigma\sigma\epsilon\epsilon}
\\ C_{\epsilon\epsilon\sigma\sigma} & C_{\epsilon\epsilon\epsilon\epsilon}\end{array}\right)_{z^{\approx 2\Delta_\sigma+2}}
= \frac{\Gamma\big(\tfrac{\beta}{2}\big)^2}{\Gamma(\beta-1)}
\left(\begin{array}{cc}
\frac{\Delta_\sigma-\frac12}{\Gamma(\Delta_\sigma)^2}\left(\frac{2}{\beta-1}\right)^{2-2\Delta_\sigma}&0
\\ 0& \frac{2}{\Gamma(\Delta_\epsilon)^2}\left(\frac{2}{\beta-1}\right)^{2-2\Delta_\epsilon} \end{array}\right)\,.
\ee
To get anomalous dimensions we look for logarithmic terms $\tfrac12\log z$.
We keep two sources: 
 identity exchange expanded to linear order in $(2\Delta_\epsilon-2\Delta_\sigma-2)$,
and $\sigma$-exchange.  Multiplying by $(M^{(0)})^{-1/2}$ on both sides to properly normalize the states,
we find that the twists of the 
$[\sigma\sigma]_1$  and $[\epsilon\epsilon]_0$ families are the eigenvalues of the following matrix
(respectively the higher and lower eigenvalues):
\be
 \tau_{\{[\sigma\sigma]_1,[\epsilon\epsilon]_0\}} = \left(\begin{array}{cc}
2\Delta_\sigma{+}2 & X \\ X &2\Delta_\epsilon \end{array}\right)
\label{tau excited}
\ee
with off-diagonal term
\be
X= \frac{4 f_{\sigma\sigma\epsilon}^2 \Gamma(\Delta_\epsilon)\Gamma(\Delta_\sigma-\Delta_\epsilon)\Gamma(\Delta_\sigma)^2}{\Gamma\big(\tfrac{\Delta_\epsilon}{2}\big)^2\Gamma\big(\tfrac{2\Delta_\sigma-\Delta_\epsilon}{2}\big)^2}
\frac{\Delta_\epsilon-\Delta_\sigma-1}{\sqrt{2\Delta_\sigma-1}}\left(\frac{2}{\beta-1}\right)^{\Delta_\sigma}\,.
\ee
To find the OPE coefficients we compare a certain derivative of the matrix of generating functions
$(2z\partial_z-\tau_{[\epsilon\epsilon]_0})C$ with the OPE:
\be
 (\tau_{[\sigma\sigma]_1}-\tau_{[\epsilon\epsilon]_0}) \left(\begin{array}{c} f_{\sigma\sigma[\sigma\sigma]_1}\\
 f_{\sigma\sigma[\epsilon\epsilon]_0}\end{array}\right)
 \raisebox{2mm}{$\left(\begin{array}{cc} f_{\sigma\sigma[\sigma\sigma]_1}\ & \ f_{\sigma\sigma[\epsilon\epsilon]_0}\end{array} \right)$}
 = (M^{(0)})^{1/2}\left[\tau_{\{[\sigma\sigma]_1,[\epsilon\epsilon]_0\}} -1\tau_{[\epsilon\epsilon]_0} \right](M^{(0)})^{1/2}\,.
\ee
The right-hand-side is an explicitly given matrix of rank 1 and so the equation allows to solve for
$f_{\sigma\sigma[\sigma\sigma]_1}$ and $f_{\epsilon\epsilon[\sigma\sigma]_1}$, up to overall sign conventions;
couplings to the $[\epsilon\epsilon]_0$ family
are obtained similarly. These approximations are plotted in fig.~\ref{fig:cheap SS1}.
Note that the accuracy is less than for $[\sigma\sigma]_0$ since the approximation
is more complicated due to the mixing yet far cruder (we did not even account for $\epsilon$ exchange).

\section{Convexity of the leading trajectory}
\label{app:convexity}

Here we give an elementary proof that the region of convergence of the Lorentzian inversion formula,
in the real $(\Delta,J$) plane, is convex.  We consider a correlator of identical operators, so
${\rm dDisc}\mathcal{G}$ is a positive-definite distribution.
Consider first the integration region where $z\ll \zb\ll 1$.
In this region, the block in eq.~(\ref{eq:inversion-integral}) has an exponential-like dependence
on $J,\Delta$: $G_{J+d-1,\Delta+1-d}\propto z^{\frac{J-\Delta}{2}+d-1}\zb^{\frac{\Delta+J}{2}}$.
The basic point is that the exponential function is convex:
\be
 e^{x A+(1-x)B} \leq x e^A +(1-x)e^B, \qquad 0\leq x\leq 1\, . \label{convexity argument}
\ee
Therefore if the integral (\ref{eq:inversion-integral}) converges at
the two points $(\Delta_A,J_A)$ and $(\Delta_B,J_B)$, it automatically
converges everywhere along the line segment joining them: the region of convergence is convex.
Increasing $J$ can only improve convergence, and adding imaginary parts does not affect convergence.

The conformal blocks which enter the Lorentzian inversion formula are more complicated functions
than exponentials, but we can apply the same logic.  A better model, which reflects the shadow-symmetry of the curve,
is a $\cosh$ function; indeed the $z\sim \zb\ll 1$ limit of the blocks, in eq.~(\ref{limit block}), is
bounded above and below by a multiple of a cosh:
\be c_1 g(z)
  \cosh\big(\tfrac12(\Delta-\tfrac{d}{2})\log\tfrac{z}{\zb}\big)
\leq
    C_{\Delta+1-d}\left(\frac{z+\zb}{2\sqrt{z\zb}}\right) 
    \leq c_2 g(z)\cosh\big(\tfrac12(\Delta-\tfrac{d}{2})\log\tfrac{z}{\zb}\big)
\ee
where $g(z)=\cosh\big(\tfrac12(1-\tfrac{d}{2})\log\tfrac{z}{\zb}\big)$ and
the constants $c_1$ and $c_2$ depend only on spacetime dimension but
work uniformly for all $\Delta, z, \zb$.
This shows that Lorentzian inversion converges
if and only if the $\cosh$ model converges, and since $\cosh$ is a sum of two exponentials
we can apply eq.~(\ref{convexity argument}).  This takes care of the region $z\sim \zb\ll1$.
The last region which can potentially
affect convergence is $z\to 0$ with $\zb\sim1$, but in this region the $z$ and $\zb$ dependence
of the block largely decouple and the dependence on the dangerous variable $z\to 0$ is again exponential as in
eq.~(\ref{convexity argument}).
We conclude that if Lorentzian inversion converges
at two real points $(\Delta_A,J_A)$ and $(\Delta_B,J_B)$, it also converges
at all points with ${\rm Re}(\Delta)$ and ${\rm Re}(J)$ on or above the line segment joining them.


\bibliographystyle{JHEP}
\bibliography{ref}

\end{document}

%% file: lumps.pdf_tex
\begingroup%
  \makeatletter%
  \providecommand\color[2][]{%
    \errmessage{(Inkscape) Color is used for the text in Inkscape, but the package 'color.sty' is not loaded}%
    \renewcommand\color[2][]{}%
  }%
  \providecommand\transparent[1]{%
    \errmessage{(Inkscape) Transparency is used (non-zero) for the text in Inkscape, but the package 'transparent.sty' is not loaded}%
    \renewcommand\transparent[1]{}%
  }%
  \providecommand\rotatebox[2]{#2}%
  \newcommand*\fsize{\dimexpr\f@size pt\relax}%
  \newcommand*\lineheight[1]{\fontsize{\fsize}{#1\fsize}\selectfont}%
  \ifx\svgwidth\undefined%
    \setlength{\unitlength}{164.96708517bp}%
    \ifx\svgscale\undefined%
      \relax%
    \else%
      \setlength{\unitlength}{\unitlength * \real{\svgscale}}%
    \fi%
  \else%
    \setlength{\unitlength}{\svgwidth}%
  \fi%
  \global\let\svgwidth\undefined%
  \global\let\svgscale\undefined%
  \makeatother%
  \begin{picture}(1,0.68648628)%
    \lineheight{1}%
    \setlength\tabcolsep{0pt}%
    \put(0,0){\includegraphics[width=\unitlength,page=1]{lumps.pdf}}%
    \put(0.85887666,0.48739256){\makebox(0,0)[lt]{\lineheight{1.25}\smash{\begin{tabular}[t]{l}$t$\end{tabular}}}}%
    \put(0.92383674,0.37041352){\makebox(0,0)[lt]{\lineheight{1.25}\smash{\begin{tabular}[t]{l}$x$\end{tabular}}}}%
    \put(0,0){\includegraphics[width=\unitlength,page=2]{lumps.pdf}}%
    \put(0.73064212,0.05296773){\makebox(0,0)[lt]{\lineheight{1.25}\smash{\begin{tabular}[t]{l}1\end{tabular}}}}%
    \put(0.00388523,0.0601888){\makebox(0,0)[lt]{\lineheight{1.25}\smash{\begin{tabular}[t]{l}3\end{tabular}}}}%
    \put(-0.00333577,0.60852703){\makebox(0,0)[lt]{\lineheight{1.25}\smash{\begin{tabular}[t]{l}2\end{tabular}}}}%
    \put(0.7326361,0.61093403){\makebox(0,0)[lt]{\lineheight{1.25}\smash{\begin{tabular}[t]{l}4\end{tabular}}}}%
  \end{picture}%
\endgroup%